%% file: Preprint-25.tex
\documentclass[12pt,oneside,openany,article]{memoir}
\usepackage{mempatch}

\nouppercaseheads
\usepackage[amsthm,thmmarks,hyperref]{ntheorem}
\usepackage{etex}
\usepackage[english]{babel}

\usepackage{caption}

\usepackage[leqno]{mathtools}

\usepackage[scr]{rsfso}
\usepackage{upgreek}
\usepackage[compress,square,comma,numbers]{natbib}
\usepackage{hypernat} 

\usepackage{sfmath}
\usepackage{wrapfig}

\usepackage{microtype}

\usepackage{array}
\usepackage{tikz}

\usepackage{hyperxmp}
\usepackage{xr-hyper}
\usepackage{nameref}
\usepackage[pdftex,bookmarks,pdfnewwindow,plainpages=false,unicode]{hyperref}

\usepackage{bookmark}
\usepackage{enumitem}
\usepackage{amssymb}

\hypersetup{
colorlinks=true,
linkcolor=black,
citecolor=black,
urlcolor=blue!80!cyan,
pdfauthor={Victor Ivrii},
pdftitle={Asymptotics of the ground state energy of heavy molecules and related topics},
pdfsubject={Sharp Spectral Asymptotics},
pdfkeywords={Microlocal Analysis,  Sharp Spectral Asymptotics},
bookmarksdepth={4}
}

\theoremstyle{plain}
\newtheorem{theorem}{Theorem}[section]

\newtheorem{proposition}[theorem]{Proposition}
\newtheorem{corollary}[theorem]{Corollary}
\newtheorem{Problem}[theorem]{Problem}

\theoremstyle{definition}

\theoremstyle{remark}
\newtheorem{remark}[theorem]{Remark}

\makeatletter
\newtheoremstyle{plainfoot}%
  {\item[\hskip\labelsep \theorem@headerfont ##1\ ##2\,\footnotemark\theorem@separator]}%
  {\item[\hskip\labelsep \theorem@headerfont ##1\ ##2\ (##3)\, \footnotemark\theorem@separator]}
\makeatother

\theoremstyle{plainfoot}
\theoremseparator{.}
\newtheorem{theorem-foot}[theorem]{Theorem}
\newtheorem{lemma-foot}[theorem]{Lemma}
\newtheorem{proposition-foot}[theorem]{Proposition}
\newtheorem{corollary-foot}[theorem]{Corollary}
\newtheorem{conjecture-foot}[theorem]{Conjecture}
\newtheorem{condition-foot}[theorem]{Condition}

\theoremstyle{plainfoot}
\theoremseparator{.}
\theorembodyfont{}
\newtheorem{definition-foot}[theorem]{Definition}
\newtheorem{Problem-foot}[theorem]{Problem}

\theoremstyle{plainfoot}
\theoremseparator{.}
\theorembodyfont{}
\theoremheaderfont{\slshape}
\newtheorem{remark-foot}[theorem]{Remark}         
\newtheorem{example-foot}[theorem]{Example}
\newtheorem{problem-foot}[theorem]{Problem}

\numberwithin{equation}{section}

\newenvironment{phantomequation}[1][]{\refstepcounter{equation}}{}

\newenvironment{claim}[1][{\textup{(\theequation)}}]{\refstepcounter{equation}\vglue10pt
\begin{trivlist}
\item[{\hskip\labelsep#1}]}{\vglue10pt\end{trivlist}}

\newcommand{\ess}{{{\mathsf{ess}}}}

\newcommand{\Specess}{\operatorname{Spec_\ess}}
\newcommand{\error}{{\mathsf{error}}}

\newcommand{\HF}{{\mathsf{HF}}}
\newcommand{\DS}{{\mathsf{DS}}}
\newcommand{\Weyl}{{\mathsf{Weyl}}}
\newcommand{\cJ}{\mathcal{J}}
\newcommand{\cH}{\mathcal{H}}
\newcommand{\cI}{\mathcal{I}}

\newcommand{\const}{\mathsf{const}}
\newcommand{\TF}{{\mathsf{TF}}}

\newcommand{\W}{{\mathsf{W}}}
\newcommand{\E}{{\mathsf{E}}}
\newcommand{\N}{{\mathsf{N}}}
\newcommand{\n}{{\mathsf{n}}}
\newcommand{\Scott}{\mathsf{Scott}}
\newcommand{\Schwinger}{\mathsf{Schwinger}}
\newcommand{\Dirac}{\mathsf{Dirac}}

\newcommand{\I}{\mathsf{I}}
\newcommand{\y}{{\mathsf{y}}}
\newcommand{\bN}{\mathbb{N}}
\newcommand{\bR}{\mathbb{R}}
\newcommand{\bC}{\mathbb{C}}

\newcommand{\fH}{\mathfrak{H}}
\newcommand{\blangle}{{\boldsymbol{\langle}}}
\newcommand{\brangle}{{\boldsymbol{\rangle}}}
\newcommand{\cE}{\mathcal{E}}
\newcommand{\cN}{\mathcal{N}}
\newcommand{\cX}{\mathcal{X}}

\newcommand{\cZ}{\mathcal{Z}}
\newcommand{\sC}{\mathscr{C}}
\newcommand{\cQ}{\mathcal{Q}}
\newcommand{\sL}{\mathscr{L}}

\newcommand{\D}{\mathsf{D}}
\newcommand{\sfH}{{\mathsf{H}}}
\newcommand{\supp}{\operatorname{supp}}
\newcommand{\Spec}{\operatorname{Spec}}
\newcommand{\Tr}{\operatorname{Tr}}
\newcommand{\tr}{\operatorname{tr}}

\newcommand{\boldupsigma}{{\boldsymbol{\upsigma}}}

\addtopsmarks{headings}{}{%
\createmark{chapter}{right}{shownumber}{}{. \ }
}
\title{Asymptotics of the ground state energy of heavy molecules and related topics}
\author{Victor Ivrii}

\begin{document}

\maketitle

{\abstract%
We consider asymptotics of the ground state energy of heavy atoms and molecules and derive it including Schwinger and Dirac corrections.  We consider also related topics:  an excessive negative charge, ionization energy and excessive negative charge when atoms can still bind into molecules.
\endabstract}

\chapter{Introduction}%
\label{sect-25-1}

The purpose of this paper and several which will follow it  is to apply semiclassical methods developed in the Book of the author \cite{futurebook} to the theory of heavy atoms and molecules. Because of this we combine our semiclassical methods with the traditional methods of that theory, mainly function-analytic.

In this paper we consider the case without magnetic field. Next papers will be devoted to the cases of the self-generated magnetic field, strong external magnetic field and hyperstrong external magnetic field and combined external and self-generated fields. Basically this paper should be considered as an introduction. This paper is a revitalization of \cite{ivrii:MQT1} but in the case without magnetic field.

We explore the ground state energy, an excessive negative charge, ionization energy and excessive negative charge when atoms can still bind into molecules.

I learned all I know about Multiparticle Quantum Theory from Michael Sigal, Elliott Lieb, Philip Solovej. Some remarks of Volker Bach were very useful for the beginner in the field. I am extremely thankful to all of them.

\section{Framework}%
\label{sect-25-1-1}

Let us consider the following operator (quantum Hamiltonian)
\begin{gather}
\mathbf{H}=\mathbf{H}_N\coloneqq   \sum_{1\le j\le N} H _{V,x_j}+\sum_{1\le j<k\le N}|x_j-x_k| ^{-1}
\label{25-1-1}\\
\shortintertext{on}
\fH= \bigwedge_{1\le n\le N} \sfH, \qquad \sfH=\sL^2 (\bR^d, \bC^q) \label{25-1-2}\\
\shortintertext{with}
H_V =D ^2-V(x)
\label{25-1-3}
\end{gather}
describing $N$ same type particles in the external field with the scalar potential $-V$ (it is more convenient but contradicts notations of the previous papers), and repulsing one another according to the Coulomb law.

Here $x_j\in \bR ^d$ and $(x_1,\ldots ,x_N)\in\bR ^{Nd}$, potential $V(x)$ is assumed to be real-valued. Except when specifically mentioned we assume that
\begin{equation}
V(x)=\sum_{1\le m\le M} \frac{Z_m }{|x-\y_m|}
\label{25-1-4}
\end{equation}
where $Z_m>0$ and $\y_m$ are charges and locations of nuclei.

Mass is equal to $\frac{1}{2}$ and the Plank constant and a charge are equal to $1$ here. The crucial question is the quantum statistics.

\begin{claim}\label{25-1-5}
We assume that the particles (electrons) are \emph{fermions\/}\index{fermion}. This means that the Hamiltonian should be considered on the \emph{Fock space\/}\index{Fock space\/} $\fH$ defined by (\ref{25-1-2}) of the functions antisymmetric with respect to all variables $(x_1,\varsigma _1),\ldots, (x_N,\varsigma _N)$.
\end{claim}

Here $\varsigma \in \{1,\ldots,q\}$ is a \emph{spin variable\/}\index{spin variable}.

\begin{remark}\label{rem-25-1-1}
\begin{enumerate}[label=(\roman*), wide, labelindent=0pt]
\item\label{rem-25-1-1-i}
Meanwhile for \emph{bosons\/} one should consider this operator on the space of symmetric functions. The results would be very different from what we will get here. Since our methods fail in that framework, we consider only fermions here.

\item\label{rem-25-1-1-ii}
In this paper we do not have magnetic field and we can assume that $q=1$; for $q\ge 1$ no modifications of our arguments is required and results are the same albeit with different numerical coefficients. In the next papers we introduce magnetic field (external or self-generated) we will be interested in $d=3$, $q=2$ and
\begin{equation}
H_{V,A}=\bigl((i\nabla -A)\cdot \boldupsigma \bigr) ^2-V(x)
\label{25-1-6}
\end{equation}
where $\boldupsigma  =(\upsigma _1,\upsigma _2,\upsigma _3)$,
$\upsigma _k$ are Pauli matrices.
\end{enumerate}
\end{remark}

Let us assume that
\begin{claim}\label{25-1-7}
Operator $\mathbf{H}$ is self-adjoint on ${\fH}$.
\end{claim}

As usual we will never discuss this assumption.

\section{Problems to consider}
\label{sect-25-1-2}

We are interested in the \emph{ground state energy\/}\index{ground state energy} $\E=\E_N$ of our system i.e. in the lowest eigenvalue of the operator $\mathbf{H}=\mathbf{H}_N$ on $\fH$:
\begin{equation}
\E\coloneqq   \inf \Spec \mathbf{H}\qquad \text{on\ \ }\fH;
\label{25-1-8}
\end{equation}
more precisely we are interested in the asymptotics of
$\E_N=\E (\underline{\y};\underline{Z};N)$ as $V$ is defined by (\ref{25-1-4}) and $N\asymp Z\coloneqq   Z_1+Z_2+\ldots+Z_M\to \infty$ and we are going to prove that\footnote{\label{foot-25-1} Under reasonable assumption
$|\y_m -\y_{m'}|\gg Z^{-\frac{1}{3}}$ for all $m\ne m'$.} $\E$ is equal to \emph{Thomas-Fermi energy\/}\index{Thomas-Fermi energy} $\cE^\TF$ with Scott and Dirac-Schwinger corrections and with $o(Z^{\frac{5}{3}})$ error.

Here we use notations $\underline{\y}= (\y_1,\ldots,\y_M)$,
$\underline{Z}= (Z_1,\ldots,Z_M)$.

We are also interested in the asymptotics for the \emph{ionization energy\/}\index{ionization energy}
\begin{equation}
\I_N \coloneqq   \E_{N-1}-\E_N.
\label{25-1-9}
\end{equation}
It is well-known (see G.~Zhislin~\cite{zhislin:spectrum}) that $\I_N>0$ as $N\le Z$ (i.e. molecule can bind at least $Z$ electrons) and we are interested in the following question: estimate \emph{maximal excessive negative charge\/}\index{maximal excessive negative charge}
\begin{equation}
\max_{N:\ \I_N>0} (N-Z)
\label{25-1-10}
\end{equation}
i.e. \emph{how many extra electrons can bind a molecule?}.

All these questions so far were considered in the framework of the fixed positions $\y_1,\ldots, \y_M$ but we can also consider
\begin{gather}
\widehat{\E}= \widehat{\E}_N= \widehat{\E}(\underline{\y};\underline{Z}; N)=
\E + U(\underline{\y};\underline{Z})\label{25-1-11}\\
\shortintertext{with}
U(\underline{\y};\underline{Z})\coloneqq   \sum_{1\le m< m'\le M}\frac {Z_mZ_{m'}}{|\y_m-\y_{m'}|}\label{25-1-12}\\
\shortintertext{and}
\widehat{\E} (\underline{Z}; N)=\inf_{\y_1,\ldots,\y_M} \widehat{\E}(\underline{\y};\underline{Z}; N) \label{25-1-13}
\end{gather}
and replace $\I_N$ by $\widehat{\I}_N = -\widehat{\E}_N + \widehat{\E}_{N-1}$ and modify all our questions accordingly. We call these frameworks \emph{fixed nuclei model\/}\index{fixed nuclei model} and \emph{free nuclei model\/}\index{free nuclei model} respectively.

In the free nuclei model we can consider two other problems:
\begin{enumerate}[label=(\alph*), wide, labelindent=0pt]
\item\label{sect-25-1-2-a}
Estimate from below \emph{minimal distance between nuclei} i.e.
\begin{equation*}
a\coloneqq \min_{1\le m < m'\le M} |\y_m-\y_{m'}|
\end{equation*}
for which such minimum is achieved;
\item\label{sect-25-1-2-b}
Estimate \emph{maximal excessive positive charge\/}\index{maximal excessive positive charge}
\begin{equation}
\max_N \bigl\{Z-N\colon   \widehat{\E} <
\min_{\substack{\\[3pt]N_1,\ldots, N_M: \\[3pt] N_1+\ldots N_M=N}}\;
\sum_{1\le m\le M} \ \E (Z_m; N_m)\bigr\}
\label{25-1-14}
\end{equation}
for which molecule does not disintegrates into atoms\footnote{\label{foot-25-2} One can ask the same question about disintegration into smaller molecules but our methods are too crude to distinguish between such questions.}.
\end{enumerate}

\section{Thomas-Fermi theory}%
\label{sect-25-1-3}
The first approximation is the Thomas-Fermi theory. Let us introduce the \emph{spacial  density\/}\index{density!spacial} of the particle with the state
$\Psi \in {\fH}$:
\begin{equation}
\rho (x)=\rho _\Psi (x)=N
\int |\Psi (x,x_2,\ldots ,x_N)| ^2 \,dx_2\cdots dx_N
\label{25-1-15}
\end{equation}
where $|\cdot|$ means a norm in $\bC^{Nq}$ and antisymmetricity of $\Psi $ implies that it does not matter what variable $x_j$ is replaced by $x$ while in the general case one should sum on $j=1,\ldots,N$. Let us write the Hamiltonian, describing the corresponding ``quantum liquid'':
\begin{gather}
\cE(\rho )=\int \tau (\rho (x)) \,dx- \int V(x)\rho (x)\, dx+\frac{1}{2}
\D(\rho ,\rho ),\label{25-1-16}\\
\shortintertext{with}
\D(\rho ,\rho )=\iint |x-y| ^{-1}\rho (x)\rho (y)\,dxdy\label{25-1-17}
\end{gather}
where $\tau $ is the energy density of a gas of noninteracting electrons. Namely,
\begin{equation}
\tau (\rho )=\sup _{w\ge 0} \bigl(\rho w-P (w)\bigr)
\label{25-1-18}
\end{equation}
is the Legendre transform of the \emph{pressure\/}\index{pressure} $P(w)$ given by the formula
\begin{equation}
P (w)=\varkappa _1 w_+ ^{\frac{d}{2}+1},\qquad
\varkappa _1=2 (2\pi)^{-d}(d+2)^{-1}\varpi_d q.
\label{25-1-19}
\end{equation}

The classical sense of the second and the third terms in the right-hand
expression of (\ref{25-1-16}) is clear and the density of the kinetic energy is given by $\tau (\rho )$ in the semiclassical approximation (see Remark~\ref{rem-25-1-2}). So, the problem is
\begin{claim}\label{25-1-20}
Minimize functional $\cE(\rho)$ defined by (\ref{25-1-16}) under restrictions:
\begin{phantomequation}\label{25-1-21}\end{phantomequation}
\begin{equation}
\rho \ge 0,\qquad \int \rho \,dx\le N. \tag*{$\textup{(\ref*{25-1-21})}_{1,2}$}
\end{equation}
\end{claim}
\vglue-10pt
The solution if exists is unique because functional $\cE(\rho )$ is strictly convex (see below). The existence and the property of this solution denoted further by $\rho ^\TF$ is known in the series of physically important cases.

\begin{remark}\label{rem-25-1-2}
If $w$ is the negative potential then
\begin{equation}
\tr e(x,x,0)\approx P'(w)
\label{25-1-22}
\end{equation}
defines the density of all non-interacting particles with negative energies at point $x$ and
\begin{equation}
\int ^0_{-\infty} \tau \,d_\tau \tr e(x,x,\tau) dx \approx -\int P(w)\,dx
\label{25-1-23}
\end{equation}
is the total energy of these particles; here $\approx$ means ``in the semiclassical approximation''.
\end{remark}

We consider in the case of $d=3$ a large (heavy) molecule with potential (\ref{25-1-4}). It is well-known\footnote{\label{foot-25-3} E.~Lieb, ``Thomas-fermi and related theories of atoms and molecules'', \cite{lieb:selecta}, pp. 263--301.} that

\begin{proposition}\label{prop-25-1-3}
\begin{enumerate}[label=(\roman*), wide, labelindent=0pt]
\item\label{prop-25-1-3-i}
For $V(x)$ given by \textup{(\ref{25-1-4})} minimization problem \textup{(\ref{25-1-20})} has a unique solution $\rho=\rho^\TF$; then denote $\cE^\TF \coloneqq   \cE (\rho^\TF)$;

\item\label{prop-25-1-3-ii}
Equality in $\textup{(\ref{25-1-21})}_2$ holds if and only if
$N\le Z\coloneqq   \sum_m Z_m$.

\item\label{prop-25-1-3-iii}
Further, $\rho^\TF$ does not depend on $N$ as $N\ge Z$.

\item\label{prop-25-1-3-iv}
Thus
\begin{equation}
\int \rho^\TF \,dx =\min (N,Z),\qquad Z\coloneqq   \sum_{1\le m\le M} Z_m.
\label{25-1-24}
\end{equation}
\end{enumerate}
\end{proposition}

\section{Main results sketched and plan of the paper}%
\label{sect-25-1-4}

In the first half of the paper we derive asymptotics for ground state energy and justify Thomas-Fermi theory.

First of all, in Section~\ref{sect-25-2} we reduce the calculation of $\E$ to calculation of $\N_1 (H_W-\nu)$ and to estimate for
$\D \bigl(e(x,x,\nu) -\rho, e(x,x,\nu)-\rho\bigr)$ where
${\N_1 (H_W-\nu)=\Tr ((H_W-\nu)^-)}$ is the sum of the negative eigenvalues of operator $H_W-\nu$ $H_W=D^2 -W$, $W=W^\TF$, $\rho=\rho^\TF$ are Thomas-Fermi potential and Thomas-Fermi density respectively (or their appropriate approximations), $\nu$ is either $\lambda_N$ ($N$-th eigenvalue of $H_W$) or its appropriate approximation and $e(x,y,\nu)$ is the Schwartz kernel of $E(\nu)$ which is the spectral projector of $H_W$.

Section~\ref{sect-25-3} is devoted to the systematic presentation of the Thomas-Fermi theory.

Further, in Section~\ref{sect-25-4} we apply our standard semiclassical arguments and calculate $\N_1 (H_W-\nu)$ and estimate
$\D \bigl(e(x,x,\nu) -\rho, e(x,x,\nu)-\rho\bigr)$ and also $|\lambda_N -\nu|$ where now $\nu$ is the chemical potential (which is the Thomas-Fermi approximation to $\lambda_N$). As a result under appropriate restrictions to $N$, $Z$ and
\begin{gather}
a\coloneqq   \min_{j\ne k}|\y_j-\y_k|\gg Z^{-\frac{1}{3}}
\label{25-1-25}\\
\shortintertext{we prove that}
\E = \cE ^\TF +\Scott+ \Dirac+\Schwinger + o(Z^{\frac{5}{3}})\label{25-1-26}\\
\shortintertext{and}
\D( \rho_\Psi - \rho^\TF, \rho_\Psi - \rho^\TF) = o(Z^{\frac{5}{3}})
\label{25-1-27}\\
\shortintertext{where}
\Scott=q \sum_{1\le m\le M} Z_m ^2
\label{25-1-28}\\
\Dirac= -\frac{9}{2}(36\pi ) ^{\frac{2}{3}}
q^{\frac{2}{3}} \int (\rho ^\TF )^\frac{4}{3}\,dx,
\label{25-1-29}\\
\Schwinger= (36\pi ) ^{\frac{2}{3}} q^{\frac{2}{3}}
\int (\rho ^\TF )^{\frac{4}{3}}\,dx
\label{25-1-30}
\end{gather}
and $\Psi$ is the ground state.

\begin{remark}\label{rem-25-1-4}
\begin{enumerate}[label=(\roman*), wide, labelindent=0pt]
\item\label{rem-25-1-4-i}
Actually we will recover even slightly better remainder estimate $O(Z^{\frac{5}{3}-\delta})$ in (\ref{25-1-26}) and (\ref{25-1-27}) as $a\ge Z^{-\frac{1}{3}+\delta_1}$.

\item\label{rem-25-1-4-ii}
Condition $a\gtrsim Z^{-\frac{1}{3}}$ bans nuclei to be so close that the repulsion energy between them be much larger than the total energy of
all the electrons. Estimates in case when this condition is violated will be also proven;

\item\label{rem-25-1-4-iii}
Keeping in mind that there is no binding in Thomas-Fermi theory (and this statement could be quantified) one gets immediately that in the free nuclei model $a\ge Z^{-\frac{5}{21}}$ and therefore remainder estimate $O(Z^{\frac{5}{3}-\delta})$ holds.

\item\label{rem-25-1-4-iv}
Due to scaling in the Thomas-Fermi theory (see proposition~\ref{prop-25-3-3})
$\cE^\TF\sim q^{\frac{2}{3}}Z^{\frac{7}{3}}=q^3 (q^{-1}Z)^{\frac{3}{7}}$, $\Scott \sim q Z^2=q^3 (q^{-1}Z)^2$, and both $\Dirac$ and $\Schwinger$ are $\sim q^{\frac{4}{3}}Z^{\frac{5}{3}}=q^3(q^{-1}Z)^{\frac{5}{3}}$.
\end{enumerate}
\end{remark}

In the second half of the paper we apply estimate (\ref{25-1-27}) to investigate negatively and positively charged systems. In Section~\ref{sect-25-5} we consider negatively charged systems and derive an upper estimate for the excessive negative charge $(N-Z)$ such that $\I_N>0$ and ionization energy $\I_N$ itself.

In Section~\ref{sect-25-6} we derive upper and lower estimates for  $\I_N+\nu$ and an upper estimate for the excessive positive charge for which in the framework of the free nuclei model $a<\infty$.\enlargethispage{2\baselineskip}

\chapter{Reduction to semiclassical theory}%
\label{sect-25-2}

To justify the heuristic formula
$\E \sim \cE^\TF =\cE(\rho ^\TF )$ and to find an error estimate let us deduce the lower and upper estimates for $\E$.

\section{Lower estimate}%
\label{sect-25-2-1}

For the lower estimate we apply the \emph{electrostatic inequality\/}\index{electrostatic inequality} due
to E.~H.~Lieb:
\begin{multline}
\sum_{1\le j<k\le N}
\int |x_j-x_k| ^{-1}|\Psi (x_1,\dots ,x_N)| ^2\, dx_1 \cdots dx_N\ge\\
\frac{1}{2}\D(\rho_\Psi ,\rho_\Psi )-C\int \rho_\Psi ^{\frac{4}{3}}(x)\, dx
\label{25-2-1}
\end{multline}
with $\rho_\Psi$ defined by (\ref{25-1-15}).

\begin{remark}\label{rem-25-2-1}
Inequality (\ref{25-2-1}  holds for all (not necessarily antisymmetric) functions $\Psi $ with $\|\Psi \|_{\sL ^2(\bR ^{3N})}=1$.
\end{remark}
Therefore
\begin{multline}
\blangle \mathsf{H}_N\Psi ,\Psi \brangle \ge
\sum_{1\le j\le N} \blangle H_{V,x_j}\Psi ,\Psi \brangle +
\frac{1}{2}\D\bigl(\rho_\Psi ,\rho_\Psi )-
C\int \rho_\Psi ^{\frac{4}{3}}(x)\,dx=\\[3pt]
\sum_{1\le j\le N}
\blangle H_{W,x_j}\Psi ,\Psi \brangle +
\frac{1}{2}\D\bigl(\rho_\Psi - \rho ,\rho_\Psi -\rho \bigr)-
\frac{1}{2}\D\bigl( \rho , \rho \bigr)-C\int \rho_\Psi ^{\frac{4}{3}}(x)\,dx
\label{25-2-2}
\end{multline}
where $\blangle \cdot,\cdot\brangle $ means the inner product in $\fH$ and $H_W$ is one-particle Schr\"odinger operator with the potential
\begin{equation}
W=V-|x| ^{-1}*\rho ,
\label{25-2-3}
\end{equation}
where $\rho$ is an arbitrary chosen real-valued non-negative function.

The physical sense of the second term in $W$ is transparent: it is a potential created by a charge $-\rho$. Skipping the positive second term in the right-hand expression of (\ref{25-2-2}) and believing that the last term is not very important for the ground state function $\Psi $\,\footnote{\label{foot-25-4} When we derive also an upper estimate for $\E$ we will get an upper estimate for this term as a bonus.} we see that we need to estimate from below the first term.

Here assumption that $\Psi $ is antisymmetric is crucial. Namely, for general (or symmetric--does not matter) $\Psi $ the best possible estimate is
$N\lambda _1$ where $\lambda _1$ is the lowest eigenvalue of $H_W$ (we always assume that there is sufficiently many eigenvalues under the bottom of the essential spectrum of $H_W$) and we cannot apply semiclassical theory. However, for antisymmetric $\Psi $ situation is rather different.

Namely, let $\lambda_1 \le \lambda_2 \le \lambda_3\le \ldots $ be negative eigenvalues of $H_W$ (on $\sfH=\sL^3(\bR^3, \bC^q)$. Then the first term in the right-hand expression of (\ref{25-2-2}) is bounded from below by
\begin{equation}
\sum_{1\le j\le N}\lambda _j=\N_1(H_W-\bar{\lambda})+\bar{\lambda}N
\label{25-2-4}
\end{equation}
where $\N(B)$, $\N_1(B)=\Tr (B^-)$ are the number and the sum of all the negative eigenvalues of operator $B$ respectively such that
$\Specess (B)\subset \bR ^+$ provided $\bar{\lambda}=\lambda _N<0$; the latter assumption is equivalent to
\begin{equation}
\N(H_W) \ge N.
\label{25-2-5}
\end{equation}
Applying the semiclassical approximation (which needs to be justified!) one gets
\begin{align}
&\N_1(H_W-\bar{\lambda})=\cN _1(H_W-\bar{\lambda})+ \error_1
\label{25-2-6}\\
\shortintertext{with}
&\cN _1(H_W- \bar{\lambda})\coloneqq   -\int P\bigl(W(x)+\bar{\lambda}\bigr)\,dx
\label{25-2-7}
\end{align}
and $\error_0$ \emph{an error in the semiclassical approximation\/} for $\N_1(H_W-\bar{\lambda})$. Therefore the lower estimate for the ground state energy is
\begin{equation}
\E \ge -\int P (W+\bar{\lambda})\,dx
+\bar{\lambda}N -\frac{1}{2}\D( \rho , \rho )-\error
\label{25-2-8}
\end{equation}
where $\error$ now includes both an estimate for
$\int \rho_\Psi ^\frac{4}{3}\,dx$ and the semiclassical remainder estimate.

Furthermore, applying a semiclassical approximation for the number
$\N(H_W-\bar{\lambda})$ of eigenvalues below $\bar{\lambda}$ (and this number should be approximately $N$) one gets an equality
\begin{align}
&N= \cN (H_W-\bar{\lambda})+\error_0\label{25-2-9}\\
\shortintertext{with}
& \cN (H_W-\bar{\lambda})\coloneqq   \int P' \bigl(W(x)+\bar{\lambda}\bigr)\,dx
\label{25-2-10}
\end{align}
and $\error_0$ \emph{an error in the semiclassical approximation\/}  for $\N(H_W-\bar{\lambda})$.

To get the best possible lower estimate  one should pick up $\rho$ delivering maximum to the functional
\begin{equation}
-\int P \bigl(W(x)+\nu\bigr)\,dx +\nu N -\frac{1}{2}\D(\rho ,\rho )
\label{25-2-11}
\end{equation}
($\nu =\bar{\lambda}$ here) under assumptions $\textup{(\ref{25-1-21})}_{1,2}$ and (\ref{25-2-3}) as we skip all the errors.

One can see that the optimal choice is the \emph{Thomas-Fermi potential $W^\TF$ and density $\rho^\TF$\/}.
\index{Thomas-Fermi!potential}\index{potential!Thomas-Fermi}\index{Thomas-Fermi!density}\index{density!Thomas-Fermi}%
The above arguments are very standard in MQT with
$\rho =\rho ^\TF, W=W ^\TF$\,\footnote{\label{foot-25-5} Or some their close approximations.} from the very beginning.

On the other hand, let us consider the Euler-Lagrange equation for
$\rho =\rho ^\TF$ under condition $\int \rho\, dx=N$:
\begin{equation}
\tau ' (\rho )-W=\nu \quad (\rho >0),
\qquad W=V-|x| ^{-1}*\rho
\label{25-2-12}
\end{equation}
with the Lagrange factor $\nu $\,\footnote{\label{foot-25-6} Called \emph{chemical potential\/}\index{chemical potential} and in contrast to $\bar{\lambda}$ belonging to Thomas-Fermi theory.}.
Expressing $\rho $ and integrating we get
\begin{equation}
N={\cN}(H_W-\nu )=\int P' \bigl(W(x)+\nu \bigr)\,dx.
\label{25-2-13}
\end{equation}
Comparing (\ref{25-2-12}) and (\ref{25-2-13}) we get that with some error
$\bar{\lambda}\sim\nu $. Substituting to the first term in (\ref{25-2-6})
$\bar{\lambda}=\nu $ and $\nu -W=-\tau '_B(\rho )$ we get the lower
estimate $\E\ge \cE ^\TF - \error $.

\begin{remark}\label{rem-25-2-2}
\begin{enumerate}[label=(\roman*), wide, labelindent=0pt]
\item\label{rem-25-2-2-i}
Instead of (\ref{25-2-6}) we will use a better estimate\footnote{\label{foot-25-7} With $\cE^\TF $ replaced by $\cE^\TF +\Scott+\Schwinger$ and with much smaller $\error_1$ than for a simple semiclassical approximation.};

\item\label{rem-25-2-2-ii}
To minimize errors we will also recalculate (\ref{25-2-4}) effectively replacing $\bar{\lambda}$ by $\nu$:
\begin{equation}
\sum_{1\le j\le N}\lambda _j =
\sum_{1\le j\le N}(\lambda _j-\nu )+\nu N\ge
\N_1(H_W-\nu )+\nu N.
\label{25-2-14}
\end{equation}
The advantage is that we even do not mess up with the semiclassical asymptotics for $\N(H_W-\nu )$. Further, one can replace here $N$ by $\int \rho ^\TF\,dx$: (these quantities fail to be equal only for $N>Z$ i.e. for $\nu =0$).
\item\label{rem-25-2-2-iii}
Recall that we assumed that $\lambda_N <0$ i.e. (\ref{25-2-5}) holds. In the opposite case
\begin{equation}
\N(H_W) < N.
\label{25-2-15}
\end{equation}
we estimate the first term in the right-hand expression of (\ref{25-2-2}) from below by $\N_1 (H_W)$ i.e. we will get the same formula but with $\nu=0$.
\end{enumerate}
\end{remark}
\enlargethispage{\baselineskip}

\section{Upper estimate}%
\label{sect-25-2-2}

To get the upper estimate one takes a test function $\Psi (x_1,\ldots ,x_N)$ which is not a ground state here but an antisymmetrization with respect to $(x_1,\ldots ,x_N)$ of the product $\phi _1(x_1)\cdots \phi _N(x_N)$ where
$\phi _1,\dots,\phi _N$ are orthonormal eigenfunctions of $H_W$ corresponding to eigenvalues $\lambda _1,\ldots ,\lambda _N$, provided $\lambda_N<0$. Namely this function minimizes the first term in the right-hand expression of (\ref{25-2-2}).

One can write
\begin{equation}
\Psi =\frac{1}{N!} \det \bigl(\phi _i(x_j)\bigr)_{i,j=1,\ldots ,N}
\label{25-2-16}
\end{equation}
and it is called the \emph{Slater determinant\/}\index{Slater determinant}. Obviously, $\|\Psi \|=1$ and
\begin{gather}
\rho _\Psi (x)=\tr e_N(x,x)
\label{25-2-17}\\
\shortintertext{where}
e_N(x,y)=\sum_{1\le j\le N} \phi _j(x) \phi^\dag _j(y)
\label{25-2-18}
\end{gather}
is the Schwartz kernel of the projector to the subspace spanned on
$\{\phi _j\}_{1\le j\le N}$.

\begin{remark}\label{rem-25-2-3}
If $q\ge 2$ then $\phi_j=\phi_j(x,\varsigma)$ and
$\Psi (x_1,\varsigma_1;\ldots ;x_N,\varsigma_N)$ is an antisymmetrization with respect to $(x_1,\varsigma_1;\ldots ;x_N,\varsigma_N)$ of the product
$\phi _1(x_1,\varsigma_1)\cdots \phi _N(x_N,\varsigma_N)$.
\end{remark}

Easy calculations show that
\begin{multline}
\blangle \mathbf{H}\Psi ,\Psi \brangle =
\sum_{1\le j\le N}\lambda _j
+\frac{1}{2}\D (\rho_\Psi -\rho , \rho_\Psi -\rho )\\
-\frac{1}{2}\D (\rho,\rho )
-\frac{1}{2}\iint |x-y| ^{-1}\tr e^\dag_N(x,y) e_N(x,y) \, dxdy.
\label{25-2-19}
\end{multline}
The first term in the right-hand expression again is equal to the middle expression in (\ref{25-2-4}) which does not exceed
\begin{equation}
\N_1 (H_W-\nu) + \nu N+ |\lambda_N-\nu| \cdot |\N(H_W-\nu)-N|.
\label{25-2-20}
\end{equation}
Really, we need to consider (non-zero) terms which do not cancel in
\begin{equation}
\sum_{j\le N}(\lambda_j-\nu)- \sum_{\lambda_j<\nu} (\lambda_j-\nu)
\label{25-2-21}
\end{equation}
and their absolute value does not exceed $|\lambda_N-\nu|$ while their number does not exceed $|\N(H_W-\nu)-N|$.

Again, discounting all the errors and considering semiclassical approximation (including $\rho_\Psi (x)\sim P'\bigl(W(x)+\nu)\bigr)$ we arrive to a functional
\begin{multline}
-\int P\bigl(W(x)+\nu\bigr) \,dx +\nu N -\frac{1}{2}\D(\rho,\rho)+\\
\frac{1}{2}\D (P'(W+\nu)-\rho, P'(W+\nu)-\rho)
\label{25-2-22}
\end{multline}
which needs to be minimized under assumptions $\textup{(\ref{25-1-21})}_{1,2}$ and (\ref{25-2-3}). This functional differs from (\ref{25-2-11}) which was minimized by the last term. One can prove that (\ref{25-2-22}) minimizes as $\rho=\rho^\TF$, $W=W^\TF$ and $\nu$ is a chemical potential. So again we may pick them (or their appropriate approximations) up from the very beginning.

Therefore in addition to a semiclassical $\error_1$\,\footref{foot-25-7} of the previous subsection we need to consider also semiclassical errors
\begin{gather}
\D \bigl( \tr e(x,x,\nu)- P'(W+\nu) ,\tr e(x,x,\nu)- P'(W+\nu)\bigr),
\label{25-2-23}\\
\D \bigl( e(x,x,\nu)- e_N(x,x) , e(x,x,\nu)- e_N(x,x)\bigr)
\label{25-2-24}\\
\intertext{where $e(x,y,\nu)$ is the Schwartz kernel of the spectral projector $\uptheta (\tau -H_W)$ of $H_W$,}
N (H_W-\nu) -\int P'(W+\nu)\,dx
\label{25-2-25}\\
\shortintertext{and}
\lambda_N-\nu .
\label{25-2-26}
\end{gather}

\begin{remark}\label{rem-25-2-4}
\begin{enumerate}[label=(\roman*), wide, labelindent=0pt]
\item\label{rem-25-2-4-i}
Recall that we assumed that $\lambda_N <0$ i.e. (\ref{25-2-5}) holds. In the opposite case (\ref{25-2-15}) selecting appropriate $\phi_j(x)$ with $j=\N(H_W)+1,\ldots ,N$ we with arbitrarily small error estimate the first term in the right-hand expression of (\ref{25-2-2}) from above by $\N_1 (H_W)$ as i.e. we will get the same formula but with $\nu=0$ and we also will need to estimate (\ref{25-2-23}) with $\nu=0$.

\item\label{rem-25-2-4-ii}
To make this case compatible with the case (\ref{25-2-5}) we will need to estimate $|\nu|$ (and $|N-Z|$) under assumption (\ref{25-2-15}); we will also compare $\cE^\TF$ calculated for such $\nu$ (or, equivalently, $N$ as they are connected) and $\nu=0$ (and $N=Z$).

\item\label{rem-25-2-4-iii}
Sure $\rho^\TF$ and $W^\TF$ depend on $\nu$ (or $N$) but we will prove that for $N-Z$ relatively small we can do all calculations as $\nu=0$ (and $N=Z$).

\item\label{rem-25-2-4-iv}
If we are interested in the estimate for
$\D(\rho_\Psi-\rho^\TF, \rho_\Psi-\rho^\TF)$ where $\Psi$ is the ground state, we do not need to calculate a semiclassical error in $\N_1 (H_W-\nu)$. In fact, we can simply stick with $\N_1 (H_W-\bar{\lambda})$ with $\bar{\lambda}=\lambda_N$ under assumption (\ref{25-2-5}) and $\bar{\lambda}=0$ otherwise. As a result in certain cases our estimate for
$\D(\rho_\Psi-\rho^\TF, \rho_\Psi-\rho^\TF)$ will be better than the error in an approximation for $\E$ and we need the former rather than the latter for the results of the second half of this paper. Especially significant the  difference will be when we introduce magnetic field.
\end{enumerate}
\end{remark}

\section{Remarks and Dirac correction}
\label{sect-25-2-3}

Now almost everything is in framework of the theory we developed; the only missing is an estimate
\begin{equation}
\int \rho _\Psi ^\frac{4}{3}dx\le CZ^{\frac{5}{3}}
\label{25-2-27}
\end{equation}
for a reasonable candidate $\Psi$ to the ground state; one can find it in E.~Lieb's \emph{Selecta\/}\footref{foot-25-3}.

However if we want a more sharp asymptotics with Dirac--Schwinger terms, we need a remainder estimate $o(Z^{\frac{5}{3}})$ or better; luckily there is improved electrostatic inequality due to Theorem 1, G.~Graf and J.~P.~Solovej~\cite{graf:solovej} (see also V.~Bach~\cite{bach}).

\begin{theorem}\label{thm-25-2-5}
Let $N\ge \epsilon Z$. Then for the ground state $\Psi $
\begin{align}
\E _{\HF}\ge \E &\ge \E_{\HF}- CZ^{\frac{5}{3}-\delta }\label{25-2-28}\\
\shortintertext{and}
\E &\ge \E _{\DS}- CZ^{\frac{5}{3}-\delta }\label{25-2-29}
\end{align}
with some exponent $\delta >0$ where
\begin{equation}
\E _{\HF} \coloneqq   \inf _\Psi E_{\HF}(\Psi),
\label{25-2-30}
\end{equation}
where in \textup{(\ref{25-2-30})} $\Psi$ runs through Slater determinants\,\footnote{\label{foot-25-8} Albeit not necessarily of eigenfunctions of $H_{W}$.} and
\begin{multline}
E_{\HF}(\Psi)\coloneqq   \sum_{1\le j\le N} \blangle H_{V,x_j}\Psi ,\Psi \brangle +
\frac{1}{2}\D\bigl(\rho_\Psi ,\rho_\Psi )-\\
\frac{1}{2}\iint |x-y| ^{-1}\tr e^\dag_N(x,y) e_N(x,y)\, dxdy,
\label{25-2-31}
\end{multline}
\begin{equation}
E _\DS \coloneqq   \sum_{j: 1\le j\le N; \lambda_j<0} \lambda_j -\frac{1}{2}\D (\rho^\TF,\rho^\TF) - \kappa_\Dirac \rho^{\TF, \frac{4}{3}}\,dx,
\label{25-2-32}
\end{equation}
$\kappa_\Dirac =(2\pi)^{-3}qc_\TF^2$, $c_\TF=(6\pi^2/q^2)^{\frac{2}{3}}$ is a Dirac constant.
\end{theorem}

Here (\ref{25-2-28})--(\ref{25-2-31}) are (1.15), (1.16), (1.8), (1.6) respectively and (\ref{25-2-32}) is a combination of (1.12) and (\ref  {25-3-30}) of this paper\footnote{\label{foot-25-9} We do not have a coefficient $\frac{1}{2}$ in the definition of $\D(.,.)$ but G.~Graf and J.~P.~Solovej~\cite{graf:solovej} have.}. Actually we need only (\ref{25-2-29}) and (\ref{25-2-32}).

As we are going to prove that the last terms in (\ref{25-2-31}) and (\ref{25-2-19}) coincide modulo $O(Z^{\frac{5}{3}-\delta})$ we made a necessary step completely.

\chapter{Thomas-Fermi theory}%
\label{sect-25-3}

Thomas-Fermi theory is well-developed in the no-magnetic-field case.
We cannot suggest any better reading than E.~Lieb's \emph{Selecta\/}\footref{foot-25-3}.

In the Thomas-Fermi theory $N$ is a real nonnegative number (not necessarily an integer).

\section{Existence}%
\label{sect-25-3-1}

Let us recall that in order to get the best lower estimate (neglecting  semiclassical errors) one needs to maximize
\begin{equation}
\Phi _* (W+\nu )\coloneqq  -\int P (W+\nu )\,dx -\frac{1}{8\pi}\|\nabla (W-V)\| ^2
\label{25-3-1}
\end{equation}
given by (\ref{25-2-11}) where we used equalities
\begin{gather}
\D(\rho ,\rho )=-(\rho ,W-V)=\frac{1}{4\pi}\|\nabla (W-V)\| ^2,\label{25-3-2}\\
\rho \coloneqq   \frac{1}{4\pi} \Delta (W-V),\label{25-3-3}
\end{gather}
$\|.\|$ means $\sL ^2$-norm and $W\to 0$ as $|x|\to \infty $.

On the other hand, to get the best possible upper estimate (neglecting semiclassical errors) one needs to minimize
\begin{gather}
\Phi ^* (\rho ',\nu )\coloneqq
\int \bigl(\tau (\rho ')-V\rho '\bigr)dx+\frac{1}{2}\D(\rho ',\rho ')
-\nu \int \rho 'dx
\label{25-3-4}
\shortintertext{where}
\rho '\coloneqq   P' (W+\nu )
\label{25-3-5}
\end{gather}
and $\tau (\rho)$ the Legendre transformation (\ref{25-1-18}) of $P$. Recall that according to (\ref{25-1-19})\begin{phantomequation}\label{25-3-6}\end{phantomequation}
\begin{gather}
P(w)= \frac{q}{15\pi^2}w_+^{\frac{5}{2}},\qquad
P'(w)= \frac{q}{6\pi^2}w_+^{\frac{3}{2}} \tag*{$\textup{(\ref*{25-3-6})}_{1,2}$}\label{25-3-6-*}\\
\shortintertext{and therefore}
\tau(\rho)= \frac{3}{5}
\bigl(6\pi^2 q^{-1}\bigr)^{\frac{2}{3}}\rho^{\frac{5}{3}}.
\label{25-3-7}
\end{gather}
\enlargethispage{2\baselineskip}

\begin{proposition}\label{prop-25-3-1}
In our assumptions for any fixed $\nu \le 0$
\begin{enumerate}[label=(\roman*), wide, labelindent=0pt]

\item\label{prop-25-3-1-i}
$\Phi_* (W+\nu )$ is a strictly concave functional.

\item\label{prop-25-3-1-ii}
$\Phi ^*(\rho )$ is a strictly convex functional.

\item\label{prop-25-3-1-iii}
$\Phi _* (W+\nu )\le \Phi ^*(\rho,\nu )$ for any $\rho \ge 0$ and $W$.

\item\label{prop-25-3-1-iv}
These extremal problems have a common solution $W$ and $\rho $ and
\begin{gather}
\rho =\frac{1}{4\pi}\Delta (W-V)=P '(W+\nu ), \label{25-3-8}\\
W=o(1)\qquad \text{as\ \ }|x|\to \infty .\label{25-3-9}
\end{gather}

\item\label{prop-25-3-1-v}
On the other hand, solution of \textup{(\ref{25-3-8})}--\textup{(\ref{25-3-9})} is the solution of the both extremal problems.

\item\label{prop-25-3-1-vi}
Neither of these problem has a solution for $\nu >0$.

\item\label{prop-25-3-1-vii}
Function
\begin{equation}
{\cN}(\nu )=\int P' (W+\nu )\, dx
\label{25-3-10}
\end{equation}
is continuous and monotone increasing at $(-\infty ,0]$ with
${\cN}(\nu )\to 0$ as $\nu \to -\infty $ and ${\cN}(0)=Z$.

\item\label{prop-25-3-1-viii}
For $\nu $ and $N$ linked by $N={\cN}(\nu )$ solutions of the
problem above coincide with $\rho ^\TF, W^\TF$ of the problem
\textup{(\ref{25-1-20})} and one can skip condition $\textup{(\ref{25-1-20})}_2$ for $N\ge Z$ and
\begin{equation}
\cE ^\TF=\Phi (W ^\TF+\nu )+\nu N=
\Phi ^*(\rho ^\TF,\nu )+\nu N.
\label{25-3-11}
\end{equation}
\end{enumerate}
\end{proposition}

\begin{proof} The proof of Statements~\ref{prop-25-3-1-i} and \ref{prop-25-3-1-ii} is obvious; therefore
both problems have unique solutions. Comparing Euler-Lagrange equations
we get that these solutions coincide which yields Statements~\ref{prop-25-3-1-iv} and \ref{prop-25-3-1-iii}.

Proof of  Statements~\ref{prop-25-3-1-v}--\ref{prop-25-3-1-viii} is also rather obvious.
\end{proof}

\begin{proposition}\label{prop-25-3-2}
For arbitrary $W$ the following estimates hold with absolute constants
$\epsilon _0>0$ and $C_0$:
\begin{multline}
\epsilon _0 \D(\rho -\rho ^\TF,\rho -\rho ^\TF)\le
\Phi_* (W ^\TF+\nu )-\Phi_* (W+\nu )\le \\
C_0 \D(\rho -\rho ',\rho -\rho ')\label{25-3-12}
\end{multline}
and
\begin{multline}
\epsilon _0 \D(\rho '-\rho ^\TF,\rho '-\rho ^\TF)\le
\Phi ^*(\rho ,\nu )-\Phi ^*(\rho ^\TF,\nu )\le\\
C_0 \D(\rho -\rho ',\rho -\rho ')\label{25-3-13}
\end{multline}
with $\rho =\frac{1}{4\pi}\Delta (W-V)$, $\rho '=P'(W+\nu )$.
\end{proposition}

\begin{proof} This proof is rather obvious as well.
\end{proof}

\section{Properties}
\label{sect-25-3-2}

\begin{proposition}\label{prop-25-3-3}
The solution of the Thomas-Fermi problem has following scaling properties
\begin{align}
W^\TF (x;\, \underline{Z};\, \underline{y};\, N;\, q)&=
q^{\frac{2}{3}} N^{\frac{4}{3}} W^\TF (q^{\frac{2}{3}} Z^{\frac{1}{3}}x;\,
N^{-1}\underline{Z};\, q^{\frac{2}{3}} N^{\frac{1}{3}}\underline{\y};\, 1;\, 1),
\label{25-3-14}\\[4pt]
\rho^\TF (x;\, \underline{Z};\, \underline{\y};\, N;\, q)&=
N^2 q^2 \rho^\TF (q^{\frac{2}{3}} Z^{\frac{1}{3}}x;\,
N^{-1}\underline{Z};\, q^{\frac{2}{3}} N^{\frac{1}{3}}\underline{\y};\, 1;\, 1),
\label{25-3-15}\\[4pt]
\cE^\TF (\underline{Z};\, \underline{y};\, N;\, q)&=
q^{\frac{2}{3}} N^{\frac{7}{3}} \cE^\TF
(N^{-1}\underline{Z};\, q^{\frac{2}{3}}N^{\frac{1}{3}}\underline{\y};\, 1;\, 1),
\label{25-3-16}\\[4pt]
\nu^\TF (\underline{Z};\, \underline{y};\, N;\, q) &=
q^{\frac{2}{3}} N^{\frac{4}{3}} \nu^\TF
(N^{-1}\underline{Z};\, q^{\frac{2}{3}} N^{\frac{1}{3}}\underline{\y};\, 1;\, 1)
\label{25-3-17}
\end{align}
where $\nu^\TF=\nu$ is the chemical potential; recall that $\underline{Z}=(Z_1,\ldots,Z_M)$ and $\underline{\y}=(\y_1,\ldots,\y_M)$ are arrays and parameter $q$ also enters into Thomas-Fermi theory.
\end{proposition}

\begin{proof}
Proof is trivial by scaling.
\end{proof}

Since we can exclude $q$ by scaling, we do not indicate dependence on it anymore. The following properties of Thomas-Fermi potential and density in the case of the single atom ($M=1$) are  well-known:

\begin{proposition}\label{prop-25-3-4}
Let $M=1$. Then the solution of the Thomas-Fermi problem has the following properties:
\begin{enumerate}[label=(\roman*), wide, labelindent=0pt]
\item\label{prop-25-3-4-i}
$W^\TF(x; Z_m,\y_m;N)$ and $\rho^\TF(x; Z_m,\y_m;N)$ are spherically symmetric (with respect to $\y_m$) and are non-increasing convex functions of $|x-\y_m|$.

\item\label{prop-25-3-4-ii}
If $N=Z_m$, then
\begin{gather}
W^\TF\asymp \min \bigl(Z_m|x-\y_m| ^{-1}, |x-\y_m| ^{-4}\bigr),
\label{25-3-18}\\
\rho^\TF\asymp \min \bigl(Z_m^{\frac{3}{2}}|x-\y_m| ^{-\frac{3}{2}},
|x-\y_m| ^{-6}\bigr)
\label{25-3-19}
\end{gather}
with the threshold at $|x-\y_m|\asymp r^*_m = Z_m^{-\frac{1}{3}}$ when $W^\TF \asymp Z_m^{\frac{4}{3}}$ and $\rho^\TF\asymp Z_m^{2}$.

\item\label{prop-25-3-4-iii}
If $\epsilon Z_m\le N<Z_m$, then
\begin{equation}
-\nu \asymp |Z_m-N|^{\frac{4}{3}}
\label{25-3-20}
\end{equation}
and \textup{(\ref{25-3-18})} holds as
$|x-\y_m|\le \bar{r}_m $ where
\begin{equation}
\bar{r}_m = -\nu^{-1}|Z_m-N|^{-1}\asymp   |Z_m-N|^{-\frac{1}{3}}
\label{25-3-21}
\end{equation}
 \underline{for atoms} denotes the \underline{exact radius of the support} of $\rho^\TF$ (see Statement~\ref{prop-25-3-4-iv}.

\item\label{prop-25-3-4-iv}
On the other hand,
\begin{equation}
W^\TF = (Z_m-N) |x-\y_m|^{-1}\qquad \text{as\ \ } |x-\y_m|\ge \bar{r}_m
\label{25-3-22}
\end{equation}
and  $\rho^\TF = 0$ as $|x-\y_m|\ge \bar{r}_m$;

\item\label{prop-25-3-4-v}
Meanwhile,
$W^\TF\asymp -\nu$ and
$\rho^\TF =O\bigl(|Z_m-N|^2)$ as $|x-\y_m|\asymp \bar{r}_m$.

\item\label{prop-25-3-4-vi}
Finally,
\begin{equation}
-\langle x-\y_m,\nabla W \rangle \asymp W.
\label{25-3-23}
\end{equation}
\end{enumerate}
\end{proposition}

Consider now the molecular case ($M\ge 2$):

\begin{proposition}\label{prop-25-3-5}
\begin{enumerate}[label=(\roman*), wide, labelindent=0pt]
\item\label{prop-25-3-5-i}
Let $M\ge 2$. Then
\begin{gather}
\nu \asymp |Z-N|^{\frac{4}{3}}
\label{25-3-24}\\
\shortintertext{and}
\sum_m \epsilon W_m^\TF (c (x-\y_m)) \le W ^\TF \le
c\sum_m W_m ^\TF (\epsilon (x-\y_m))
\label{25-3-25}
\end{gather}
where $Z=Z_1+\ldots +Z_M$ and $W_m ^\TF$ denotes an atomic Thomas-Fermi potential with the charge $Z_m$ located at\/ \ $0$ and the same chemical potential $\nu $. Here $\epsilon $ and $c$ depend only on $M$.

\item\label{prop-25-3-5-ii}
In particular, if $N<Z$ and $|x-\y_m|\ge c\bar{r}_m$ for all $m=1,\ldots,M$, then
\begin{gather}
W^\TF (x)\asymp \sum_m |Z-N| |x-\y_m|^{-1}
\label{25-3-26}\\
\shortintertext{and}
\rho^\TF (x)=0.
\label{25-3-27}
\end{gather}
\end{enumerate}
\end{proposition}

\begin{proof}
Proof is due to the comparison arguments of E.~Lieb, J.~P.~Solovej and J.~Yngvarsson~\cite{LSY1,LSY2}.
\end{proof}

\begin{proposition}\label{prop-25-3-6}
\begin{enumerate}[label=(\roman*), wide, labelindent=0pt]
\item\label{prop-25-3-6-i}
Let $N\le Z$ and
\begin{align}
&\ell (x)=\frac{1}{2}\min_m|x-\y_m|,
\label  {25-3-28}\\
&\zeta (x) = \bigl(W^\TF (x)\bigr)^{\frac{1}{2}}.
\label  {25-3-29}
\end{align}
Then
\begin{multline}
\zeta (x) \le \bar{\zeta}(x) = \\
C\left\{\begin{aligned}
&Z^{\frac{1}{2}}\ell (x)^{-\frac{1}{2}}\qquad
&&\text{as\ \ } \ell (x)\le Z^{-\frac{1}{3}},\\
&\ell(x)^{-2} \qquad
&&\text{as\ \ } Z^{-\frac{1}{3}}\le \ell (x)\le |Z-N|^{-\frac{1}{3}},\\
&|Z-N|^{\frac{1}{2}}\ell(x)^{-\frac{1}{2}} \qquad
&&\text{as\ \ } \ell (x)\ge |Z-N|^{-\frac{1}{3}};
\end{aligned}\right.
\label  {25-3-30}
\end{multline}
$\zeta(x)$ and $\bar{\zeta}(x)$ are both $\ell$-admissible and
\begin{equation}
|D ^\alpha W ^\TF(x)\bigr|\le C_\alpha \zeta (x) ^2\,\ell (x) ^{-|\alpha |}
\qquad \forall \alpha : |\alpha |\le 3,
\label  {25-3-31}
\end{equation}
and
\begin{multline}
|D ^\alpha W ^\TF(x)- D^\alpha W ^\TF (y)|\le
C_3 \zeta (x) ^2 \ell (x) ^{-\frac{7}{2}} |x-y| ^{\frac{1}{2}}\\
\forall x,y: |x-y|\le \epsilon \ell (x)
\label  {25-3-32}
\end{multline}
\item\label{prop-25-3-6-ii}
Unless $\zeta(x)\asymp(-\nu )^{\frac{1}{2}}$
estimates \textup{(\ref{25-3-32})} hold for all $\alpha $.
\end{enumerate}
\end{proposition}

\begin{proof}
This proof is rather obvious corollary of the
Euler-Lagrange equation.
\end{proof}

\begin{remark}\label{rem-25-3-7}
Let
\begin{equation}
Z_m \asymp N \qquad \forall m.
\label  {25-3-33}
\end{equation}
Then $\zeta (x)\asymp \bar{\zeta}(x)$.
\end{remark}

\begin{theorem-foot}\footnotetext{\label{foot-25-10} Theorem 1 of R.~Benguria \cite{benguria}; we combine two last statements in \ref{thm-25-3-8-iii}.}\label{thm-25-3-8}
Consider $\cE^\TF $ and
\begin{gather}
\widehat\cE^\TF \coloneqq   \cE^\TF + U,
\label  {25-3-34}\\
U= U(Z_1,\ldots, Z_M; \y_1,\ldots, \y_M)=
\sum_{1\le m< m'\le M}\frac {Z_mZ_{m'}}{|\y_m-\y_{m'}|}.
\label  {25-3-35}
\end{gather}
Select a nucleus $\y_m$ and a unit vector $\n$ such that
\begin{equation}
\langle \y_k-\y_m, \n\rangle \le 0 \qquad \forall k
\label  {25-3-36}
\end{equation}
and plug $\y_m +\alpha \n$ instead of $\y_m$ into $\cE^\TF$ and into $\widehat\cE^\TF$\,\footnote{\label{foot-25-11} So all other nuclei are confined in half-space and $\y_m$ moves away outside.}.
Then
\begin{enumerate}[wide, labelindent=0pt, label=(\roman*)]
\item\label{thm-25-3-8-i}
$\widehat\cE^\TF_\alpha$ is a non-increasing function of $\alpha\ge 0$;
\item\label{thm-25-3-8-ii}
$\cE^{\TF}_\alpha$ is a non-increasing function of $\alpha\ge 0$;
\item\label{thm-25-3-8-iii}
For fixed $\alpha>0$ both
$\widehat\cE^\TF_\alpha- \widehat\cE^\TF_0$ and
$\cE^{\TF}_\alpha- \cE^{\TF}_0$ are non-decreasing functions of $N$.
\end{enumerate}
\end{theorem-foot}
\enlargethispage{\baselineskip}

Equality
\begin{gather}
\nu = \frac{\partial \cE^\TF}{\partial N}
\label  {25-3-37}\\
\intertext{implies that \ref{thm-25-3-8-iii} is equivalent to}
\nu _\alpha \ge \nu _0.
\label  {25-3-38}
\end{gather}

\begin{theorem-foot}\footnotetext{\label{foot-25-12} (1.8)--(1.9) of H.~Brezis and E.~Lieb \cite{brezis:lieb}.}\label{thm-25-3-9}
\begin{enumerate}[label=(\roman*), wide, labelindent=0pt]
\item\label{thm-25-3-9-i}
For fixed $Z_1,\ldots,Z_M; \y_1,\ldots,\y_M$ and $N =Z$
\begin{equation}
\lambda^7 \widehat\cE^\TF(\underline{Z}; \lambda\underline{y};N)=
\widehat\cE^\TF (\lambda^3\underline{Z}; \underline{\y};\lambda^3N)
\label  {25-3-39}
\end{equation}
is positive non-decreasing function of $\lambda>0$ and has a finite limit as
$\lambda\to +\infty$.
\item\label{thm-25-3-9-ii}
This limit does not depend on $Z_1,\ldots, Z_M$.
\end{enumerate}
\end{theorem-foot}

\begin{remark}\label{rem-25-3-10}
One can observe easily that the same scaling property without assumption $N=Z$ holds for $\cE^\TF$ and $U$ as well.
\end{remark}

These two theorems and remark imply immediately

\begin{proposition}\label{prop-25-3-11}
Let assumption \textup{(\ref  {25-3-33})} be fulfilled. Then
\begin{gather}
\widehat\cE^\TF (\underline{Z}; \underline{\y}; N) -
\min_{\substack{N_1,\ldots, N_M: \\[2pt] N_1+\ldots N_M=N}}\ \sum_{1\le m\le M}
\ \cE^\TF (Z_m; N_m) \ge \epsilon \min(a^{-7},Z^{\frac{7}{3}})
\label  {25-3-40}\\
\shortintertext{where}
a=\frac{1}{2}\min_{m< m'} |\y_m-\y_{m'}|.
\label  {25-3-41}
\end{gather}
\end{proposition}

\begin{proof}
In virtue of theorem~\ref{thm-25-3-8}\ref{thm-25-3-8-i} it suffices to prove proposition for $M=2$ (all other nuclei could be pulled to infinity), and in virtue of theorem~\ref{thm-25-3-8}\ref{thm-25-3-8-iii} it suffices to prove proposition for $Z=N$.

Then the proof is due to theorem~\ref{thm-25-3-9}\ref{thm-25-3-8-i} and (\ref  {25-3-33}), which provides uniformity.
\end{proof}

\begin{remark}\label{rem-25-3-12}
In virtue of (\ref  {25-3-37}) the minimum (with respect to $N_1,\ldots, N_M$) in the sum in the right hand expression is reached when $\nu_j =\nu_k$ for all $j,k$. The same is true for a system of isolated molecules.
\end{remark}

\begin{proposition}\label{prop-25-3-13}
Let $\cQ$ denote Thomas-Fermi excess energy which is the left-hand expression of \textup{(\ref  {25-3-40})}. Then
\begin{equation}
\D(\rho ^\TF - \bar{\rho}^\TF, \rho ^\TF - \bar{\rho}^\TF)\le C\cQ, \qquad
\bar{\rho}^\TF\coloneqq   \sum_{1\le m\le M}\rho^\TF _m.
\label  {25-3-42}
\end{equation}
\end{proposition}

\begin{proof} We follow ``non-binding'' proof due to Baxter (see E.~Lieb \emph{Selecta\/}\footref{foot-25-3}).

According to Baxter's lemma there exist $g$, $0\le g\le \rho^\TF $ and
$h=\rho^\TF -g$ such that $g*|x| ^{-1}=V_1$ a.e. when $h>0$ and
$g*|x| ^{-1}\le V_1$ a.e. when $h=0$. Here $V_m=Z_m|x-\y_m|^{-1}$.

Let $\alpha =\int g\,dx$, $\beta =\int h\,dx$ and let $\cE^\TF_1$, $\cE^\TF_{(1)}$ be Thomas-Fermi energies for the first atom and for the rest of molecule respectively and $\rho^\TF _1$, $\rho^\TF_{(1)}$ be corresponding Thomas-Fermi densities. Then
\begin{multline}
\min_{N _1+N '\le N}
\Bigl(\cE^\TF_1 (N_1)+\cE^\TF_{(1)}(N')\Bigr)\le
\cE _1(\alpha )+ \cE _{(1)}(\beta )\le \\[5pt]
\cE _1(g)+ \cE _{(1)}(h) -\epsilon \D(g-\rho^\TF _1,g-\rho^\TF _1)-
\epsilon \D(h-\rho ^\TF_{(1)}, h-\rho ^\TF_{(1)})\le \\[5pt]
\cE (g+h)+\int h(V_1-g*|x| ^{-1})\,dx - \int (V_1-|g*|x| ^{-1})\,d\upmu_{(1)} \\[5pt]
- \epsilon \D(g-\rho^\TF _1,g-\rho^\TF _1)-
\epsilon \D(h-\rho^\TF_{(1)}, h-\rho^\TF_{(1)})
\label  {25-3-43}
\end{multline}
where $\upmu_1$ and $\upmu_{(1)}$ are measures with the densities respectively
$Z_1\updelta (x-\y_1)$ and
$\sum_{2\le m\le M}Z_m\updelta (x-\y_m)$ and we used the superadditivity of $\tau(\rho)=\rho^{\frac{5}{3}}$. The last expression does not exceed
\begin{equation}
\cE ^\TF-\epsilon \D(g-\rho^\TF _1,g-\rho^\TF _1)-
\epsilon \D(h-\rho^\TF_{(1)}, h-\rho^\TF_{(1)}).
\label  {25-3-44}
\end{equation}
Using induction with respect to $M$ we arrive to
\begin{multline}
\D(\rho ^\TF-\rho^\TF _1-\rho^\TF_{(1)},
\rho^\TF -\rho^\TF _1-\rho^\TF_{(1)})\le \\
2\D(g-\rho^\TF _1,g-\rho^\TF_1)+2\D(h-\rho^\TF_{(1)}, h-\rho^\TF_{(1)})
\le C\cQ
\label  {25-3-45}
\end{multline}
and finally to (\ref  {25-3-42}).
\end{proof}

\begin{Problem}\label{Problem-25-3-14}
Find the stronger lower bound in (\ref  {25-3-40}) as $N<Z$. Would be the left-hand expression $\asymp \min(a^{-7}+|Z-N|^2a^{-1},Z^{\frac{7}{3}})$?
\end{Problem}

\chapter{Application of semiclassical methods}%
\label{sect-25-4}

\section{Asymptotics of the trace}
\label{sect-25-4-1}

In this subsection we calculate asymptotics of $\Tr ((H_W-\nu)^-)$. Here we need to consider both inner and outer zones.

An \emph{inner zone\/}\index{zone!inner} (near nucleus $\y_m$) is a ball where $V_j=Z_m|x-\y_m|^{-1}$ dominates $W-V_m$. For a single nucleus ($M=1$) it is defined by
\begin{gather}
|x-\y_m|\le \epsilon Z_m^{-\frac{1}{3}}\label{25-4-1}\\
\intertext{but in the case $M\ge 2$ there are another restrictions}
|x-\y_m|\le \epsilon \min_{m'\ne m}
\Bigl(Z_m (Z_m+Z_{m'})^{-1} |\y_m-\y_{m'}| \Bigr)\label{25-4-2}\\
\intertext{and}
|x-\y_m|\le Z_m \nu ^{-1}\label{25-4-3}
\intertext{but we shrink this zone to}
|x-\y_m|\le r_m \coloneqq   \epsilon
\min \bigl( Z_m Z^{-1}a, Z_m^{-\frac{1}{3}}\bigr).
\label{25-4-4}
\end{gather}

Let us consider contribution of the zone $\cX_m$ described by (\ref{25-4-4}) to $\N_1 (H_W-\nu)=\Tr ((H_W -\nu)^-)$, both into the principal part of asymptotics and the remainder. Let $\psi_m$ be a partition element concentrated in $\cX_m$ and equal to $1$ in $\{x\colon   |x-\y_m|\le \frac{1}{2}r_m\}$. Then, according to Theorem~\ref{book_new-thm-12-5-8},
\begin{equation}
\Tr \bigl((H_W-\nu)^- \psi_m\bigr) =
\int \Weyl_1(x)\psi_m (x)\,dx + \Scott_m + O(R_m)
\label{25-4-5}
\end{equation}
where $\Weyl_1(x)$ and $\Scott_m$ are calculated for the case $q=1$ and then multiplied by $q$\,\footnote{\label{foot-25-13} As operator $H_{w}-\nu$ is nothing but $q$ copies of $\Delta - (W+\nu)$.}:
\begin{equation}
\Weyl_1 (x)\coloneqq -\frac{q}{15\pi^2} \bigl(W(x)+\nu\bigr)_+^{\frac{5}{2}}
\label{25-4-6}
\end{equation}
while $R_m$ is $C\zeta^2 (\zeta \ell) = C\zeta^3\ell$ calculated on its border i.e.
\begin{equation}
R_m= C Z_m^{\frac{5}{3}} + C Z_m ^{\frac{3}{2}} r_m^{-\frac{1}{2}} \le
CZ^{\frac{5}{3}} + CZ^{\frac{3}{2}} a^{-\frac{1}{2}}.
\label{25-4-7}
\end{equation}
Really, one needs just to rescale $x\mapsto (x-\y_m)r_m^{-1}$ and
$\tau \mapsto \tau Z_m^{-1}r_m$ and introduce a semiclassical parameter
$\hbar = Z_m^{\frac{1}{2}}r_m^{\frac{1}{2}}$.

\begin{remark}\label{rem-25-4-1}
\begin{enumerate}[label=(\roman*), wide, labelindent=0pt]
\item\label{rem-25-4-1-i}
Clearly, these arguments work only if $r_m \ge Z_m^{-1}$ (i.e. {$Z_m^2 \ge a^{-1}Z$}).

\item\label{rem-25-4-1-ii}
On the other hand, if $Z_m^2 \ge a^{-1}Z$ but $a\ge Z^{-1}$ we define $r_m= a^{\frac{1}{2}}Z^{-\frac{1}{2}}$ and we do not include $\Scott_m$\,\footnote{\label{foot-25-14} However it will be less than the remainder estimate, so we can include it into the principal part of asymptotics anyway.} into the principal expression; moreover, in this case we include $\cX_m$ into a singular zone and use variational methods to estimate its contribution into the principal  part of asymptotics; it will not exceed
$CZ a^{-1}\le C Z^{\frac{3}{2}}a^{-\frac{1}{2}}$.

\item\label{rem-25-4-1-iii}
Furthermore, if $a\le Z^{-1}$, we set $r_m = Z^{-1}$ and we do not include any $\Scott_m$\,\footref{foot-25-14} into the principal  part of asymptotics and include all $\cX_m$ into singular zones; using variational methods we estimate their contributions into the principal part of asymptotics by $CZ^2$.
\end{enumerate}
\end{remark}

Therefore we conclude that
\begin{claim}\label{25-4-8}
The total contribution of all inner zones into remainder does not exceed the right-hand expression of \textup{(\ref{25-4-7})} as
$a\ge Z^{-1}$ and $CZ^2$ as $a\le Z^{-1}$.
\end{claim}

Let us consider contributions of the \emph{outer zone\/} $\cX_0$ which is complimentary to the union of all inner zones. Then
\begin{equation}
\Tr \bigl((H_W-\nu)^- \psi_0\bigr) =
\int \Weyl_1(x)\psi_0 (x)\,dx + O(R_0)
\label{25-4-9}
\end{equation}
with $\Weyl_1(x)$ defined by (\ref{25-4-6}) where
\begin{gather}
R_0= \int_{\cX_0} C\bar{\zeta} (x) ^3 \ell ^{-2}\, dx
\label{25-4-10}\\
\shortintertext{and}
\bar{\zeta}= \left\{\begin{aligned}
&Z^{\frac{1}{2}} \ell^{-\frac{1}{2}} \qquad
&&\text{as\ \ } \ell \le Z^{-\frac{1}{3}},\\
&\ell^{-2}\qquad &&\text{as\ \ } \ell \ge Z^{-\frac{1}{3}}.
\end{aligned}\right.
\label{25-4-11}
\end{gather}
We justify (\ref{25-4-9})--(\ref{25-4-11}) a bit later by an appropriate partition of unity. One can see easily that the contribution of
the zone $\{x\colon   \ell(x)\le Z^{-\frac{1}{3}}\}$ into expression (\ref{25-4-10}) does not exceed the same expression as in (\ref{25-4-8}) and that the contribution of the zone
$\{x\colon   \ell(x)\ge Z^{-\frac{1}{3}}\}$ into (\ref{25-4-10}) does not exceed $CZ^{\frac{5}{3}}$. Then we arrive to

\begin{theorem}\label{thm-25-4-2}
Let $W=W^\TF$, $N\asymp Z$, $W=W^\TF$. Then
\begin{gather}
\Tr \bigl((H_W-\nu)^-\bigr) = \int \Weyl_1 (x)\,dx + \sum_{1\le m\le M} \Scott_m +O(R), \label{25-4-12}\\
\shortintertext{with}
R\coloneqq   \left\{\begin{aligned}
&C Z^{\frac{5}{3}} + CZ^{\frac{3}{2}}a^{-\frac{1}{2}}\qquad
&&\text{as \ \ }a\ge Z^{-1},\\
&CZ^2 \qquad &&\text{as \ \ }a\le Z^{-1}.
\end{aligned}\right.
\label{25-4-13}
\end{gather}
\end{theorem}

\begin{proof}
\begin{enumerate}[label=(\roman*), wide, labelindent=0pt]
\item\label{pf-25-4-2-i}
Consider $\nu =0$ first. Then we just apply $\ell$-admissible partition of unity. Sure, $\zeta \ell\le 1$ as $\ell\ge 1$ but we can deal with it \underline{either} by taking $\zeta=\ell^{-1}$ here \underline{or} considering it as a singular zone and applying here variational estimate as well.

\item\label{pf-25-4-2-ii}
Variational estimate works for $\nu<0$ as well; furthermore, zone
$\{x\colon   W(x)\le (1-\epsilon )\nu, \ell \ge |\nu|^{-\frac{1}{2}}\}$ is classically forbidden.

\item\label{pf-25-4-2-iii}
However as $\nu \le -c$ we have a little problem as $W^\TF$ is not very smooth, it is only $\sC^{\frac{7}{2}}$ as $W \asymp -\nu $. The best\footnote{\label{foot-25-15} From the point of view of generalization to the case when magnetic field is present and it is not too weak, so magnetic version of $W^\TF$ has multiple singularities.} way to deal with it is to take $\varepsilon\ell$-mollification with $\varepsilon = \hbar^{1-\delta}$, $\hbar=(\zeta \ell)^{-1}$, use rough microlocal analysis of Section~\ref{book_new-sect-4-6} and the bracketing; it will bring an approximation error not exceeding $C\varepsilon ^{\frac{7}{2}}\hbar^{-3}|\nu|$ which does not exceed $|\nu|=O(|Z-N|^{\frac{4}{3}})$. We leave easy details to the reader.
\end{enumerate}
\end{proof}

Now we arrive to the lower estimate for $\E$:
\enlargethispage{\baselineskip}

\begin{corollary}\label{cor-25-4-3}
Let $N\asymp Z$. Then
\begin{equation}
\E \ge \cE^\TF +\Scott - CR
\label{25-4-14}
\end{equation}
with $R$ defined by \textup{(\ref{25-4-13})}.
\end{corollary}

\begin{proof}
We know from Subsection~\ref{sect-25-2-1} that
\begin{equation}
\E \ge \N_1(H_W-\nu)+\nu N -\frac{1}{2}\D(\rho,\rho) - CZ^{\frac{5}{3}}
\label{25-4-15}
\end{equation}
as $W$ is given by (\ref{25-2-3}). In virtue of Theorem~\ref{thm-25-4-2}
\begin{gather}
\E \ge \int \Weyl_1(x) \,dx + \nu \int \Weyl(x) \,dx -
\frac{1}{2}\D(\rho,\rho) +\Scott - CR\label{25-4-16}\\
\intertext{where we also plugged instead of $N$ as $N<Z$ (and $\nu<0$)}
N = \int \Weyl(x) \,dx\label{25-4-17}\\
\shortintertext{with}
\Weyl(x) =\frac{q}{6\pi^2} \bigl(W(x)+\nu\bigr)_+^{\frac{3}{2}}.\label{25-4-18}
\end{gather}
One can check easily that three first terms in the right-hand expression of (\ref{25-4-18}) constitute exactly $\Phi_* (W+\nu)$ coinciding with $\cE^\TF$ as $W=W^\TF$.
\end{proof}

\begin{remark}\label{rem-25-4-4}
\begin{enumerate}[label=(\roman*), wide, labelindent=0pt]
\item\label{rem-25-4-4-i}
As $a\le Z^{-\frac{1}{3}}$  using the same method one can prove a slightly better remainder estimate--with $Z^{\frac{3}{2}}a^{-\frac{1}{2}}$ replaced by
\begin{equation}
\sum_{1\le m<m'\le M} \min \Bigl(
(Z_m+Z_{m'})^{\frac{3}{2}} |x_m-\y_{m'}|^{-\frac{1}{2}}, (Z_m+Z_{m'})^2\Bigr)
\label{25-4-19}
\end{equation}
allowing to lighter nuclei to be closer one to another.
\item\label{rem-25-4-4-ii}
To improve this estimate further, allowing lighter nuclei to be closer to heavier, ones one needs to improve Theorem~\ref{book_new-thm-12-5-8} which seems to be too difficult task for a such little gain.
\end{enumerate}
\end{remark}

\section{Upper estimate for $\E$}
\label{sect-25-4-2}

Recall that in virtue of Subsection~\ref{sect-25-2-2}
\begin{multline}
\E \le \N_1(H_W-\nu)+\nu N -\frac{1}{2}\D(\rho,\rho) +\\
|\lambda_N-\nu| \cdot |\N(H_W-\nu)-N| +
\frac{1}{2}\D (\tr e_N(x,x)-\rho, \tr e_N(x,x)-\rho\bigr)
\label{25-4-20}
\end{multline}
and thus we need to estimate two last terms in the right-hand expression.

\subsection{Estimating \texorpdfstring{$|\lambda_N-\nu|$}{|\textlambda\_N-\textnu|} }
\label{sect-25-4-2-1}
First,  we need to estimate $|\lambda_N-\nu|$. We will use the heuristic equality $\N(H_W-\lambda_N)\approx N$ or more precisely two inequalities
\begin{phantomequation}\label{25-4-21}\end{phantomequation}
\begin{gather}
\N(H_W-\lambda_N-0) \le N \le \N(H_W-\lambda_N+0)
\tag*{$\textup{(\ref{25-4-21})}_{1,2}$}\\
\shortintertext{and equality}
\int \Weyl(x)\,dx =\min (N,Z)
\label{25-4-22}
\end{gather}
where the right inequality $\textup{(\ref{25-4-21})}_{2}$ is valid only if $\lambda_N<0$ i.e. $\N(H_W)\ge N$.

\subsubsection{Case $\lambda_N < \nu $.} Then we will use $\textup{(\ref{25-4-21})}_{2}$ and to calculate $\N(H_W-\lambda_N+0)$ we will use semiclassical approximation:
\begin{gather}
\N(H_W-\lambda_N+0)=\int \Weyl (x,\lambda_N)\, dx + O(R_0)
\label{25-4-23}
\intertext{with the semiclassical error}
R_0 = \int \zeta^2 \ell^{-1}\, dx
\label{25-4-24}
\intertext{with the integral not exceeding}
CZ \int_{\{|x|\le Z^{-\frac{1}{3}}\}} |x|^{-2}\,dx +
C \int_{\{|x|\ge Z^{-\frac{1}{3}}\}} |x|^{-5}\,dx \asymp CZ^{\frac{2}{3}};
\notag
\end{gather}
again we need to consider separately the case of $N\ge Z$ and $\nu=0$, when integral (\ref{25-4-24}) is taken over $\bR^3$, and the case of $N<Z$ and
$\nu <0$, when this integral should be taken over
$\{x\colon \ell(x)\le C(Z-N)^{-\frac{1}{3}}\}$; in the latter case to cover non-smoothness we consider an approximation (mollification) of $W$ and an approximation error $\varepsilon^{\frac{7}{2}}\hbar^{-3}\ll 1$. Therefore
\begin{equation}
\int \Weyl (x,\lambda_N)\, dx \ge N - CZ^{\frac{2}{3}}.
\label{25-4-25}
\end{equation}
Comparing with (\ref{25-4-22}) we conclude that
\begin{equation}
\int \Bigl( \bigl(W(x)+\nu\bigr)_+^{\frac{3}{2}} - \bigl(W(x)+\lambda_N\bigr)_+^{\frac{3}{2}}\Bigr)\, dx \le CZ^{\frac{2}{3}}.
\label{25-4-26}
\end{equation}
Here an integrand is non-negative (since $\lambda_N<\nu $). Further, one can see easily that the main contribution to this integral is delivered by the zone
$\{x\colon   \ell(x)\asymp |\lambda_N|^{-\frac{1}{4}}\}$\,\footnote{\label{foot-25-16} Provided
$|\lambda_N |\le C_0Z^{\frac{4}{3}}$. On the other hand, if $|\lambda_N |\ge C_0Z^{\frac{4}{3}}$ one can see easily that (\ref{25-4-26}) is $\asymp Z$ because $|\nu|\le c_0Z^{\frac{4}{3}}$ due to $N\asymp Z$.} and the whole integral is $\asymp |\lambda_N|^{-\frac{1}{4}}|\nu-\lambda_N|$. Therefore (\ref{25-4-26}) yields that
\begin{gather}
|\lambda_N-\nu| \le C(|\nu|+|\lambda_N-\nu| )^{\frac{1}{4}} Z^{\frac{2}{3}},
\notag\\
\intertext{which is equivalent to}
|\lambda_N-\nu| \le CZ^{\frac{8}{9}}+ C|\nu |^{\frac{1}{4}} Z^{\frac{2}{3}} \asymp CZ^{\frac{8}{9}} + C(Z-N)_+ ^{\frac{1}{3}} Z^{\frac{2}{3}} \le CZ.
\label{25-4-27}
\end{gather}
In particular,
\begin{claim}\label{25-4-28}
If $|\nu |\le c_0 Z^{\frac{8}{9}}$ (i.e. $(Z-N)_+\le c_1 Z^{\frac{2}{3}}$), then $|\lambda_N |\le C_0 Z^{\frac{8}{9}}$
\end{claim}
and also
\begin{equation}
|\lambda_N-\nu| \cdot |\N(H_W-\nu)-N| \le
CZ\cdot Z^{\frac{2}{3}}=CZ^{\frac{5}{3}}.
\label{25-4-29}
\end{equation}

\subsubsection{Case $\lambda_N\ge \nu$.}
Then we will use the left inequality $\textup{(\ref{25-4-21})}_1$, but if
$\N (H_W)<N$ then integral (\ref{25-4-24}) is diverging.

To avoid all related difficulties we will consider first the case when we necessarily conclude that $|\lambda_N|\ge (1-\epsilon)|\nu|$. To do so observe that even if the main contribution to the integral (\ref{25-4-26})\,\footnote{\label{foot-25-17} Now an integrand is non-positive.} is delivered by the zone
$\{x\colon  \ell (x)\asymp (Z-N) |\lambda_N|^{-1}\asymp |\nu|^{\frac{3}{4}}|\lambda_N|^{-1}\}$, we will ignore this observation  and consider a larger zone
$\{x\colon  \ell (x)\le C_0|\nu|^{-\frac{1}{4}}\}$ instead of
$\{x\colon  \ell (x)\le C_0|\nu|^{\frac{3}{4}}|\lambda_N|^{-1}\}$\,\footnote{\label{foot-25-18} Obviously these zones coincides as $|\nu|\asymp |\lambda_N|$.}.

One can prove easily that
\begin{claim}\label{25-4-30}
The contribution of the zone $\{x\colon \ell (x) \le C_0|\nu|^{-\frac{1}{4}}\}$ to the semiclassical remainder, when calculating $\N(H_W-\lambda_N)$, does not exceed $CZ^{\frac{2}{3}}$.
\end{claim}
Therefore we arrive to the estimate
\begin{equation}
|\lambda_N-\nu|\le C|\nu|^{\frac{1}{4}} Z^{\frac{2}{3}}\asymp
C(Z-N)_+^{\frac{1}{3}}Z^{\frac{2}{3}}\le CZ
\label{25-4-31}
\end{equation}
(cf. (\ref{25-4-27})). In particular, we conclude that
\begin{claim}\label{25-4-32}
If $|\nu|\ge CZ^{\frac{8}{9}}$ (i.e. $(Z-N)\ge CZ^{\frac{2}{3}}$), then $|\lambda_N|\asymp |\nu|$
\end{claim}
(cf. (\ref{25-4-28})). Then one can easily recover (\ref{25-4-27}) completely. Since $|\N(H_W-\nu)-N|\le CZ^{\frac{2}{3}}$ we arrive to (\ref{25-4-29}).
\enlargethispage{\baselineskip}

\subsubsection{Case $|\nu|\le \eta= CZ^{\frac{8}{9}}$.}
This case (i.e. $(Z-N)\le \eta ^{\frac{3}{4}}=CZ^{\frac{2}{3}}$) is the most important one. The easiest way to tackle it is to pick up $\rho^\TF$ and $\W^\TF$ calculated as if $\nu=0$ i.e. $Z=N$; that means the change of the test function $\Psi$ in the upper estimate.

We need to modify an upper estimate of $\sum_{1\le j\le N} \lambda_j$. To do this we note that
\begin{equation}
\lambda_N\ge -\eta
\label{25-4-33}
\end{equation}
and
\begin{claim}\label{25-4-34}
A number of eigenvalues between $\lambda_N$ (if $\lambda_N<0$) and $0$ does not exceed $CZ^{\frac{2}{3}}+ C\eta ^{\frac{3}{4}}$,
\end{claim}
which can be proven easily by our standard methods. Therefore
\begin{equation}
\sum_{1\le j\le N} \lambda_j \le \Tr (H_W ^- ) +
C\eta \bigl(Z^{\frac{2}{3}}+ \eta ^{\frac{3}{4}} \bigr)
\label{25-4-35}
\end{equation}
and the last term is less than $CZ^{\frac{5}{3}}$ as long as
$\eta \le CZ^{\frac{20}{21}}$ which is fulfilled.

In this last case\footnote{\label{foot-25-19} Pending an analysis of the next subsubsection.} we arrive to an upper estimate with $\cE^\TF$ calculated as if $\nu=0$ i.e. $\cE^\TF (\underline{Z};\underline{\y}; Z)$ which is less than
$\cE^\TF (\underline{Z};\underline{\y}; N)$. Actually the difference between these two is $\asymp |\nu|^{\frac{7}{4}}\le |\eta|^{\frac{7}{4}}$.

\subsection{Estimating $\D$-term}
\label{sect-25-4-2-2}

We need to estimate the last term in the right-hand expression of (\ref{25-4-20}); we estimate it by
\begin{multline}
C_0\D \bigl (\tr e(x,x,\nu)- P'(W+\nu), \tr e(x,x,\nu)- P'(W+\nu)\bigr)\\
+C_0\D \bigl (\tr e(x,x,\nu)- \tr e_N(x,x), \tr e(x,x,\nu)- \tr e_N(x,x)\bigr)\ \ \\
+C_0 \D \bigl(\rho -P'(W+\nu), \rho -P'(W+\nu)\bigr),
\label{25-4-36}
\end{multline}
where the last term vanishes as $\rho=\rho^\TF$, $W=W^\TF$; however, for some technical reasons we want to avoid this assumption.

\subsubsection{Estimating the first term.}\label{sect-25-4-2-2-1}
To estimate the first term in (\ref{25-4-36}) we apply the semiclassical asymptotics
\begin{equation}
\tr e(x,x,\nu) = \Weyl (x) + O(\zeta^2 \ell^{-1}),
\label{25-4-37}
\end{equation}
where $\Weyl (x) = P'(W(x)+\nu)$ and therefore this term does not exceed
\begin{equation}
\iint \zeta (x)^2 \zeta(y)^2 \ell (x)^{-1} \ell(y)^{-1}|x-y|^{-1}\,dx dy.
\label{25-4-38}
\end{equation}
Estimating this integral by the double sum of the integrals over domains
$\{(x,y)\colon  \ell(x)\le Z^{-\frac{1}{3}}, \ell(y)\le Z^{-\frac{1}{3}}\}$ and
$\{(x,y)\colon  \ell(x)\ge Z^{-\frac{1}{3}}, \ell(y)\ge Z^{-\frac{1}{3}}\}$ we get
\begin{gather}
C Z^2 \iint _{\{|x|\le Z^{-\frac{1}{3}}, |y|\le Z^{-\frac{1}{3}}\}}
|x|^{-2} |y|^{-2} |x-y|^{-1}\,dx dy
\label{25-4-39}\\
\shortintertext{and}
C \iint _{\{|x|\ge Z^{-\frac{1}{3}}, |y|\ge Z^{-\frac{1}{3}}\}}
|x|^{-3} |y|^{-3} |x-y|^{-1}\,dx dy
\label{25-4-40}
\end{gather}
respectively, and rescaling we get the same integrals but both with the ``threshold'' $1$ rather than $Z^{-\frac{1}{3}}$ and both with factor $Z^{\frac{5}{3}}$ rather than $Z^2$ or $1$ respectively; one can see easily that both obtained integrals (without factor $Z^{\frac{5}{3}}$) are $\asymp 1$.  Therefore, expression (\ref{25-4-38}) is $O(Z^{\frac{5}{3}})$.

Therefore we proved

\begin{proposition}\label{prop-25-4-5}
As $W=W^\TF$ the first term in \textup{(\ref{25-4-36})} does not exceed $CZ^{\frac{5}{3}}$.
\end{proposition}

\begin{remark}\label{rem-25-4-6}
\begin{enumerate}[label=(\roman*), wide, labelindent=0pt]
\item\label{rem-25-4-6-i}
For $N\ge Z$ and $\nu =0$ we used that $\zeta \le C_1\ell^{-2}$ for
$\ell \ge Z^{-\frac{1}{3}}$.

\item\label{rem-25-4-6-ii}
For $N<Z$ and $\nu <0$ we used that zone
$\{x\colon  \ell \ge C (Z-N)^{-\frac{1}{3}}\}$ is classically forbidden
($W+\nu <0 $ there) and therefore integral is taken over
zone $\{x\colon  \ell \le C (Z-N)^{-\frac{1}{3}}\}$ where $\zeta \le C_1\ell^{-2}$.\enlargethispage{\baselineskip}

\item\label{rem-25-4-6-iii}
As $N<Z$ $W^\TF$ is not very smooth near $W+\nu =0$ but one can handle it by rescaling arguments.

\item\label{rem-25-4-6-iv}
Alternatively (preferably\footref{foot-25-15}) one can replace $W^\TF$ by its mollification $W^\TF_\varepsilon$.
\end{enumerate}
\end{remark}

\subsubsection{Estimating the second term.}\label{sect-25-4-2-2-2}
Consider now the second term in (\ref{25-4-36}). Due to the arguments of the previous paragraph modulo $O(Z^{\frac{5}{3}})$ one can rewrite it as
\begin{equation}
C_1 \D \bigl( P'(W+\lambda_N)-P'(W+\nu),P'(W+\lambda_N)-P'(W+\nu)\bigr).
\label{25-4-41}
\end{equation}
Really, if we replace $\tr e(x,x,\nu)$ and $\tr e_N (x,x)$ by $\Weyl (x,\nu)$ and $\Weyl(x,\lambda_N)$ respectively we make a semiclassical errors estimated by $Z^{\frac{5}{3}}$ provided either $\lambda_N<\nu$ or $\lambda_N\asymp \nu$ which is always the case unless $|\nu|\le CZ^{\frac{8}{9}}$ but in this case we just ``cheat'' resetting everything to the case $\nu=0$.

Let us estimate (\ref{25-4-41}). According to the previous Subsubsection~\ref{sect-25-4-2-1}.1 there are two cases:

\begin{enumerate}[label=(\roman*), wide, labelindent=0pt]
\item\label{sect-25-4-2-2-2-i}
$N\ge Z - CZ^{\frac{2}{3}}$, in which case $|\nu|\le CZ^{\frac{8}{9}}$ and
$|\lambda_N|\le CZ^{\frac{8}{9}}$, and (\ref{25-4-41}) does not exceed
\begin{align*}
C|\nu -\lambda_N|^2 &\iint _{\{\ell(x)\le L, \ell(y)\le L\}}
|x-y|^{-1}\zeta(x)\zeta(y) \,dxdy \\
+C &\iint _{\{\ell(x)\ge L, \ell(y)\ge L\}}
|x-y|^{-1}\zeta(x)^3\zeta(y)^3 \,dxdy
\end{align*}
with $L=|\lambda_N-\nu|^{-\frac{1}{4}}$; one can calculate easily that both terms are of the magnitude $C|\lambda_N-\nu|^{\frac{7}{4}}\le CZ^{\frac{14}{9}}\ll Z^{\frac{5}{3}}$.

\item\label{sect-25-4-2-2-2-ii}
$N\le Z - CZ^{\frac{2}{3}}$, in which case (\ref{25-4-41}) does not exceed
\begin{equation*}
C|\nu -\lambda_N|^2 \iint _{\{\ell(x)\le L, \ell(y)\le L\}}
|x-y|^{-1}\zeta(x)\zeta(y) \,dxdy
\end{equation*}
with $L= |\nu|^{-\frac{1}{4}}$ and this integral does not exceed $|\lambda_N-\nu|^2 |\nu|^{-\frac{1}{4}}$ which due to (\ref{25-4-27}) does not exceed $C(Z-N)^{\frac{1}{3}} Z^{\frac{4}{3}}\le CZ^{\frac{5}{3}}$.
\end{enumerate}

\subsubsection{Estimating the third term.}\label{sect-25-4-2-2-3}
The third term in (\ref{25-4-36}) does not exceed
\begin{multline*}
C_1 \D \bigl(\rho-\rho^\TF, \rho-\rho^\TF\bigr)+\\
C_1 \D \bigl(P'(W+\nu)-P'(W^\TF +\nu), P'(W+\nu)-P'(W^\TF +\nu)\bigr)
\end{multline*}
and we leave to the reader an easy proof that both terms here are $O(Z^{\frac{5}{3}-\delta})$ as $W$ is a described mollification of $W^\TF$.

\subsection{Finally, a theorem}
\label{sect-25-4-2-3}

As we finished an upper estimate for $\E$ we arrive to the following main result:

\begin{theorem}\label{thm-25-4-7}
Let $N\asymp Z$ and let $\Psi$ be a ground state. Then
\begin{equation}
|\E - \cE^\TF - \Scott |\le C\left\{\begin{aligned}
&Z^{\frac{5}{3}} + a^{-\frac{1}{2}}Z^{\frac{3}{2}}\qquad
&&\text{as \ \ } a\ge Z^{-1},\\
&Z^2 \qquad
&&\text{as \ \ } a\le Z^{-1},
\end{aligned}\right.
\label{25-4-42}
\end{equation}
where $a$ is the minimal distance between nuclei.
\end{theorem}

\section{Improved asymptotics}
\label{sect-25-4-3}

So far as $a\ge Z^{-\frac{1}{3}}$ we recovered only $O(Z^{\frac{5}{3}})$ for both error estimate in $\E$ and (as a coproduct, see Subsection~\ref{sect-25-4-4}) for $\D(\rho_\Psi-\rho^\TF, \rho_\Psi-\rho^\TF)$.

Our purpose is to improve them to $o(Z^{\frac{5}{3}})$ (or slightly better)
as $a\gg Z^{-\frac{1}{3}}$ and recover the Schwinger and Dirac terms. To do so in the lower estimate for $\E$ one just need an improved electrostatic inequality (see Theorem~\ref{thm-25-2-5}) and also improved semiclassical estimates in $\Tr ((H_W-\nu)^-)$ and $\N (H_W-\nu)$.

For the improved upper estimate we will need also to improve estimate
$\frac{1}{2}\D( \rho_\Psi- \rho^\TF, \D( \rho_\Psi- \rho^\TF)$ for the test function $\Psi$ and apply an estimate
\begin{multline}
|\frac{1}{2} \iint |x-y|^{-1} \tr e _N(x,y) e^\dag _N(x,y)\, dx dy +\\
\kappa_\Dirac \int \rho^{\TF\,\frac{4}{3}}(x)\,dx| \le Z^{\frac{5}{3}-\delta}
\label{25-4-43}
\end{multline}
which is due to Theorem~\ref{book_new-thm-6-4-17}\footnote{\label{foot-25-20} This theorem implies the above estimate with $\delta=1$ which is definitely overkill for our purposes.}.

\begin{remark}\label{rem-25-4-8}
\begin{enumerate}[label=(\roman*), wide, labelindent=0pt]
\item\label{rem-25-4-8-i}
Only contributions to the remainder of zones
\begin{equation}
\{x\colon  Z^{-\frac{1}{3}-\delta_1} \le |x-\y_m| \le Z^{-\frac{1}{3}} b^{\delta_1}\}
\label{25-4-44}
\end{equation}
with $b\coloneqq \min(a Z^{\frac{1}{3}}, 1)$ (where now we  assume that $b\ge 1$) should be considered because the contributions of both zones $\{x\colon  \ell(x) \le Z^{-\frac{1}{3}}b^{-\delta_1}\}$ and
$\{x\colon  \ell(x) \le Z^{-\frac{1}{3}}b^{\delta_1}\}$
are $O(Z^{\frac{5}{3}}b^{-\delta})$.

\item\label{rem-25-4-8-ii}
Only $m$ with $Z_m\ge Z b^{-\delta_2}$ should be considered because contributions of other $m$ are are $O(Z^{\frac{5}{3}}b^{-\delta})$ as well.
\end{enumerate}
\end{remark}

\begin{proposition}\label{prop-25-4-9}
Let $\vartheta$ be a (small) parameter such that
$b^{-\delta_2}\le \vartheta \le 1$; consider $m$ with
$Z_m\ge Z\vartheta^{\delta_1}$. Let $\phi(x)= \psi (r^{-1}(x-\y_m))$ with
$\psi \in \sC_0^\infty (B(0,1))$ and $r$ defined below.

\begin{enumerate}[label=(\roman*), wide, labelindent=0pt]
\item\label{prop-25-4-9-i}
Further, let
$(Z-N)^{-\frac{1}{3}}\ge r=Z ^{-\frac{1}{3}}\vartheta ^{-1}$. Then inequalities
\begin{multline}
\bigl|\int \phi (x) \int_{-\infty } ^\lambda \bigl(e (x,x, \lambda ')-
\Weyl (x,\lambda ')\bigr)\,dxd\lambda '\\
-\Scott - \Schwinger \bigr|\le
CZ ^{\frac{5}{3}}\vartheta ^\delta,
\label{25-4-45}
\end{multline}
\begin{equation}
\bigl|\int \phi (x) \bigl(e (x,x, \lambda )-
\Weyl (x,\lambda )\bigr)\,dx\bigr|\le
CZ ^{\frac{2}{3}}\vartheta ^\delta
\label{25-4-46}
\end{equation}
\vglue-15pt\noindent
and
\vglue-15pt
\begin{multline}
\D\Bigl(\phi \bigl(e(x,x,\lambda )-\Weyl (x,\lambda )\bigr),
\phi \bigl(e(x,x,\lambda )-\Weyl (x,\lambda )\Bigr)\bigr)\le\\
CZ ^{\frac{5}{3}}\vartheta ^\delta
\label{25-4-47}
\end{multline}
hold with some exponent $\delta >0$ for all $\lambda \le 0$ and for
$\phi $ which is $r$-admissible.

\item\label{prop-25-4-9-ii}
On the other hand, let $(Z-N)^{-\frac{1}{3}}\le r$. Let $W$ be a constructed above mollification of $W ^\TF$. Then
\begin{enumerate}[label=(\alph*), leftmargin=*]
\item\label{prop-25-4-9-iia}
Estimates \textup{(\ref{25-4-45})}--\textup{(\ref{25-4-47})} hold for all
$\lambda \le \nu $.

\item\label{prop-25-4-9-iib}
Further, estimates \textup{(\ref{25-4-46})}, \textup{(\ref{25-4-47})} hold
for $\lambda \in [\nu ,0]$ such that $\N(H-\lambda )\le N$.

\item\label{prop-25-4-9-iic}
Furthermore, in this last case
\begin{equation}
|\lambda -\nu |\le CZ ^{\frac{2}{3}}(Z-N)^{\frac{1}{3}}\vartheta ^\delta .
\label{25-4-48}
\end{equation}
\end{enumerate}
\end{enumerate}
\end{proposition}

To prove these statements we need to study behavior of the Hamiltonian trajectories. First we want to prove that in the indicated zone $W^\TF$ is a weak perturbation of $W_m^\TF$ which is a single atom Thomas-Fermi potential with $Z_m$ and with $\nu_m=\nu$.

\begin{proposition}\label{prop-25-4-10}
In the framework of Proposition~\ref{prop-25-4-9} in $B(\y_m, r)$
\begin{equation}
|D^\alpha (W^\TF -W_m^\TF)|\le
c_\alpha W_m^\TF |x-\y_m|^{-|\alpha|}\vartheta^\delta.
\label{25-4-49}
\end{equation}
This estimate holds for all $\alpha$ as $W^\TF /(-\nu)$ is disjoint from $1$; otherwise it holds for $|\alpha|\le 3$ and
\begin{multline}
|D^\alpha (W^\TF -W_m^\TF)(x) - D^\alpha (W^\TF -W_m^\TF)(y)|\le \\
c W_m^\TF |x-\y_m|^{-\frac{7}{2}}|x-y|^{\frac{1}{2}}\vartheta^\delta
\label{25-4-50}
\end{multline}
for $|\alpha|=3$ and $|x-\y_m|\asymp |y-y_m|\asymp (Z-N)^{-\frac{1}{3}}$.
\end{proposition}

\begin{proof}
An easy proof based on the variational approach is left to the reader.
\end{proof}

Next, let us consider a manifold
$\Sigma _\lambda =\{(x,\xi)\colon H(x,\xi )=\lambda \}$ with the classical Hamiltonian
$H(x,\xi )=|\xi| ^2 -W(x)$, and let us introduce a measure $\upmu _\lambda $
with the density $dxd\xi :dH$ on $\Sigma _\lambda $; this measure is invariant with respect to the Hamiltonian flow with the Hamiltonian $H(x,\xi )$. Note that $\upmu _\lambda (\Sigma _\lambda )\asymp Z $.

\begin{proposition}\label{prop-25-4-11}
In the framework of Proposition~\ref{prop-25-4-9} there exists a set
$\Sigma '_{\lambda ,\vartheta }\subset \Sigma _\lambda $ such that
\begin{equation}
\upmu _\lambda (\Sigma '_{\lambda ,\vartheta })\le C\vartheta ^\delta Z
\label{25-4-51}
\end{equation}
and through each point $(x,\xi )$ belonging to
$\vartheta (Z ^{-\frac{1}{3}},Z ^{\frac{2}{3}})$-vicinity of
$\Sigma _\lambda \setminus \Sigma '_{\lambda ,\vartheta }$ in $T^*\bR^3$ there
passes a Hamiltonian trajectory $(x(t),\xi (t))$ of $H$ of the length
$T=Z ^{-1}\vartheta ^{-\delta }$ along which
\begin{gather}
\bigl|D_{(xZ ^{\frac{1}{3}},\xi Z ^{-\frac{2}{3}})} ^\alpha
\bigl(x(t)Z ^{\frac{1}{3}},\xi (t) Z ^{-\frac{2}{3}}\bigr)\bigr|\le
C\vartheta ^{-K}\qquad \forall \alpha :|\alpha |\le k
\label{25-4-52}\\
\shortintertext{and}
|x(t)-x(0)|Z ^{\frac{1}{3}}+|\xi (t)-\xi (0)|Z ^{-\frac{2}{3}}\ge
\vartheta ^K|t|Z,
\label{25-4-53}
\end{gather}
where $m$ is arbitrary and $K,\delta $ depend on $k$.
\enlargethispage{\baselineskip}
\end{proposition}

\begin{proof}
We will just sketch the proof.

\begin{enumerate}[label=(\roman*), wide, labelindent=0pt]
\item\label{pf-25-4-11-i}
Consider first the case $M=1$. Then the the classical dynamical system is completely integrable since both the angular momentum and the energy are preserved.

First, let us include into $\Sigma '_{\lambda ,\vartheta }$ all the
points $(x,\xi )$ with trajectories not residing over
$\{\vartheta ^{\delta'}Z^{-\frac{1}{3}} \le \ell (x)\le Z^{-\frac{1}{3}}\vartheta ^{-\delta '}\}$
for all $t:|t|\le T$\,\footnote{\label{foot-25-21} I.e. trajectories, leaving this ``comfort zone'' for some $t:|t|\le T$.}.

Further, if $(Z-N)_+\ge Z\vartheta ^{\delta ''}$ we also include into
$\Sigma '_{\lambda ,\vartheta }$ all the points $(x,\xi )$ with the trajectories not residing over $B\bigl(\y_1, (1-\vartheta ^{\delta '})\bar{r}\bigr)$; recall that for atoms $\bar{r}$ is an \emph{exact\/} radius of $\supp (\rho^\TF)$. Then
(\ref{25-4-51})--(\ref{25-4-52}) hold and we reduce $\delta ,\delta '$ if necessary.

It is known that not all the Hamiltonian trajectories are closed (see Appendix~\ref{sect-25-A-2}). Then one can prove easily that adding to
$\Sigma '_{\lambda ,\vartheta }$ the set satisfying (\ref{25-4-51}) we can fulfill (\ref{25-4-51}) and (\ref{25-4-53}) as well\footnote{\label{foot-25-22} It is important that in the  classical dynamics is completely integrable.}.  One can find the similar arguments in the proof of Theorem~\ref{book_new-thm-7-4-12}.

\item\label{pf-25-4-11-ii}
The general case $M\ge 2$ is due of this particular one, Proposition~\ref{prop-25-4-10} and trivial perturbation arguments.
\end{enumerate}\end{proof}

\begin{proof}[Proof of Proposition~\ref{prop-25-4-9}]
Now  estimates (\ref{25-4-45}) and (\ref{25-4-46}) follow from Propositions~\ref{prop-25-4-10} and~\ref{prop-25-4-11} and the standard arguments. Note that if  $(Z-N)_+\ge Z\vartheta^{\delta''}$ we need to mollify $W^\TF$ in the standard way.

To prove estimate (\ref{25-4-47}) we can use decomposition (\ref{book_new-16-3-1}) and apply to the contribution of zone
$\{(x,y)\colon  |x-y|\ge Z^{-\frac{1}{2}}\vartheta^{\delta'}\}$ the same standard arguments. Meanwhile one can notice that the contribution of the zone $\{(x,y)\colon  |x-y|\ge Z^{-\frac{1}{2}}\vartheta^{\delta'}\}$ is $O(Z^{\frac{5}{3}-\delta})$.
\end{proof}

Combining all these improvements we arrive to

\begin{theorem}\label{thm-25-4-12}
Let $N\asymp Z$. Let $a\ge Z^{-\frac{1}{3}}$ and let $\Psi$ be a ground state. Then
\begin{equation}
|\E - \cE^\TF - \Scott-\Dirac -\Schwinger|\le
CZ^{\frac{5}{3}} \bigl(Z^{-\delta }+ (aZ^{\frac{1}{3}})^{-\delta}\bigr).
\label{25-4-54}
\end{equation}
\end{theorem}

\section{Corollaries and discussion}
\label{sect-25-4-4}

\subsection{Estimates for \texorpdfstring{$\D(\rho_\Psi-\rho^\TF, \rho_\Psi-\rho^\TF)$}{D(\textrho \_\textPsi- \textrho \^{}TF, \textrho \_\textPsi- \textrho \^{}TF)} }
\label{sect-25-4-4-1}

Recall that in the lower estimate there was in the left-hand expression a non-negative term $\frac{1}{2}\D(\rho_\Psi-\rho^\TF, \rho_\Psi-\rho^\TF)$ which we so far just dropped. Then in the frameworks of Theorems~\ref{thm-25-4-7} and ~\ref{thm-25-4-12}  we conclude that this term does not exceed the right-hand expressions of (\ref{25-4-42}) and (\ref{25-4-54})  respectively.

However, we can do better in the case $a\le Z^{-\frac{1}{3}}$. Indeed,  recall that the term $a^{-\frac{1}{2}}Z^{\frac{3}{2}}$ in the remainder estimate (\ref{25-4-42}) appeared \emph{only\/} because we replaced $\Tr( (H_W-\nu)^-)$ by its Weyl approximation (with correction terms) which we by no means need for estimating this term since $\Tr ((H_W-\nu)^-)$ was present in both lower and upper estimates. Then we arrive to the following
\begin{theorem}\label{thm-25-4-13}
Let $N\asymp Z$ and let $\Psi$ be a ground state. Then
\begin{multline}
\D(\rho_\Psi -\rho^\TF, \rho_\Psi -\rho^\TF)\le CQ\coloneqq   \\
C\left\{\begin{aligned}
&Z^{\frac{5}{3}} \qquad
&&\text{if \ \ } a \le Z^{-\frac{1}{3}},\\
&Z^{\frac{5}{3}} \bigl(Z^{-\delta }+ (aZ^{\frac{1}{3}})^{-\delta}\bigr)\qquad
&&\text{if \ \ }a \ge Z^{-\frac{1}{3}}.
\end{aligned}\right.
\label{25-4-55}
\end{multline}
\end{theorem}

\subsection{Estimates for distance between nuclei in the free nuclei model}
\label{sect-25-4-4-2}

Let us estimate from below the distance between nuclei in the stable molecule in the free nuclei model (with the full energy optimized with respect to the position of the nuclei).

\begin{theorem}\label{thm-25-4-14}
Let $M\ge 2$ and let condition \textup{(\ref{25-3-33})} be fulfilled. Assume that
\begin{multline}
\E (\underline{Z}; \underline{\y};N) +
\sum_{1\le m<m'\le M} Z_jZ_k|\y_m-\y_{m'}| ^{-1}\le\\
\min_{\substack {\\[2pt]N_1,\dots,N_M: \\[3pt]
N_1+\ldots+N_M=N}}
\sum_{1\le m\le M} \E (Z_m;N_m)
\label{25-4-56}
\end{multline}
Then
\begin{gather}
|\y_{m}-\y_{m'}|\ge Z^{-\frac{5}{21}+\delta_1}\qquad \forall m\ne m'
\label{25-4-57}\\
\shortintertext{and}
|\widehat{\E}^\TF (\underline{Z};N)- \widehat\cE^\TF (\underline{Z};N) -\Scott - \Dirac -\Schwinger |\le CZ^{\frac{5}{3}-\delta}.
\label{25-4-58}
\end{gather}
\end{theorem}

\begin{proof}
Note first that $|\E|\le CZ ^{\frac{7}{3}}$ and in virtue of Theorem~\ref{thm-25-4-7} we can replace $\E$ by $\cE^\TF$ with $O(Z^2)$ error:
\begin{multline}
\cE^\TF (\underline{Z}; \underline{\y};N) +
\sum_{1\le m<m'\le M} Z_jZ_k|\y_m-\y_{m'}| ^{-1}\le\\
\min_{\substack {\\[2pt]N_1,\dots,N_M: \\[3pt]
N_1+\ldots+N_M=N}}
\sum_{1\le m\le M} \cE^\TF (Z_m;N_m) - CZ^2;
\label{25-4-59}
\end{multline}
which in virtue of Proposition~\ref{prop-25-3-11} is impossible unless $a^{-7}\le CZ^2$ i.e. $a \ge \epsilon Z^{-\frac{2}{7}}\gg  Z^{-\frac{1}{3}}$.

Then, again in virtue of Theorem~\ref{thm-25-4-7} we can replace $\E$ by $\cE^\TF+\Scott $ with $O(Z^{\frac{5}{3}})$ error. Let us take into account that for molecule $\Scott$ equals the sum of $\Scott_m$. Therefore in (\ref{25-4-59}) we can replace $CZ^2$ by $CZ^{\frac{5}{3}}$. Applying again Proposition~\ref{prop-25-3-11} we conclude that
$a\ge \epsilon Z^{-\frac{5}{21}}$.

Let us improve this estimate further. In virtue of Theorem~\ref{thm-25-4-12} we can replace $\E$ by $\cE^\TF+\Scott +\Dirac+\Schwinger$ with $O(Z^{\frac{5}{3}-\delta_2})$ error. However one needs to compare Dirac--Schwinger correction for the molecule with the sums of such corrections for separate atoms:

\begin{proposition}\label{prop-25-4-15}
If $a\ge Z^{-\frac{1}{3}+\delta_1}$ and
\begin{gather}
\cE^\TF (\underline{Z}; \underline{\y};N) =
\sum_{1\le m\le M} \cE^\TF (Z_m;N_m) + O(Z^{\frac{7}{3}-\delta_1})
\label{25-4-60}
\intertext{where $N=N_1+\ldots+N_M$, then}
\int (\rho^\TF)^{\frac{4}{3}}\,dx =
\sum_{1\le m\le M} \int (\rho_m^ \TF)^{\frac{4}{3}}\,dx + O(Z^{\frac{5}{3}-\delta_2}),
\label{25-4-61}
\end{gather}
where $\rho^\TF _m=\rho^\TF (x-\y_m; Z_m; N_m)$ are atomic Thomas-Fermi densities.
\end{proposition}
\enlargethispage{\baselineskip}

Proof of this Proposition~\ref{prop-25-4-15} will be provided immediately. Therefore $\Dirac$ and $\Schwinger$ correction terms for a molecule are equal to the sums of $\Dirac_m$ and $\Schwinger_m$ correction terms respectively with $O(Z^{\frac{5}{3}-\delta_2})$ error and
in (\ref{25-4-59}) we can replace $CZ^2$ by $CZ^{\frac{5}{3}-\delta_2}$.

Applying Proposition~\ref{prop-25-3-11} again we conclude that
$a\ge Z^{\frac{5}{21}+\delta}$.
\end{proof}

\begin{proof}[Proof of Proposition~\ref{prop-25-4-15}]
Note that the left-hand expression of (\ref{25-3-42}) is equal to
\begin{equation}
\| \nabla \bigl(W^\TF-\bar{W}^\TF\bigr)\|^2 \qquad\text{with\ \ }
\bar{W}^\TF \coloneqq   \sum_{1\le m\le M} W^\TF_m
\label{25-4-62}
\end{equation}
and therefore this expression is less than $CQ \le CZ^{\frac{7}{3}-\delta_1}$ as well. Then since  $a\ge Z^{-\frac{1}{3}+\delta_1}$ we conclude that if we restrict the norm to the zone $\{x: |x-\y_m|\le Z^{-\frac{1}{3}+\delta}\}$, we can replace $\bar{\rho}^\TF$ and $\bar{W}^\TF$ by $\rho_m^\TF$ and  $W_m^\TF$ respectively in (\ref  {25-3-42}), (\ref{25-4-62}).

Using Thomas-Fermi equations we conclude that in this zone
\begin{equation}
|D^\alpha (W^\TF -W^\TF_m) |\le C_\alpha W_m^\TF \ell^{-|\alpha|} Z^{-\delta_2} \qquad \forall \alpha :|\alpha|\le 2
\label{25-4-63}
\end{equation}
and then
$|(\rho^\TF)^{\frac{4}{3}}-(\rho_m^\TF)^{\frac{4}{3}}|\le C(\rho_m^\TF)^{\frac{4}{3}}Z^{-\delta_2} $ which implies (\ref{25-4-61}) because
$\int (\rho_m^\TF)^{\frac{4}{3}}\,dx \asymp Z^{\frac{5}{3}}$ and contributions of zone $\{x:\ell(x)\ge Z^{-\frac{1}{3}+\delta}\}$ to each integral is $O(Z^{\frac{5}{3}-\delta})$.
\end{proof}

The following problem seems to be very challenging:

\begin{Problem}\label{Problem-25-4-16}
In the framework of the free nuclei model consider the case when assumption (\ref{25-3-33}) is violated, i.e. when some nuclei are much lighter than the others. We do not need this assumption for Theorems~\ref{thm-25-4-7},~\ref{thm-25-4-12} or~\ref{thm-25-4-13} but we need to estimate from below the minimal distance between nuclei $a$.

We cannot do this without some generalization of Proposition~\ref{prop-25-3-11}, which we definitely do not expect to survive in its current form without assumption (\ref{25-3-33}). It would be unrealistic to expect any estimate from below for $a_m\coloneqq \min_{m'\ne m}|\y_m-\y_{m'}|$ without some estimate from below for $Z_m$.
\end{Problem}

The following problem seems to be reasonably challenging:

\begin{Problem}\label{Problem-25-4-17}
\begin{enumerate}[label=(\roman*), wide, labelindent=0pt]
\item\label{Problem-25-4-17-i}
Let us discuss the case $M=2$ and $Z_2\ll Z_1$, $a\le Z^{-\frac{1}{3}}$. Then there is an unpleasant remainder $O(a^{-\frac{1}{2}}Z^{\frac{3}{2}})$ in the trace asymptotics. Let us discuss how one can improve it.

Let us observe that in $\bR^3\setminus B(\y_1, C b)$  we have
$W^\TF_1 \gg W^\TF_2$, where
$b =\min (Z_2^{-\frac{1}{3}+\delta}, aZ_2Z^{-1})$. One can expect that we can modify $W^\TF$ in $B(\y_1, C r_2)$ to $W$  so that the dynamical systems corresponding to Hamiltonians $H_W$ and $H_{W_1}$ would be close; then for $H_W$ we would be able to recover trace asymptotics with the remainder estimate $O(Z^{\frac{5}{3}-\delta})$.

Meanwhile, if  $b\ge Z_2^{-1}$, i.e. $a\ge Z_2^{-2}Z$,
the contribution of $B(\y_1, C r_2)$ to the trace remainder would be $O(b^{-\frac{1}{2}}Z_2^{\frac{3}{2}})=
O(Z_2^{\frac{5}{3}-\delta}) + O(a^{-1/2}Z_2 Z^{1/2})$ and we would improve $O(a^{-\frac{1}{2}}Z^{\frac{3}{2}})$ to
$O(Z^{\frac{5}{3}-\delta} + a^{-\frac{1}{2}}Z_2 Z^{\frac{1}{2}})$.

On the other hand, if  $b\le Z_2^{-1}$, i.e. $a\le Z_2^{-2}Z$,
the contribution of $B(\y_1, C r_2)$ to the trace remainder would be
$O(Z_2^2)$, so we would get the remainder $O(Z^{\frac{5}{3}-\delta} + Z_2 ^2)$.

In particular, we get $O(Z^{\frac{5}{3}-\delta})$ provided $Z_2\le Z^{\frac{5}{6}-\delta}$.

\item\label{Problem-25-4-17-ii}
Generalize for $M\ge 2$.
\end{enumerate}
\end{Problem}

\chapter{Negatively charged systems}
\label{sect-25-5}

In this section we consider the case $N\ge Z$ and provide upper estimates for excessive negative charge $(N-Z)$ if $\I_N>0$ and for ionization energy $\I_N\coloneqq \E_{N-1}-\I_N$.

\section{Estimates of the correlation function}
\label{sect-25-5-1}
First of all, we provide some estimates which will be used for both negatively and positively charged systems. Let us consider the ground-state function
$\Psi (x_1,\varsigma_1;\ldots;x_N,\varsigma_N)$ and the corresponding density $\rho _\Psi (x)$.

The crucial role plays estimate (\ref{25-4-55})
$\D\bigl(\rho _\psi -\rho ^\TF,\rho _\psi -\rho ^\TF\bigr)\le CQ$
of Theorem~\ref{thm-25-4-13} and the difference between upper and lower bounds for $\E_N$ (with $\N_1(H_W-\nu)+\nu N$ not replaced by its semiclassical approximation).

Let us consider the \emph{classical density\/}\index{density!classical} of the electron system
\index{density!classical!smeared}
\begin{gather}
\varrho _{\underline{x}}(x)=
\sum_{1\le j\le N} \updelta (x-x_j)
\label{25-5-1}
\intertext{and the \emph{smeared classical density\/}}
\varrho _{\underline{x},\varepsilon}(x)=
\varrho _{\underline{x}} * \phi _\varepsilon=
\sum_{1\le j\le N} \phi _\varepsilon (x-x_j)
\label{25-5-2}
\end{gather}
where $\varepsilon $ will be chosen later; here
$(\underline{x},\underline{\varsigma})=
(x_1,\varsigma_1;\ldots, x_N,\varsigma_N)\in (\bR^3 \times \bC^q)^N$,
\begin{claim}\label{25-5-3}
$\phi _\varepsilon (z)=\varepsilon ^{-3}\phi (z\varepsilon ^{-1})$,
$\phi \in \sC_0 ^\infty \bigl(B(0,\frac{1}{2})\bigr)$ is a spherically symmetric, non-negative function such that  $\int  \phi (x)\,dx =1$.
\end{claim}
Then
$\int \phi _\varepsilon (x)f(x)\,dx = f(0)+ O(\varepsilon ^2)$ for any
$f\in \sC ^2$.  Let us consider
\begin{equation}
K_N(\underline{x})\coloneqq
\frac{1}{2}\D\bigl(\varrho _{\underline{x}}(\cdot )-\rho(\cdot ),
\varrho _{\underline{x}}(\cdot )-\rho (\cdot )\bigr)
\label{25-5-4}
\end{equation}
where $\rho=\frac{1}{4\pi}\Delta (W-V)$, $W$ is either $W^\TF$ if $\nu=0$ or a ``good'' approximation for it, constructed in the previous section.

Using Newton's screening theorem\footnote{\label{foot-25-23} That uniformly charged sphere $S(0,r)$ creates potential $v(x)= -q \min(|x|^{-1},r^{-1})$ where $q$ is the total charge of the sphere.} we conclude that
\begin{equation}
\sum_{1\le j< k\le N}|x_j-x_k| ^{-1}\ge
\frac{1}{2}\D\bigl(\varrho _{\underline{x}}(\cdot ),
\varrho _{\underline{x}}(\cdot ) \bigr) - C\varepsilon ^{-1}N
\label{25-5-5}
\end{equation}
where the last term estimates
\begin{equation*}
\sum_{1\le j\le N} \D \bigl(\phi _\varepsilon (x-x_j), \phi _\varepsilon (x-x_j)\bigr).
\end{equation*}

On the other hand,
\begin{gather}
\frac{1}{2}\D\bigl(\varrho _{\underline{x}}(\cdot ),
\varrho _{\underline{x}}(\cdot ) \bigr)=
\int \varrho _{\underline{x}}(|x| ^{-1}*\rho)\,dx +
K_N(\underline{x}) -\frac{1}{2}\D(\rho, \rho)
\label{25-5-6}\\
\intertext{and therefore}
\mathbf{H}_N\ge \sum_{1\le j\le N} H_{W_\varepsilon}(x_j) + K_N(\underline{x})-
\frac{1}{2}\D(\rho, \rho) - C\varepsilon ^{-1}N
\label{25-5-7}
\end{gather}
on $(\sL ^2(\bR^3 ,\bC^q ))^N$ where $W_\varepsilon$ is the \emph{smeared potential\/}\index{potential!smeared}:
\begin{equation}
W_\varepsilon (x)=V(x)-\phi _\varepsilon *|x| ^{-1}*\rho.
\label{25-5-8}
\end{equation}

Observe that the smeared potential does not depend on $\underline{x}$
and is defined via $\rho$ rather than $\rho _\Psi $.

Also let us define
\begin{gather}
N_x(x_2,\dots,x_N)\coloneqq   \sum_{2\le j\le N}(\chi _x*\phi _\varepsilon)(x_j)
\label{25-5-9}\\
\shortintertext{and}
\bar{N}_x \coloneqq  \int \rho(y)\chi _x(y)dy
\label{25-5-10}
\end{gather}
with $\chi _x(y)\coloneqq   \chi (x,y)$, $\chi \in \sC ^\infty (\bR ^6)$.

Furthermore, let us consider function $\theta \in \sC^\infty (\bR^3)$ such that
\begin{equation}
0\le \theta \le 1,\qquad
|\nabla ^\alpha \theta^{\frac{1}{2}}  |\le c_\alpha b ^{-|\alpha |} \quad
 \forall \alpha.
\label{25-5-11}
\end{equation}
Finally, consider
\begin{equation}
\cJ \coloneqq   |\int \Bigl(\rho _\Psi ^{(2)}(x,y) -
\rho(y)\rho_\Psi (x)\Bigr)\theta (x)\chi (x,y)\,dxdy|.
\label{25-5-12}
\end{equation}
Obviously
\begin{multline*}
\cJ \le \int \rho _\Psi ^{(2)}(x,y)
\bigl|\chi _x * \phi _\varepsilon -\chi _x(y)\bigr| \theta (x)\,dxdy \\
\begin{aligned}
&+ N\int |\Psi (x,x_2,\dots,x_N)| ^2
\bigl|N_x(x_2,\dots,x_N)-\bar{N}_x\bigr|\theta (x)\, dxdx_2\cdots dx_N \\[5pt]
& \le CN\varepsilon ^s\|\nabla ^{s+1}_y\chi \|_{\sL ^\infty}\Theta
\end{aligned}\\
+\Bigl(N\int |\Psi (x,x_2,\dots,x_N)| ^2|N_x-{\bar N}_x | ^2
\theta (x)\, dxdx_2\cdots dx_N\Bigr) ^{\frac{1}{2}}\Theta ^{\frac{1}{2}}
\end{multline*}
where $\rho ^{(2)}_\Psi (\cdot,\cdot)$ is the \emph{quantum correlation function,\/}\index{quantum correlation function}
\begin{gather}
\rho ^{(2)}_\Psi (x,y)\coloneqq   N(N-1)\int |\Psi (x,y,x_3,\ldots,x_N)| ^2\, dx_3\cdots dx_N, \label{25-5-13}\\
\int \rho ^{(2)}_\Psi (x,y)dy =(N-1)\rho _\Psi (x),\label{25-5-14}\\
\Theta =\Theta_\Psi \coloneqq  \int \theta (x)\rho _\Psi (x)dx\label{25-5-15}
\end{gather}
and we used Cauchy-Schwartz inequality.

Since
\begin{gather}
N_x(x_2,\ldots,x_N)-\bar{N}_x=
\int \bigl(\sum_{2\le j\le N}\phi _\varepsilon (y-x_j)-
\rho(y)\bigr)\chi (x,y)\,dxdy,\label{25-5-16}\\
\intertext{we again from Cauchy-Schwartz inequality conclude that}
\bigl|N_x-\bar{N}_x | ^2\le
C\|\nabla _y\chi \| ^2_{\sL ^2}\cdot K_{N-1} (x_2,\ldots,x_N).
\label{25-5-17}
\end{gather}
Note that
\begin{multline}
\sum_{1\le j\le N}\blangle \mathbf{H}_N \theta^{\frac{1}{2}} (x_j)\Psi ,
\theta^{\frac{1}{2}} (x_j)\Psi \brangle =\\
\E_N \int \int \theta (x)\rho _\Psi (x)\,dx +
\int |\nabla \theta^{\frac{1}{2}}  | ^2(x)\rho _\Psi (x)\,dx
\label{25-5-18}
\end{multline}
and then (\ref{25-5-7}) yields that
\begin{multline}
\E_N\int \theta (x)\rho _\Psi (x)\, dx\ge
-\int |\nabla \theta^{\frac{1}{2}} | ^2(x)\rho _\Psi (x)\, dx\\
\shoveright{+\sum_{1\le j,k\le N}
\blangle H_{W_\varepsilon}(x_k) \theta^{\frac{1}{2}} (x_j)\Psi ,\theta  (x_j)\Psi\brangle}\\ + \int K_N(\underline{x})\theta (x_1)|\Psi (x_1,\ldots,x_N)| ^2\,dx_1\ldots dx_N
-\frac{1}{2}\D(\rho,\rho)\Theta - C\varepsilon ^{-1}N\Theta .
\label{25-5-19}
\end{multline}
Also note that
\begin{claim}\label{25-5-20}
The sum of $(N-1)$ lowest eigenvalues of $H_{W_\varepsilon}$ on ${\cH}$ is greater than $\bigl(\nu N+\N_1(H_{W_\varepsilon}-\nu )\bigr)$.
\end{claim}

Then the second term in the right-hand expression of (\ref{25-5-19}) is bounded from below by  $\bigl(\nu N+\N_1(H_{W_\varepsilon}-\nu )\bigr)\Theta$, while the left-hand expression is $\E_N\Theta$. Therefore assembling terms proportional to $\Theta$ we conclude that
\begin{multline}
S\Theta + \int |\nabla \theta^{\frac{1}{2}}|^2 \rho_\Psi \,dx \ge
\sum_j \blangle H_{W_\varepsilon,x_j}\theta^{\frac{1}{2}}(x_j)\Psi ,
\theta^{\frac{1}{2}}(x_j)\Psi \brangle \\
+N\int K_N(x_1,\ldots,x_N)\theta (x_1)|\Psi (x_1,\ldots,x_N)|^2\,
dx_1\cdots dx_N
\label{25-5-21}
\end{multline}
with
\begin{equation}
S \coloneqq \E_N -\nu N-\N_1(H_{W_\varepsilon}-\nu ) +
\frac{1}{2}\D(\rho ,\rho).
\label{25-5-22}
\end{equation}
Due to the non-negativity of operator $D_x^2$, the last term in (\ref{25-5-21}) is greater than $-CT\Theta $ with
\begin{equation}
T= \sup _{\supp(\theta)} W;
\label{25-5-23}
\end{equation}
so we arrive to
\begin{multline}
N\int K_N(x_1,\ldots,x_N)\theta (x_1)|\Psi (x_1,\ldots,x_N)| ^2\,
dx_1\cdots dx_N \le \\
C\bigl(S+T+\varepsilon ^{-1}N\bigr)\Theta +P
\label{25-5-24}
\end{multline}
with
\begin{equation}
P= \int |\nabla \theta^{\frac{1}{2}}|^2 \rho_\Psi \,dx.
\label{25-5-25}
\end{equation}
Combining this inequality with (\ref{25-5-13}), (\ref{25-5-17}) we conclude that
\begin{multline}
\cJ \le C\sup_x \|\nabla _y\chi _x\|_{\sL^2(\bR ^3)}
\Bigl(\bigl( S +\varepsilon ^{-1}N +T \bigr)\Theta +
P \Bigr) ^{\frac{1}{2}}\Theta ^{\frac{1}{2}} \\
+C\varepsilon N\|\nabla _y\chi \|_{\sL ^\infty }\Theta
\label{25-5-26}
\end{multline}
due to obvious estimate
\begin{equation}
K_{N-1}(x_2,\ldots,x_N)\le 2K_N(x_1,\ldots,x_N)+\varepsilon ^{-1}N.
\label{25-5-27}
\end{equation}

Now we want to estimate $S$ from above and for this we need an upper estimate for $\E_N$. Recall that due to the arguments of Subsection~\ref{sect-25-2-2}
$S\le CQ$ provided we manage to prove that

\begin{claim}\label{25-5-28}
Expressions (\ref{25-2-23})--(\ref{25-2-26}) satisfy the same estimates as before if we plug $W_\varepsilon$ instead of $W$.
\end{claim}

So, we need to calculate both semiclassical errors (which are calculated exactly as for $W=W^\TF$) and the principal parts, and in calculations of
\begin{gather}
\D\bigl(P'(W_\varepsilon +\nu)-\rho^\TF, P'(W_\varepsilon +\nu)-\rho^\TF\bigr)
\label{25-5-29}\\
\shortintertext{and}
\D(\rho_\varepsilon -\rho^\TF, \rho_\varepsilon -\rho^\TF)
\label{25-5-30}
\end{gather}
 an error is $O(\varepsilon^2 Z^3)$ due to estimates
\begin{multline}
|D^\alpha(W-\W_\varepsilon)|\le \\
C\left\{\begin{aligned}
&Z\ell^{-1-|\alpha|}  (1+ \ell \varepsilon^{-1}) ^{-2}\quad &&\forall \alpha \quad&&\text{if\ \ } \ell\le \epsilon Z^{-\frac{1}{3}}\\
& \varepsilon^2\ell^{-6-|\alpha|}  \quad &&\forall \alpha  \quad&&\text{if\ \ } \ell\ge \epsilon Z^{-\frac{1}{3}},
\ell\not \asymp\bar{r},\\
&\varepsilon^2\ell^{-6-|\alpha|} \quad &&\forall \alpha:|\alpha|\le \frac{3}{2}\quad&&\text{if\ \ } \ell\asymp\bar{r}.
\end{aligned}\right.
\label{25-5-31}
\end{multline}

Then one can prove easily that the sum of these two expressions (\ref{25-5-29}) and (\ref{25-5-30})  does not exceed $CQ+ CZ^3\varepsilon^2 + CZ^2\varepsilon$, and this estimate cannot be improved. Choosing $\varepsilon \le Z^{-\frac{2}{3}}$ we estimate these two terms by $CZ^{\frac{5}{3}}$.

Under this restriction a smearing error in the principal part of the asymptotics of $\int e(x,x,\lambda)\,dx$, namely
\begin{equation}
|\int  \bigl(P'(W_\varepsilon +\nu)-P'(W +\nu)\bigr)\,dx|,
\label{25-5-32}
\end{equation}
does not exceed $CZ^{\frac{3}{2}}\varepsilon ^{\frac{3}{2}}=O(Z^{\frac{1}{2}})$ which is less than the semiclassical error. Then $S\le CQ$.

So, the following proposition is proven:

\begin{proposition}\label{prop-25-5-1}
If $\theta ,\chi $ are as above  then
\begin{multline}
\cJ=|\int \Bigl(\rho _\Psi ^{(2)}(x,y)-
\rho(y)\rho _\Psi (x)\Bigr)\theta (x)\chi (x,y)\,dxdy |\le \\[3pt]
C\sup_x \|\nabla _y\chi _x\|_{\sL ^2(\bR ^3)}
\Bigl((Q+\varepsilon ^{-1}N+T)^{\frac{1}{2}}\Theta + P^{\frac{1}{2}}\Theta^{\frac{1}{2}}\Bigr) +
C \varepsilon N\|\nabla _y\chi \|_{\sL ^\infty }\Theta
\label{25-5-33}
\end{multline}
with $\Theta=\Theta_\Psi$ defined by \textup{(\ref{25-5-15})} and $T,P$ defined by \textup{(\ref{25-5-23})}, \textup{(\ref{25-5-25})} respectively and arbitrary
$\varepsilon \le Z^{-\frac{2}{3}}$.
\end{proposition}

\section{Excessive negative charge}
\label{sect-25-5-2}
Let us select $\theta=\theta_b$
\begin{equation}
\supp (\theta)  \subset \{x:\ \ell (x)\ge b\}.
\label{25-5-34}
\end{equation}
Observe  that $\mathbf{H}_N\Psi =\E_N\Psi $ yields
\begin{multline}
\E_N\int \rho _\Psi (x)\ell (x)\theta (x)\,dx=
\sum_j \blangle \Psi ,\ell (x_j)\theta (x_j)\mathbf{H}_N\Psi \brangle =\\[3pt]
\sum_j \blangle \ell (x_j) ^{\frac{1}{2}}\theta^{\frac{1}{2}} (x_j) \Psi ,
\ell (x_j)^{\frac{1}{2}}\theta^{\frac{1}{2}}(x_j)\mathbf{H}_N\Psi \brangle
- \sum_j
\|\nabla \bigl(\theta^{\frac{1}{2}} (x_j) \ell (x_j)^{\frac{1}{2}}\bigr)\Psi\|^2
\label{25-5-35}
\end{multline}
and isolating the contribution of $j$-th electron in $j$-th term we get
\begin{multline}
\E_N\int \rho _\Psi (x)\ell (x)\theta \,dx\ge
\E_{N-1}\int \rho _\Psi (x)\ell (x)\theta (x)\,dx+ \\[3pt]
\sum_j\blangle \Psi ,\ell (x_j)\theta (x_j)
\Bigl(-V(x_j)+ \sum_{k: k\ne j}|x_j-x_k| ^{-1}\Bigr)\Psi \brangle -
\sum_j
\|\nabla \bigl(\theta^{\frac{1}{2}} (x_j) \ell (x_j)^{\frac{1}{2}}\bigr)\Psi\|^2
\label{25-5-36}
\end{multline}
due to non-negativity of operator $D_x^2$.

Now let us select $b$ to be able to calculate the magnitude of $\Theta$. Observe that
\begin{multline}
|\int \theta (x) \bigl(\rho_\Psi (x)-\rho (x)\bigr)\,dx | \le
C \D(\rho_\Psi-\rho, \rho_\Psi-\rho)^{\frac{1}{2}}
\|\nabla \theta^{\frac{1}{2}} \| \asymp\\
C \D(\rho_\Psi-\rho, \rho_\Psi-\rho)^{\frac{1}{2}} b^{\frac{1}{2}}\le CQ^{\frac{1}{2}} b^{\frac{1}{2}}
\label{25-5-37}
\end{multline}
and
\begin{gather}
\int \theta (x) \rho (x) \,dx  \asymp b^{-3}
\label{25-5-38}\\
\shortintertext{as long as}
Z^{-\frac{1}{3}}\le b\le \epsilon (Z-N)_+^{-\frac{1}{3}}
\label{25-5-39}
\end{gather}
because $\rho \asymp |x|^{-6}$ as
$Z^{-\frac{1}{3}}\le |x|\le c_0 (Z-N)_+^{-\frac{1}{3}}$. Note that the right-hand expression of (\ref{25-5-38}) is larger than the  right-hand expression of (\ref{25-5-38}) as $b\ge C_0Q^{-\frac{1}{7}}$. Therefore let us pick up
\begin{gather}
b\coloneqq \epsilon_0 Q^{-\frac{1}{7}};\label{25-5-40}\\
\intertext{it does not conflict with (\ref{25-5-39}) provided}
N\ge Z- C_0 Q^{\frac{3}{7}}\label{25-5-41}\\
\shortintertext{and then}
\Theta \asymp Q^{\frac{3}{7}}.
\label{25-5-42}
\end{gather}
Then the same arguments imply that the last term in the right-hand expression of (\ref{25-5-36}) does not exceed $C b^{-1}\Theta$; using inequality
\begin{equation}
\int \rho _\Psi (x)\ell (x)\theta  (x)dx \ge b\Theta
\label{25-5-43}
\end{equation}
we conclude that
\begin{multline}
b\I_N\Theta _b\le \int \theta (x)V(x)\ell (x)\rho _\Psi (x)\,dx\\
\shoveright{-\int\rho _\Psi ^{(2)}(x,y)\ell (x)|x-y| ^{-1}\theta (x)\,dxdy +
Cb ^{-1}\Theta=}\\
\begin{aligned}=&\int \theta (x)V(x)\ell (x)\rho _\Psi (x)\,dx\\
-&\int \rho _\Psi ^{(2)}(x,y)\ell (x)|x-y| ^{-1}
\bigl(1-\theta (y)\bigr)\theta (x)\,dxdy \\
-&\int \rho _\Psi ^{(2)}(x,y)\ell (x)|x-y| ^{-1}\theta (y)\theta (x)\,dxdy + Cb ^{-1}\Theta .\end{aligned}
\label{25-5-44}
\end{multline}
Denote by $\cI_1$, $\cI_2$, and $\cI_3$ the first, second and third terms in the right-hand expression of (\ref{25-5-44}) respectively. Symmetrizing $\cI_3$ in the right-hand expression of (\ref{25-5-44}) with respect to $x$ and $y$ we see that
\begin{equation*}
\cI_3=-\frac{1}{2}
\int \rho _\Psi ^{(2)}(x,y)\bigl(\ell(x)+\ell(y)\bigr)
|x-y| ^{-1}\theta (y)\theta (x)\,dxdy
\end{equation*}
and using inequality
$\ell (x)+\ell (y)\ge \min_j(|x-\y_j|+|y-\y_j|)\ge |x-y|$ we conclude that this term is less than
\begin{multline}
-\frac{1}{2}\int \rho _\Psi ^{(2)}(x,y)\theta (y)\theta (x)\,dxdy=\\[5pt]
-\frac{1}{2}(N-1) \int \rho _\Psi (x)\theta (x)\,dx +
\frac{1}{2}
\int \rho _\Psi ^{(2)}(x,y)\bigl(1-\theta (y)\bigr) \theta (x)\, dxdy.
\label{25-5-45}
\end{multline}
Here the first term is exactly $-\frac{1}{2}(N-1)\Theta$; replacing
$\rho _\Psi ^{(2)}(x,y)$ by $\rho (y)\rho_\Psi(x)$ we get
\begin{gather}
\frac{1}{2}\int \bigl(1-\theta (y)\bigr)\rho(y)\,dy \times \Theta
\label{25-5-46}\\
\shortintertext{with an error}
\frac{1}{2}
\int \bigl(\rho _\Psi ^{(2)}(x,y)-\rho(y)\rho_\Psi (x)\bigr)
\bigl(1-\theta  (y)\bigr) \theta (x)\, dxdy
\label{25-5-47}
\end{gather}
which we estimate using Proposition~\ref{prop-25-5-1} with
$\chi(x,y) = 1-\theta (y)$. Then
$\|\nabla_y \chi_x\|_{\sL^2}\asymp b^{\frac{1}{2}}$,
$\|\nabla_y \chi\|_{\sL^\infty}\asymp b^{-1}$, $T\asymp b^{-4}$,
$P\asymp b^{-1}\Theta$ and picking up $\varepsilon=Z^{-\frac{2}{3}}$ we conclude that an error (\ref{25-5-47}) is less than
$Cb^{\frac{1}{2}}Q^{\frac{1}{2}} \Theta \asymp CQ^{\frac{3}{7}} \Theta$ and then we conclude that
\begin{equation}
\cI_3\le -\frac{1}{2}(N-Z)_+\Theta + CQ^{\frac{3}{7}} \Theta
\label{25-5-48}
\end{equation}
because $\int \rho \theta \,dy\asymp Q^{\frac{3}{7}}$ and
$\int  \rho(y)\,dy =\min (Z,N)$.

On the other hand,
\begin{equation}
\cI_2\le -\int \rho _\Psi ^{(2)}(x,y)\ell (x)|x-y| ^{-1}
\bigl(1-\bar{\theta} (y)\bigr)\theta (x)\,dxdy
\label{25-5-49}
\end{equation}
with $\bar{\theta}=\theta _{b(1-\epsilon )}$ and replacing
$\rho _\Psi ^{(2)}(x,y)$ by $\rho (y)\rho_\Psi(x)$ we get
\begin{equation}
 -\int \rho _\Psi (x) \rho (y) \ell (x)|x-y| ^{-1}
\bigl(1-\bar{\theta } (y)\bigr)\theta (x)\,dxdy
\label{25-5-50}
\end{equation}
and we estimate an error in the same way by $CQ^{\frac{3}{7}}\Theta$.

Therefore
\begin{multline}
\cI_1+\cI_2\le \int \theta (x)V(x)\ell (x)\rho _\Psi (x)\,dx-\\[3pt]
\shoveright{\int \rho _\Psi (x) \ell (x)\rho (y) |x-y| ^{-1}
\bigl(1-\bar{\theta} (y)\bigr)\theta (x)\,dxdy +
CQ^{\frac{3}{7}}\Theta=}\\
\shoveleft{\int \theta  (x)W(x)\ell (x)\rho _\Psi (x)\,dx }\\
+ \int \rho (y)\rho _\Psi (x)\ell (x)|x-y| ^{-1}
\bar{\theta}(y) \theta (x)\,dxdy+CQ^{\frac{3}{7}}\Theta
\label{25-5-51}
\end{multline}
due to $V-W= |x|^{-1}*\rho$. Since $W\ell \le Cb^{-3}$ we can skip the first term in the right-hand expression. Furthermore, as
\begin{equation}
\int \rho (y) |x-y|^{-1}\bar{\theta}(y)\,dy \asymp \Theta \asymp Q^{\frac{3}{7}}
\label{25-5-52}
\end{equation}
we can skip the second term as well.

Adding $\cI_3$ to and multiplying by $\Theta^{-1}$ we arrive to
\begin{equation}
b\I_N\le  -\frac{1}{2} (N-Z)_+  + CQ ^{\frac{3}{7}}
\label{25-5-53}
\end{equation}
which implies immediately

\begin{theorem}\label{thm-25-5-2}
Let condition \textup{(\ref  {25-3-33})} be fulfilled.
\begin{enumerate}[label=(\roman*), wide, labelindent=0pt]
\item\label{thm-25-5-2-i}
In the fixed nuclei model let \ $\I_N>0$. Then
\begin{equation}
(N-Z)_+\le CQ^{\frac{3}{7}} =CZ^{\frac{5}{7}}
\left\{\begin{aligned}
&1 \qquad &&\text{if\ \ } a\le Z^{-\frac{1}{3}},\\
&Z^{-\delta}+ (aZ^{\frac{1}{3}})^{-\delta}
&&\text{if\ \ } a\ge Z^{-\frac{1}{3}}.
\end{aligned}\right.
\label{25-5-54}
\end{equation}
\item\label{thm-25-5-2-ii}
In particular, for a single atom and for molecule with
$a\ge Z^{-\frac{1}{3}+\delta}$
\begin{equation}
(N-Z)_+\le Z^{\frac{5}{7}-\delta'}.
\label{25-5-55}
\end{equation}
\item\label{thm-25-5-2-iii}
In the free nuclei model let \ $\widehat{\I}_N>0$. Then estimate \textup{(\ref{25-5-55})} holds.
\end{enumerate}
\end{theorem}

\section{Estimate for ionization energy}
\label{sect-25-5-3}

Recall that as $N<Z$ we assumed that $N\ge Z-CQ^{\frac{3}{7}}$ (see (\ref{25-5-41})) and $b= Q^{-\frac{1}{7}}$. Then (\ref{25-5-53}) also implies $\I_N \le CQ^{\frac{4}{7}}$ and we arrive to

\begin{theorem}\label{thm-25-5-3}
Let condition \textup{(\ref  {25-3-33})} be fulfilled and let
$N\ge Z-C_0 Z^{\frac{5}{7}}$. Then
\begin{enumerate}[label=(\roman*), wide, labelindent=0pt]
\item\label{thm-25-5-3-i}
In the framework of fixed nuclei model
\begin{equation}
\I_N \le  CZ^{\frac{20}{21}}.
\label{25-5-56}
\end{equation}
\item\label{thm-25-5-3-ii}
In the framework of free nuclei model with $N\ge Z-C_0 Z^{\frac{5}{7}-\delta}$
\begin{equation}
\widehat{\I}_N \le Z^{\frac{20}{21}-\delta'}.
\label{25-5-57}
\end{equation}
\end{enumerate}
\end{theorem}

\begin{remark}\label{rem-25-5-4}
\begin{enumerate}[label=(\roman*), wide, labelindent=0pt]
\item\label{rem-25-5-4-i}
Classical theorem of G.~Zhislin~\cite{zhislin:spectrum} implies that the system can bind at least $Z$ electrons; the proof is based on the demonstration that the energy of the system with $N<Z$ electrons plus one electron on the distance $r$ is increasing as $r\to  +\infty$ because potential created by the system with $N<Z$ electrons behaves as $(Z-N)|x|^{-1}$ as $|x|\to \infty$;

\item\label{rem-25-5-4-ii}
In the proof of Theorem~\ref{thm-25-5-2}\ref{thm-25-5-2-iii} and~\ref{thm-25-5-3}\ref{thm-25-5-3-ii} we note that tearing off one electron in free nuclei model is easier than in the fixed nuclei model.
\end{enumerate}
\end{remark}

The following problem does not look extremely challenging:

\begin{Problem}\label{Problem-25-5-5}
In  Theorems~\ref{thm-25-5-2}\ref{thm-25-5-2-i},\ref{thm-25-5-2-ii} and~\ref{thm-25-5-3}\ref{thm-25-5-3-i} get rid off condition (\ref{25-3-33}).
\end{Problem}

\chapter{Positively charged systems}
\label{sect-25-6}

In this section we consider the case of positively charged system with
$Z-N\ge C_0Q^{\frac{3}{7}}$ with sufficiently large $C_0$.

First let us find asymptotics of the ionization energy; the principal term will be $-\nu$ but we need to estimate a remainder.

\section{Estimate from above for ionization energy}
\label{sect-25-6-1}

\begin{wrapfigure}[12]{r}[10pt]{5.5truecm}
\begin{tikzpicture}[scale=5]
\clip (1.2,-1.65) rectangle (2.4,-.7);
\draw[very thick] plot [domain=1:2] function {-4/(x**2+1)} node [right] {$-W(x)$};
\draw  (1,-1)--(2,-1) node [right] {$\nu $};
\draw (1,-1.2) --(2,-1.2) node [right] {$\nu-2\upsilon$};
\draw [dashed] (1.53,-2)--(1.53,-1);
\draw [thick, dotted] (1.63,-2)--(1.63,-1);
\draw [ultra thick, |-> ](1.53,-1.5)--(2,-1.5);
\node at (1.8,-1.5) {\colorbox{white}{$\operatorname{supp}\theta$}};
\draw [ultra thick, |-> ](1.63,-1.4)--(2,-1.4);
\node at (1.8,-1.4) {\colorbox{white}{$\theta=1$}};
\end{tikzpicture}
\end{wrapfigure}
As $M=1$ construction is well-known: let us pick up function $\theta$ such that  $\theta =1$ as  $|x-\y_m| \ge \bar{r}-\beta$  and $\theta =0$ as
$|x-\y_m|\le \bar{r}-2\beta$ where $\bar{r}$ is an exact radius of support $\rho^\TF$. Here $\beta\ll \bar{r}$. As $M\ge 1$ let us pick instead
\begin{equation}
\theta (x) = f^2 \bigl(\upsilon^{-1}[W(x)+\nu]\bigr)
\label{25-6-1}
\end{equation}
where $f\in \sC^\infty (\bR)$, supported in $(-\infty, 2)$ and equal $1$ in $(-\infty,1)$, $\upsilon \le \nu$.

We will assume that
\begin{equation}
a\ge \bar{r}= C_1 (Z-N)^{-\frac{1}{3}}
\label{25-6-2}
\end{equation}
with sufficiently large $C_1$; we will discuss dropping this assumption later. Then as we know that $\rho^\TF$ is supported in $c\bar{r}$-vicinity of nuclei, we conclude that ``atoms'' are rather disjoint.

One can see easily that then as $W+\nu \approx 0$,
\begin{gather}
|\nabla W|\asymp \bar{r}^{-5}\label{25-6-3};\\
\intertext{then the width of the zone $\{x\colon  0\le  W(x)+\nu \le 2\upsilon \}$ is $\asymp \upsilon |\nabla W|^{-1}\asymp \beta= \upsilon \bar{r}^5$ and}
\Theta^\TF \coloneqq   \int \theta (x)\rho^\TF\, dx \asymp
\upsilon ^{\frac{3}{2}} \times \beta \bar{r}^2=
\upsilon ^{\frac{5}{2}} \bar{r}^7
\label{25-6-4}
\end{gather}
while $\|\nabla \theta\|\asymp \beta^{-\frac{1}{2}}\bar{r}=
\upsilon^{-\frac{1}{2}} \bar{r}^{-\frac{3}{2}}$ and therefore to ensure that $\Theta$ has the same magnitude (\ref{25-6-4}) we pick up the smallest $\upsilon$ such that
$\upsilon ^{\frac{5}{2}} \bar{r}^7
\ge C \upsilon^{-\frac{1}{2}} \bar{r}^{-\frac{3}{2}} Q^{\frac{1}{2}}$ i.e.
\begin{gather}
\upsilon\coloneqq C_2 \bar{r}^{-\frac{17}{6}}Q^{\frac{1}{6}}\qquad (\iff  \beta= C_2 \bar{r}^{\frac{13}{6}} Q^{\frac{1}{6}});
\label{25-6-5}\\
\shortintertext{then}
\Theta \asymp \bar{r}^{-\frac{1}{12}}Q^{\frac{5}{12}}
\label{25-6-6}
\end{gather}
Then (\ref{25-5-15}) is fulfilled. Note that  $\upsilon\le \nu=\bar{r}^{-4} $ iff $(Z-N)\ge Q^{\frac{3}{7}}$ exactly as we assumed.

Then (\ref{25-5-44}) is replaced by
\begin{multline}
\I_N \int \ell(x) \rho_\Psi (x)\theta (x)\,dx \le
\int \theta (x)V(x)\ell (x)\rho _\Psi (x)\,dx \\
-\int \rho _\Psi ^{(2)}(x,y)
\ell (x)|x-y| ^{-1}\theta (x)\, dxdy + C\beta^{-2}\bar{r}\Theta
\label{25-6-7}
\end{multline}
where $C\beta^{-2}\bar{r}\Theta $ estimates the last term in the right-hand expression of (\ref{25-5-36}). Then
\begin{multline}
\I_N \int \ell(x) \rho_\Psi (x)\theta (x)\,dx \le
\int \theta (x)V(x)\ell (x)\rho _\Psi (x)\,dx \\
\begin{aligned}
-&\int \Bigl(\rho _\Psi ^{(2)}(x,y)-\rho _\Psi (x)\rho (y)\Bigr)
\ell (x)|x-y| ^{-1}\theta (x)\, dxdy \\
-&\int \rho _\Psi (x)\rho (y)\ell (x)|x-y| ^{-1}\theta (x)\, dxdy + C\beta^{-2}\bar{r}\Theta .
\end{aligned}
\label{25-6-8}
\end{multline}

Let us estimate from above
\begin{multline}
-\int \Bigl(\rho _\Psi ^{(2)}(x,y)-\rho _\Psi(x)\rho(y)\Bigr)
\ell (x)|x-y| ^{-1}\theta (x)\, dxdy\le \displaybreak\\
-\int \Bigl(\rho _\Psi ^{(2)}(x,y)-\rho _\Psi (x)\rho (y)\Bigr)
\bigl(1-\omega  (x,y)\bigr)\ell (x)|x-y| ^{-1}\theta (x)\,dxdy \\
+\int \rho_\Psi (x)\rho(y)\omega  (x,y) \ell(x) |x-y|^{-1}\theta (x)\,dxdy
 \label{25-6-9}
\end{multline}
with $\omega=\omega_\gamma$: $\omega =0$ for $|x-y|\ge 2\gamma$ and $\omega =1$ for  $|x-y|\le \gamma$, $\gamma \ge \beta$.

To estimate the first term in the right-hand expression one can apply Proposition~\ref{prop-25-5-1}. In this case
$\|\nabla_y\chi \|_{\sL^2} \asymp C\bar{r}\gamma^{-\frac{1}{2}}$,
$\|\nabla_y\chi \|_{\sL^\infty} \asymp \bar{r} \gamma^{-2}$
and plugging $P=\beta^{-2}\Theta$ and $T=|\nu|$, $\varepsilon =Z^{-\frac{2}{3}}$ we conclude that this term does not exceed
\begin{equation}
C\bar{r}\bigl(\gamma^{-\frac{1}{2}}\bar{Q}^{\frac{1}{2}} + CZ^{\frac{2}{3}} \gamma^{-2}\bigr) \Theta, \qquad \bar{Q}=\max(Q,Z^{\frac{5}{3}}).
 \label{25-6-10}
\end{equation}
Consider the second term in the right-hand expression of (\ref{25-6-9}). Note that
\begin{equation*}
\int \rho (y) \omega (x,y) |x-y|^{-1}\,dy \le
C(\bar{r}^{-\frac{15}{2}}\gamma^{\frac{7}{2}}+\upsilon ^{\frac{3}{2}}\gamma^2)
\end{equation*}
since $\rho (y) \asymp \bigl(W(y)+\nu)^{\frac{3}{2}}$ and
$|\nabla W|\asymp \bar{r}^{-5}$; then this term does not exceed
$C(\bar{r}^{-\frac{15}{2}}\gamma^{\frac{7}{2}}+
\upsilon ^{\frac{3}{2}}\gamma^2)\bar{r}\Theta$.

Adding to (\ref{25-6-10}) we get
\begin{equation*}
C\Bigl(\gamma^{-\frac{1}{2}}\bar{Q}^{\frac{1}{2}} +
CZ^{\frac{2}{3}} \gamma^{-2}
+\bar{r}^{-\frac{15}{2}}\gamma^{\frac{7}{2}} +
\upsilon ^{\frac{3}{2}}\gamma^2\Bigr) \bar{r}\Theta.
\end{equation*}
Optimizing with respect to  $\gamma=\bar{Q}^{\frac{1}{5}}\upsilon^{-\frac{3}{5}}$
we get $C\upsilon^{\frac{3}{10}}\bar{Q}^{\frac{2}{5}}\bar{r}\Theta$ and (\ref{25-6-8}) becomes
\begin{multline}
(\I_N+\nu) \int \ell(x) \rho_\Psi (x)\theta (x)\,dx \le\\
\int \theta (x)\bigl(W(x)+\nu\bigr)\ell (x)\rho _\Psi (x)\,dx  +
C \upsilon^{\frac{3}{10}} Q^{\frac{2}{5}}\bar{r}\Theta \le\\
C\bigl(\upsilon + \upsilon^{\frac{3}{10}} \bar{Q}^{\frac{2}{5}}\bigr)\bar{r}\Theta .
\label{25-6-11}
\end{multline}
where we took into account that $V-|x|^{-1}*\rho =W$, and that
$W+\nu \le \upsilon $ on $\supp(\theta)$.


Since a factor at $(\I_N+\nu) $ in the left-hand expression of (\ref{25-6-11})  is obviously $\asymp \bar{r}\Theta$ we arrive to
$(\I_N+\nu)  \le  C\upsilon + \upsilon^{\frac{3}{10}} \bar{Q}^{\frac{2}{5}}$.

Recalling definition (\ref{25-6-4}) of $\upsilon$ we arrive to an upper estimate  in Theorem~\ref{thm-25-6-3} below:
\begin{equation}
\I_N +\nu \le C\upsilon = CQ^{\frac{1}{6}}(Z-N)^{\frac{17}{18}}.
\label{25-6-12}
\end{equation}
Really, this is true if $Q\ge Z^{\frac{5}{3}}$ (we are interested in the general approach) and in this case
$\upsilon \ge \upsilon^{\frac{3}{10}} Q^{\frac{2}{5}}$. On the other hand, if $Q=Z^{\frac{5}{3}}(aZ^{\frac{1}{3}})^{-\delta}$, $aZ^{\frac{1}{3}}\ge 1$, then we get an extra term $CQ^{\frac{1}{20}}(Z-N)^{\frac{1}{5}}Z^{\frac{2}{3}}$ but we can skip it decreasing unspecified exponent $\delta>0$ in the definition of $Q$.

\begin{remark}\label{rem-25-6-1}
\begin{enumerate}[label=(\roman*), wide, labelindent=0pt]
\item\label{rem-25-6-1-i}
We will prove the same estimate from below in the next Subsection~\ref{sect-25-6-2}.

\item\label{rem-25-6-1-ii}
Note that the relative error in the estimates is
$\upsilon/\nu = (Z-N)^{-\frac{7}{2}}Q^{\frac{1}{6}}$.

\item\label{rem-25-6-1-iii}
In the proof we used not assumption (\ref{25-6-2}) itself but its corollary (\ref{25-6-3}). If we do not have such assumption instead of equality (\ref{25-6-4}) we have inequality
$ \Theta^\TF \gtrsim \upsilon^{\frac{5}{2}}\bar{r}^7$ (but probably equality still holds). However we have a problem to estimate $|\Theta - \Theta^\TF|$: namely, we need to estimate the first factor  in the product
$\|\nabla \theta^{\frac{1}{2}}\| \D(\rho-\rho^\TF,\rho-\rho^\TF)^{\frac{1}{2}}$.

Let us select $f$ in (\ref{25-6-1}) such that $|f'|\le cf^{1-\delta/2}$ with arbitrarily small $\delta>0$. Then
\begin{multline*}
b^2\|\nabla \theta^{\frac{1}{2}}\|^2 \le  C\int_\cZ \theta^{1-\delta}\,dx \le
C(\int_\cZ \theta \,dx )^{1-\delta}C(\int_\cZ 1 \,dx )^{\delta} \le\\
C(\Theta^\TF )^{1-\delta}\upsilon ^{-\frac{3}{2}(1-\delta)}\bar{r}^{3\delta}
\end{multline*}
with $\cZ =\supp (\nabla \theta)$ and $\beta=\bar{r}^{\frac{13}{6}}Q^{\frac{1}{6}}$ and an error is less than $\epsilon \Theta^\TF$ provided
\begin{equation*}
\upsilon = C_2\bar{r}^{-\frac{17}{6}}Q^{\frac{1}{6}} \times
(\bar{r} ^7Q)^{-\delta_1}
\end{equation*}
which leads to a marginally larger error.

Estimate $P$ could be done in the same manner but here slight increase of it does not matter.

\item\label{rem-25-6-1-iv}
The same arguments of \ref{rem-25-6-1-iii} could be applied to the proof of the lower estimate in the next subsection despite rather different definition of $\Theta_\Psi$ by (\ref{25-6-18}).

\item\label{rem-25-6-1-v}
In vrirtue of Theorem~\ref{thm-25-6-4} stable molecules do not exist in the free nuclei model as $Z-N\ge CQ^{\frac{3}{7}}$ and in atomic case $\widehat{\I}_N=\I_N$.
\end{enumerate}
\end{remark}

\begin{Problem}\label{Problem-25-6-2}
Consider $f$ such that $|f'|\le f g(f)$ where $ g^{-1}(t)\in \sL^1 $ and further improve remainder estimate without assumption (\ref{25-6-3}).
\end{Problem}

\section{Estimate from below for ionization energy}
\label{sect-25-6-2}

Now let us prove estimate $\I_N +\nu$ from below. Let
$\Psi =\Psi _N(x_1,\ldots, x_N)$ be the ground state for $N$ electrons, $\|\Psi\|=1$; consider an antisymmetric test function
\begin{multline}
\tilde{\Psi}=\tilde{\Psi}(x_1,\ldots, x_{N+1})=
\Psi (x_1,\ldots, x_N)u(x_{n+1})-\\
\sum_{1\le j\le N}
\Psi (x_1,\ldots, x_{j-1},x_{N+1},x_{j+1},\ldots, x_N)u(x_j).
\label{25-6-13}
\end{multline}
Then
\begin{multline*}
\E_{N+1}\|\tilde{\Psi}\| ^2\le
\blangle \mathbf{H}_{N+1}\tilde{\Psi},\tilde{\Psi} \brangle =
N\blangle \mathbf{H}_{N+1}\Psi u,\tilde{\Psi}\brangle =\\[5pt]
N\blangle \mathbf{H}_N\Psi u,\tilde{\Psi}\brangle +
N\blangle H_{V,x_{N+1}}\Psi u,\tilde{\Psi}\brangle +
N\blangle \sum_{1\le i\le N}|x_i-x_{N+1}| ^{-1}\Psi u, \tilde{\Psi}\brangle =\\
\shoveleft{\quad (\E_N-\nu' )\|\tilde{\Psi}\| ^2 +
N\blangle H_{W+\nu' ,x_{N+1}}\Psi u,\tilde{\Psi}\brangle}\\[5pt]
+N\blangle
\bigl(\sum_{1\le i\le N}|x_i-x_{N+1}| ^{-1}- (V-W)(x_{N+1})\bigr) \Psi u, \tilde{\Psi}\brangle
\end{multline*}
and therefore
\begin{multline}
N ^{-1}(\I_{N+1}+\nu')\|\tilde{\Psi}\| ^2\ge
-\blangle H_{W+\nu',x_{N+1}}\Psi u,\tilde{\Psi}\brangle\\
-\blangle \bigl(\sum_{1\le i\le N}|x_i-x_{N+1}| ^{-1} - (V-W)(x_{N+1})\bigr)
\Psi u, \tilde{\Psi}\brangle
\label{25-6-14}
\end{multline}
with $\nu'\ge \nu $ to be chosen later. One can see easily that
\begin{multline}
N ^{-1}\|\tilde{\Psi}\| ^2=\|\Psi \| ^2\cdot \|u\| ^2-\\
N\int \Psi (x_1,\ldots,x_{N-1},x)\Psi ^\dag (x_1,\ldots,x_{N-1},y)
u(y)u^\dag(x) \, dx_1\cdots dx_{N-1}\, dxdy
\label{25-6-15}
\end{multline}
where $ ^\dag$ means a complex or Hermitian conjugation.

Note that every term in the right-hand expression in (\ref{25-6-14}) is the sum of two terms: one with $\tilde{\Psi}$ replaced by
$\Psi (x_1,\ldots,x_N)u(x_{N+1})$ and another with $\tilde{\Psi}$ replaced by $-N\Psi (x_1,\ldots,x_{N-1},x_{N+1})u(x_N)$. We call these terms \emph{direct\/}\index{direct term} and \emph{indirect\/}{indirect term} respectively.

Obviously, in the direct and indirect terms $u$ appears as
$|u(x)| ^2\,dx$ and as $u(x) u ^\dag (y)\,dxdy$ respectively, multiplied by some kernels.

\begin{wrapfigure}[11]{r}[10pt]{5.5truecm}
\begin{tikzpicture}[scale=5]
\clip (1.2,-1.65) rectangle (2.8,-.7);
\draw[very thick] plot [domain=1:2] function {-4/(x**2+1)} node [right] {$-W(x)$};
\draw  (1,-.9)--(2,-.9) node [right] {$\nu '$};
\draw (1,-1.2) --(2,-1.2) node [right] {$\nu$};
\draw [dashed] (1.63,-2)--(1.63,-.8);
\draw [thick, dotted] (1.73,-2)--(1.73,-.8);
\draw [ultra thick, |-> ](1.63,-1.5)--(2.1,-1.5);
\node at (1.9,-1.5) {\colorbox{white}{$\operatorname{supp}\theta$}};
\draw [ultra thick, |-> ](1.73,-1.4)--(2.1,-1.4);
\node at (1.9,-1.4) {\colorbox{white}{$\theta=1$}};
\end{tikzpicture}
\end{wrapfigure}

Recall that $u$ is an arbitrary function. Let us take
$u(x)=\theta^{\frac{1}{2}} (x)\phi_j(x)$ where $\phi _j$ are orthonormal eigenfunctions of $H_{W+\nu }$ and $\theta (x)$ is $\beta$-admissible function supported in $\{x\colon   - \upsilon \ge W(x)+\nu \ge \frac{2}{3} \nu \}$ and equal $1$ in $\{x\colon   |-2\upsilon \ge W(x)+\nu \ge \frac{1}{2}\nu \}$, satisfying (\ref{25-5-11}), and  $\beta=\upsilon \bar{r}^5$.

Let us substitute it into (\ref{25-6-14}), multiply by
$\varphi (\lambda _jL ^{-1})$ and take sum with respect to $j$. We get the same expressions with $|u(x)| ^2\,dx$ and $u(x)u ^\dag(y)\,dxdy$ replaced by $F(x,x)\,dx$ and $F(x,y)\,dxdy$ respectively with
\begin{equation}
F(x,y)=\int \varphi (\lambda L ^{-1})\,d_\lambda e(x,y,\lambda ).
\label{25-6-16}
\end{equation}
Here $\varphi (\tau )$ is a fixed $\sC ^\infty $ non-negative function equal to $1$ as $\tau \le \frac{1}{2}$ and equal to $0$ as $\tau \ge 1$ and
$L =\nu'-\nu =6\upsilon$.

Under described construction and procedures the direct term generated by
$N ^{-1}\|\tilde{\Psi}\| ^2$ is
\begin{gather}
\int \theta (x) \varphi (\lambda L ^{-1})\,d_\lambda e(x,x,\lambda )\,dx
\label{25-6-17}\\
\intertext{and applying the semiclassical approximation we get}
\Theta_\Psi
\coloneqq   \int \varphi (\lambda L ^{-1})\,d_\lambda P'(W+\nu -\lambda )\,dx.
\label{25-6-18}
\end{gather}
Therefore under assumptions (\ref  {25-3-33}) and (\ref{25-6-2})\,\footnote{\label{foot-25-24} Or rather its corollary (\ref{25-6-3}).} the remainder estimate is
$C \hbar^{-1}\bar{r}^2 b^{-2}= C \upsilon^{\frac{1}{2}} \bar{r}^2b^{-1}=
C \upsilon^{-\frac{1}{2}} \bar{r}^{-3}$; one can prove it easily by partition of unity on $\supp (\theta)$ and applying the semiclassical asymptotics with effective semiclassical parameter
$\hbar= 1/(\upsilon^{\frac{1}{2}} b)=\upsilon^{-\frac{3}{2}}\bar{r}^{-5}$.

On the other hand, indirect term generated by $N ^{-1}\|\tilde{\Psi}\| ^2$ is
\begin{multline}
-N\int \theta^{\frac{1}{2}} (x)\theta^{\frac{1}{2}} (y)
\Psi (x_1,\ldots,x_{N-1},x) \Psi ^\dag (x_1,\ldots,x_{N-1},y) \times \\
F(x,y)\, dxdydx_1\cdots dx_{N-1}
\label{25-6-19}
\end{multline}
and since the operator norm of $F(.,.,.)$ is $1$ the absolute value of this term does not   exceed
\begin{multline}
N\int\theta(x) |\Psi (x_1,\ldots,x_{N-1},x)| ^2\,dx =
\int \theta(x) \rho _\Psi (x)\,dx\le\\
\int \theta(x) \rho(x)\,dx + CQ^{\frac{1}{2}} \|\nabla \theta^{\frac{1}{2}}\|
\label{25-6-20}
\end{multline}
where $\rho^\TF=0$ on $\supp (\theta)$ and under assumption (\ref{25-6-2})\,\footref{foot-25-24}
$\|\nabla \theta^{\frac{1}{2}} \|\asymp b^{-\frac{1}{2}}\bar{r}\asymp \zeta^{-\frac{1}{2}}\bar{r}^{-\frac{3}{2}}$.

Recall that $P' (W ^\TF+\nu )=\rho ^\TF$. We will take $\nu'=\nu + L$ large enough to keep $\Theta_\Psi$ larger than all the remainders including those due to replacement $W$ by $W ^\TF$ and $\rho$ by $\rho ^\TF$ in the expression above. One can see easily that
\begin{equation}
\Theta_\Psi \asymp \hbar^{-3}\times b^{-2}\bar{r}^2 \asymp \upsilon^{\frac{3}{2}}b\bar{r}^2 \asymp
\upsilon ^{\frac{5}{2}}\bar{r}^7.
\label{25-6-21}
\end{equation}

Therefore
\begin{claim}\label{25-6-22}
If $\upsilon = C_0 \bar{r}^{-\frac{17}{6}}Q^{\frac{1}{6}}$ and
$Z-N\ge C_0 Z^{\frac{5}{7}}$ the total expression generated by
$N ^{-1}\|\tilde{\Psi}\| ^2$ is greater than $\epsilon \Theta $ with
$\Theta = \upsilon^{\frac{5}{2}}\bar{r}^7$.
\end{claim}

Now let us consider direct terms in the right-hand expression of (\ref{25-6-14}). The first of them is
\begin{multline}
-\int \theta^{\frac{1}{2}} (x)\varphi (\lambda L ^{-1})\,
d_\lambda \bigl(H_{W+\nu',x}\theta^{\frac{1}{2}}(x)
e(x,y,\lambda )\bigr)_{y=x}\,dx=\\
\shoveleft{-\int \theta (x)\varphi (\lambda L ^{-1})\,d_\lambda
\bigl(H_{W+\nu',x}e(x,y,\lambda )\bigr)_{y=x}\,dx}\\
\shoveright{-\frac{1}{2}\int \varphi (\lambda L ^{-1})
[[H_W,\theta^{\frac{1}{2}}  ],\theta ^{\frac{1}{2}} ] \,
d_\lambda e(x,x,\lambda )\ge }\\
\int \theta (x) (\nu'-\nu -\lambda ) \varphi (\lambda L ^{-1})
\,d_\lambda e(x,x,\lambda )\,dx
-C\int |\nabla \theta^{\frac{1}{2}} |^2 e(x,x,\nu')dx.
\label{25-6-23}
\end{multline}
Note that the absolute value of last term in the right-hand expression of (\ref{25-6-23}) does not exceed
$Cb^{-1}\bar{r}^2 L^{\frac{3}{2}}\asymp C\upsilon^{\frac{1}{2}}\bar{r}^{-3} \ll C\upsilon \Theta$.

The second direct term in the right-hand expression is
\begin{multline}
-\int \theta (x)\Bigl(\rho _\Psi * |x| ^{-1}-(V-W)(x)\Bigr)F(x,x)\,dx=\\ -\D\bigl(\rho _\Psi - \bar{\rho}, \theta (x)F(x,x)\bigr) \ge \\
-C\D\bigl(\rho _\Psi - \rho,\rho _\Psi -\rho\bigr)^{\frac{1}{2}} \cdot \D\Bigl(\theta^{\frac{1}{2}} F(x,x ),
\theta^{\frac{1}{2}} F(x,x ))\Bigr) ^{\frac{1}{2}}\ge
-CQ ^{\frac{1}{2}}\bar{r}^{\frac{15}{2}} \upsilon ^{\frac{5}{2}}
\label{25-6-24}
\end{multline}
provided $V-W=|x| ^{-1}*\rho$ with
$\D(\rho-\rho ^\TF,\rho-\rho ^\TF)\le CQ$; the absolute value of this term is $\ll \upsilon \Theta$.

Further, the first indirect term in the right-hand expression of (\ref{25-6-14})~is
\begin{multline}
-N\int \theta^{\frac{1}{2}} (y)\Psi (x_1,\ldots,x_{N-1},x) \Psi ^\dag (x_1,\ldots,x_{N-1},y) \times \\
\shoveright{\varphi (\lambda L ^{-1})\,d_\lambda
\bigl(H_{W+\nu',x}\theta^{\frac{1}{2}} (x)e(x,y,\lambda )\bigr)\,
dxdydx_1\cdots dx_{N-1}=}\\
\shoveleft{-N\int \theta^{\frac{1}{2}}(y)\theta^{\frac{1}{2}}(x)
\Psi (x_1,\ldots,x_{N-1},x)
\Psi ^\dag (x_1,\ldots,x_{N-1},y) \times }\\
\shoveright{\varphi (\lambda L ^{-1})(\nu'-\nu -\lambda )\,d_\lambda e(x,y,\lambda )\, dxdydx_1\cdots dx_{N-1}}\\
\shoveleft{-N\int \theta^{\frac{1}{2}}(y) \Psi (x_1,\ldots,x_{N-1},x)
\Psi ^\dag (x_1,\ldots,x_{N-1},y) \times }\\
\varphi (\lambda L ^{-1})[H_{W,x},\theta^{\frac{1}{2}}(x)]\,
d_\lambda e(x,y,\lambda )\,dxdydx_1\cdots dx_{N-1}.
\label{25-6-25}
\end{multline}
Note that one can rewrite the sum of the first terms in the right-hand expressions in (\ref{25-6-23}) and (\ref{25-6-25}) as
$\sum_j \varphi (\lambda _jL ^{-1}) (\nu '-\nu -\lambda_j)\|\hat{\Psi}_j\| ^2$
with
\begin{equation*}
\hat{\Psi}_j(x_1,\ldots,x_{N-1})\coloneqq
\int \Psi(x_1,\ldots,x_{N-1},x) \theta^{\frac{1}{2}}(x)\phi_j(x)\,dx
\end{equation*}
 and therefore this sum  is non-negative.

One can prove easily that the absolute value of the second term in (\ref{25-6-25}) is less than
\begin{equation*}
C\upsilon ^{\frac{1}{2}}b^{-1}\int\rho _\Psi (y)\theta^{\frac{1}{2}} (y)\,dy \le
C\upsilon^{-\frac{1}{2}}\bar{r}^{-5}\Theta \ll \upsilon \Theta.
\end{equation*}
Therefore
\begin{claim}\label{25-6-26}
The sum of the first direct and indirect terms in the right-hand expression of (\ref{25-6-14}) is greater than $-C\upsilon \Theta$.
\end{claim}

Finally, we need to consider the second indirect term generated by the right-hand expression of (\ref{25-6-14})
\begin{multline}
-\int \Bigl(\sum_{1\le i\le N}|y-x_i| ^{-1}-(V-W)(y)\Bigr)\times\\[5pt]
\shoveright{\Psi (x_1,\ldots, x_N)\Psi ^\dag(x_1,\ldots,x_{N-1},y)
\theta^{\frac{1}{2}} (x_N)\theta^{\frac{1}{2}} (y)F(x_N,y)
\,dx_1\cdots dx_Ndy=}\displaybreak\\
\shoveleft{ -\int \Bigl(|y| ^{-1}*\rho _{\underline{x}}(y) -(V-W)(y)\Bigr)
\Psi (x_1,\ldots, x_N)\Psi ^\dag(x_1,\ldots,x_{N-1},y) \times}\\
\shoveright{\theta^{\frac{1}{2}} (x_N)\theta^{\frac{1}{2}} (y) F(x_N,y)\,dx_1\cdots dx_Ndy}\\[5pt]
\shoveleft{-\int \Bigl(\sum_{1\le i\le N}|y-x_i| ^{-1}-
|y| ^{-1}*\rho _{\underline{x}}(y)\Bigr)
\Psi (x_1,\ldots, x_N)\Psi ^\dag(x_1,\ldots,x_{N-1},y) \times}\\
\theta^{\frac{1}{2}} (x_N)\theta^{\frac{1}{2}} (y)F(x_N,y)\,dx_1\cdots\, dx_Ndy;
\label{25-6-27}
\end{multline}
recall that
$\rho _{\underline{x}}$ is a smeared density, $\underline{x}=(x_1,\ldots, x_N)$.

Since $|y| ^{-1}*\rho _{\underline{x}}(y) -(V-W)(y)=
|y|^{-1}*(\rho _{\underline{x}} -\rho)$, the first term in the right-hand expression is equal to
\begin{multline}
\int \theta^{\frac{1}{2}} (x_N)\Psi (x_1,\ldots, x_N)\times \\
\D_y\Bigl( \rho _{\underline{x}} (y) -\rho (y) , F(x_N,y,\lambda )
\theta^{\frac{1}{2}} (y) \Psi (x_1,\ldots,x_{N-1},y) \Bigr)\,dx_1\cdots dx_N;
\label{25-6-28}
\end{multline}
and its absolute value does not exceed
\begin{multline}
\biggl(N\int \D\bigl(\rho _{\underline{x}} (\cdot)-\rho(\cdot),
\rho _{\underline{x}}(\cdot)- \rho(\cdot)\bigr) |\Psi (x_1,\ldots, x_N)|^2
\theta (x_N)\,dx_1\cdots dx_N\biggr) ^{\frac{1}{2}}\times \\
\shoveleft{N^{-\frac{1}{2}}
\biggl(\D_y\Bigl(F(x_N,y,\lambda )\theta^{\frac{1}{2}} (y)\Psi(x_1,\ldots,x_{N-1},y),} \\
F(x_N,y,\lambda ) \theta^{\frac{1}{2}} (y)\Psi (x_1,\ldots,x_{N-1},y) \Bigr)
\,dx_1\cdots dx_N \biggr) ^{\frac{1}{2}}.
\label{25-6-29}
\end{multline}

Due to  estimate (\ref{25-5-24}) and definition (\ref{25-5-4}) as the first factor in (\ref{25-6-29}) does not exceed
$\bigl((Q+T+\varepsilon^{-1}N)\Theta +P\bigr)^{\frac{1}{2}}$
where we assume that $\varepsilon \le Z^{-\frac{2}{3}}$ and
$\Theta\asymp b^{-\frac{1}{2}}Q^{\frac{1}{2}}\bar{r}$ is now an upper estimate for $\int \theta (y)\rho_\Psi (y)\,dy$-like expressions; due to our choice of $\upsilon$ it coincides with
$\Theta =\upsilon^{\frac{5}{2}}\bar{r}^7$.

Then according to (\ref{25-5-25}) $P\asymp  Cb^{-2}\Theta\ll Q\Theta $
and according to (\ref{25-5-23}) $T\ll Q$ and therefore in all such inequalities we may skip $P$ and $T$ terms; so we get
$C(Q+\varepsilon^{-1}N)^{\frac{1}{2}}\Theta^{\frac{1}{2}}$.

Meanwhile  the second factor in (\ref{25-6-29}) (without square root) is equal to
\begin{multline*}
N^{-1}\int L^{-2}\varphi'(\lambda L^{-1})\varphi'(\lambda' L^{-1})
|y-z|^{-1} \underbracket{e(x_N,y,\lambda)}\theta^{\frac{1}{2}} (y)\Psi(x_1,\ldots,x_{N-1},y)\times\\
\underbracket{e(x_N,z,\lambda')}\theta^{\frac{1}{2}} (z)\Psi^\dag (x_1,\ldots,x_{N-1},z)
\,dy dz\,dx_1\cdots dx_{N-1} \underbracket{dx_N}\, d\lambda d\lambda';
\end{multline*}
after integration by $x_N$  we get instead of marked terms
$e(y,z,\lambda)$ (recall that $e(.,.,.)$ is the Schwartz kernel of projector and we keep $\lambda<\lambda'$) and then integrating with respect to $\lambda'$ we arrive to
\begin{multline*}
N^{-1}\int
|y-z|^{-1} F(y,z) \theta^{\frac{1}{2}} (y)\Psi(x_1,\ldots,x_{N-1},y)\times \\
\theta^{\frac{1}{2}} (z)\Psi^\dag (x_1,\ldots,x_{N-1},z)
\,dy dz\,dx_1\cdots dx_{N-1}
\end{multline*}
where now $F$ is defined by (\ref{25-6-16}) albeit with $\varphi^2$ instead of $\varphi$. This latter expression does not exceed
\begin{multline}
N^{-1}\iint
|y-z|^{-1} |F(y,z)| \theta^{\frac{1}{2}} (y)|\Psi(x_1,\ldots,x_{N-1},y)|^2 \times\\
\,dy dz\,dx_1\cdots dx_{N-1}.
\label{25-6-30}
\end{multline}
Then due to proposition \ref{prop-25-A-3} expression $\int |y-z|^{-1}|F(y,z)|\,dz$  does not exceed $Cb^{-1}\hbar^{-1}\asymp \upsilon^{\frac{1}{2}}$, and thus expression (\ref{25-6-30}) does not exceed
$CZ^{-2} \upsilon^{\frac{1}{2}} \Theta$ and therefore the second factor in (\ref{25-6-29}) does not exceed
$CN^{-1}\upsilon^{\frac{1}{4}} \Theta^{\frac{1}{2}}$ and the whole expression
(\ref{25-6-29}) does not exceed
\begin{equation*}
C(Q+\varepsilon^{-1}N)^{\frac{1}{2}}\Theta^{\frac{1}{2}} \times
N^{-1}\upsilon^{\frac{1}{4}} \Theta^{\frac{1}{2}}=CN^{-1}(Q+\varepsilon^{-1}N)^{\frac{1}{2}}\upsilon^{\frac{1}{4}}\Theta
\end{equation*}
and then
\begin{claim}\label{25-6-31}
As $\varepsilon \ge Z^{-1}\upsilon^{-\frac{3}{2}}$ the first term in the  right-hand expression of (\ref{25-6-27}) does not exceed $C\upsilon \Theta$.
\end{claim}
Further, we need to estimate the second term in the right-hand expression of (\ref{25-6-27}). It can be rewritten in the form
\begin{multline}
\sum_{1\le i\le N} \int U(x_i,y)
\Psi (x_1,\ldots, x_N)\Psi ^\dag(x_1,\ldots,x_{N-1},y)
\theta^{\frac{1}{2}} (x_N)\theta^{\frac{1}{2}} (y)\times\\
F(x_N,y)\,dx_1\cdots dx_Ndy
\label{25-6-32}
\end{multline}
where $U(x_i,y)$ is the difference between potential generated by the charge $\updelta (x-x_i)$ and the same charge smeared; note that $U(x_i,y)$ is supported in $\{(x_i,y)\colon  |x_i-y|\le \varepsilon \}$. Let us estimate the $i$-th term in this sum with $i<N$ first; multiplied by $N(N-1)$, it does not exceed
\begin{multline}
N\biggl(\int |U(x_i,y)| ^2 |\Psi (x_1,\ldots, x_N)| ^2
\theta^{\frac{1}{2}} (x_N)\theta^{\frac{1}{2}} (y) |F(x_N,y)|\,
dx_1\cdots dx_Ndy\biggr) ^{\frac{1}{2}}\times \\
 N\biggl(\int \omega (x_i,y) |\Psi (x_1,\ldots, x_{N-1},y)| ^2
\theta^{\frac{1}{2}} (x_N)\theta^{\frac{1}{2}} (y)|F(x_N,y)|
\,dx_1\cdots dx_Ndy\biggr) ^{\frac{1}{2}};
\label{25-6-33}
\end{multline}
here $\omega $ is $\varepsilon $-admissible and supported in
$\{(x_i,y)\colon   |x_i-y|\le 2\varepsilon \}$ function. Due to Proposition~\ref{prop-25-A-3} in the second factor
$\int \theta^{\frac{1}{2}} (x_N)|F(x_N,y)|\,dx_N\le C$ and therefore the whole second factor does not exceed
\begin{equation}
C\Bigl(\underbracket{\int\theta^{\frac{1}{2}} (x)\omega (x,y)
\rho _\Psi^{(2)} (x,y)\,dx dy} \Bigr) ^{\frac{1}{2}}
\label{25-6-34}
\end{equation}
where we replaced $x_i$ by $x$. According to Proposition~\ref{prop-25-5-1} in the selected expression one can replace $\varrho^{(2)}_\Psi (x,y)$ by $\rho_\Psi(x)\rho(y) $ with an error which does not exceed
\begin{equation*}
C\Bigl(\sup_x \|\nabla _y\chi _x\|_{\sL^2(\bR ^3)}
\bigl( Q +\varepsilon ^{-1}N  \bigr)^{\frac{1}{2}}
+C\varepsilon N\|\nabla _y\chi \|_{\sL ^\infty }\Bigr)\Theta
\end{equation*}
which as we plug
$\sup_x \|\nabla _y\chi _x\|_{\sL^2(\bR ^3)}\asymp \varepsilon^{\frac{1}{2}}$, $\|\nabla _y\chi \|_{\sL ^\infty }\asymp \varepsilon^{-1}$ becomes
$CN\Theta$.
Meanwhile, consider
\begin{equation}
\int |U(x_i,y)|^2 \theta ^{\frac{1}{2}}(y) |F(x_N,y)|\,dy.
\label{25-6-35}
\end{equation}
 Again due to Proposition~\ref{prop-25-A-3} it does not exceed
\begin{equation*}
C\upsilon^{\frac{3}{2}}\int |U(x_i,y)|^2 \theta ^{\frac{1}{2}}(y) \bigl(|x_N-y|\upsilon^{\frac{1}{2}}+1\bigr)^{-s}\,dy
\end{equation*}
and this integral should be taken over $B(x_i,\varepsilon)$, with
$|U(x_i,y)|\le |x_i-y|^{-1}$, so (\ref{25-6-35}) does not exceed
$C\varepsilon   \upsilon^{\frac{3}{2}} \omega' (x_i,x_N)$ with
$\omega '(x,y)=\bigl(1+\upsilon^{\frac{1}{2}}|x-y|\bigr) ^{-s}$ (provided $\varepsilon \le \upsilon^{-\frac{1}{2}}$ which will be the case).
Therefore  the first factor in (\ref{25-6-33}) does not exceed
\begin{equation}
C \varepsilon^{\frac{1}{2}} \upsilon ^{\frac{3}{4}}
\Bigl(\underbracket{\int\theta^{\frac{1}{2}} (x)\omega '(x,y)
\rho _\Psi^{(2)} (x,y)\,dx dy }\Bigr) ^{\frac{1}{2}}.
\label{25-6-36}
\end{equation}
Therefore in selected expression one can replace $\rho _\Psi^{(2)} (x,y)$ by $\rho_\Psi(x)\rho(y) $ with an error which does not exceed what we got before but with $\varepsilon$ replaced by $\upsilon^{-\frac{1}{2}}$, i.e. also $CN\Theta$.

However in both selected expressions replacing $\rho _\Psi^{(2)} (x,y)$ by $\rho_\Psi(x)\rho(y) $ we get just $0$. Therefore expression (\ref{25-6-33}) does not exceed $C\varepsilon ^{\frac{1}{2}}\upsilon^{\frac{3}{4}}Z\Theta$ which does not exceed $C\upsilon \Theta$ provided
$\varepsilon \le C\upsilon^{\frac{1}{2}}Z^{-2}$.

So, we have two restriction to $\varepsilon$ from above: the last one and $\varepsilon\le Z^{-\frac{2}{3}}$ and one can see easily that both of them are  compatible with with restriction to $\varepsilon$ in (\ref{25-6-31}).

Finally, consider term in (\ref{25-6-32}) with $i=N$ (multiplied by $N$):
\begin{equation}
N \int U(x_N,y) |\Psi (x_1,\ldots, x_N)|^2
\theta^{\frac{1}{2}} (x_N)\theta^{\frac{1}{2}} (y)
F(x_N,y)\,dx_1\cdots dx_Ndy;
\label{25-6-37}
\end{equation}
due to Cauchy inequality it does not exceed
\begin{multline}
N \Bigl(\int |x_N-y |^{-2}|\Psi (x_1,\ldots, x_N)|^2
\theta^{\frac{1}{2}} (x_N)\theta^{\frac{1}{2}} (y)\,\,dx_1\cdots dx_Ndy\Bigr)^{\frac{1}{2}}\times \\
N \Bigl(\int |F(x_N,y)|^2|\Psi (x_1,\ldots, x_N)|^2
\theta^{\frac{1}{2}} (x_N)\theta^{\frac{1}{2}} (y)\,\,dx_1\cdots dx_Ndy\Bigr)^{\frac{1}{2}}
\label{25-6-38}
\end{multline}
where both integrals are taken over $\{|x_N-y|\le \varepsilon\}$ and integrating with respect to $y$ there we get that it does not exceed
\begin{equation*}
C\varepsilon ^{\frac{1}{2}}\Theta^{\frac{1}{2}} \times \upsilon^{\frac{3}{4}}\varepsilon^{\frac{3}{2}}\Theta^{\frac{1}{2}}=
C\upsilon^{\frac{3}{4}}\varepsilon^2 \Theta \ll \upsilon \Theta.
\end{equation*}
Therefore the right-hand expression in (\ref{25-6-14}) is $\ge -C \upsilon \Theta$ and recalling that $\nu'-\nu = O(\upsilon)$ we recover an lower estimate in Theorem~\ref{thm-25-6-3} below:
\begin{equation}
\I_N +\nu \ge -C\upsilon = CQ^{\frac{1}{6}}(Z-N)^{\frac{17}{18}}.
\label{25-6-39}
\end{equation}

Combining with a lower estimate (\ref{25-6-12}) and recalling estimate (\ref{25-4-5}) for $Q$ we arrive to

\begin{theorem}\label{thm-25-6-3}
 Let condition \textup{(\ref  {25-3-33})}  be fulfilled and let
 $N\le Z-C_0 Q^{\frac{3}{7}}$ with $Q\le C_1 Z^{\frac{5}{3}}$.

Then in the framework of fixed nuclei model under assumption \textup{(\ref{25-6-2})}
\begin{equation}
|\I_N +\nu |\le C (Z-N)^{\frac{17}{18}}Z^{\frac{5}{18}}
\left\{\begin{aligned}
&1 \qquad &&\text{as\ \ } a\le Z^{-\frac{1}{3}},\\
&Z^{-\delta}+ (aZ^{\frac{1}{3}})^{-\delta}
&&\text{as\ \ } a\ge Z^{-\frac{1}{3}}.
\end{aligned}\right.
\label{25-6-40}
\end{equation}
\end{theorem}

\section{Estimate for excessive positive charge}
\label{sect-25-6-3}

To estimate excessive positive charge when molecules can still exist in free nuclei model we apply arguments of section 5 of B.~Ruskai and J.~P.~Solovej~\cite{ruskai:solovej}. In view of Theorem~\ref{thm-25-4-14} it is sufficient to consider the case
\begin{equation}
a=\min_{j<k} |\y_j-\y_k|\ge C_0\bar{r}.
\label{25-6-41}
\end{equation}
Therefore in Thomas-Fermi theory $\rho^\TF$ is supported in separate ``atoms''.

Let us consider $a$-admissible functions $\theta _m(x)$, supported in
$B(\y_m, \frac{1}{3}a)$ for $m=1,\ldots , M$ and in
$\{ |x-\y_m|\ge \frac{1}{4} a\ \forall m=1,\ldots , M\}$ for $m=0$, such that
\begin{equation}
\theta _0 ^2+\ldots +\theta _M ^2=1.
\label{25-6-42}
\end{equation}
Then for the ground state $\Psi $
\begin{equation}
\E_N=\blangle \mathbf{H}\Psi ,\Psi \brangle =
\sum_\alpha \blangle \theta _\alpha \mathbf{H}\theta _\alpha \Psi ,\Psi \brangle -
\sum _{\alpha,j} \|(\nabla _j\theta _\alpha) \Psi \| ^2
\label{25-6-43}
\end{equation}
with the sum over of $(M+1)$-cluster decompositions
$\alpha =(\alpha _0,\ldots,\alpha _M)$ of $\{1,\ldots,N\}$ and
$\theta _\alpha (\underline{x})=
\prod_{0\le m\le M}\prod_{i\in \alpha _m} \theta _m(x_i)$; $j=1,\ldots, N$. Then for any given $\alpha $
\begin{equation}
\mathbf{H}=\sum _{0\le m\le M} \mathbf{H}_{\alpha _m}+J_\alpha
\label{25-6-44}
\end{equation}
with the \index{Hamiltonian!cluster}\emph{cluster Hamiltonians\/}  $\mathbf{H}_{\alpha _m}$, involving only potential of $m$-th nucleus (no nucleus potential as $m=0$) and only electrons belonging to $\alpha _m$ and therefore satisfying
\begin{equation}
\mathbf{H}_{\alpha _m}\ge \E_{\mathbf{at}} (N_m(\alpha ),Z_m),\qquad
\mathbf{H}_{\alpha _0}\ge 0,
\label{25-6-45}
\end{equation}
and with the \index{Hamiltonian!intercluster}\emph{intercluster Hamiltonian\/} (actually, just potentials)
\begin{multline}
J_\alpha =\sum_{1\le m\le M}\sum_{i\notin \alpha _m}-Z_m|x_i-\y_m| ^{-1}\\
+\sum_{m<l}\sum_{i\in \alpha _m,j\in \alpha _l}|x_i-x_j| ^{-1}+
\sum_{m<l} Z_mZ_l|\y_l-\y_m| ^{-1}.
\label{25-6-46}
\end{multline}
Let us note that
\begin{equation}
\sum_\alpha \theta _\alpha ^2 J_\alpha = \sum_{0\le m<l\le M}J_{ml}
\label{25-6-47}
\end{equation}
with $J_{ml}$ given by (32)--(33) of Ruskai--Solovej~\cite{ruskai:solovej} if $m,l>0$ and $m=0$ respectively:
\begin{multline}
J_{ml}=Z_mZ_l|\y_m-\y_l| ^{-1} - Z_m\sum_i \theta _l(x_i) ^2|x_i-\y_m| ^{-1}-\\
Z_l\sum_i \theta _m(x_i) ^2|x_i-\y_l| ^{-1} +
\sum_{i\ne j}\theta _m(x_i) ^2\theta _l(x_j) ^2|x_i-x_j| ^{-1},
\label{25-6-48}
\end{multline}
and
\begin{equation}
J_{0l} = \sum_i \theta _0(x_i) ^2
\Bigl(-Z_l|x_i-\y_l| ^{-1}+ \sum_j \theta _l(x_j) ^2|x_i-x_j| ^{-1}\Bigr).
\label{25-6-49}
\end{equation}
Then we recover (35) of Ruskai--Solovej~\cite{ruskai:solovej}
\begin{multline}
\blangle J_{ml}\Psi ,\Psi \brangle = Z_mZ_l|\y_m-\y_l| ^{-1}-
Z_l\int \rho _\Psi (x)\theta _m(x) ^2|x-\y_l| ^{-1}\,dx -\\
Z_m\int \rho _\Psi (x)\theta _l(x) ^2|x-\y_m| ^{-1}\,dx+
\int \rho _\Psi ^{(2)}(x,y)\theta _m(x)\theta _l(y)\,dx dy.
\label{25-6-50}
\end{multline}
Applying Proposition~\ref{prop-25-5-1} and estimate (\ref{25-4-55}) (replacing first $\theta _m$ with $m=1,\ldots, M$ by $\tilde{\theta}_m$ supported in $B(\y_m, c\bar{r})$ and estimating an error), we conclude that
\begin{multline}
\int \rho _\Psi (x)\theta _m(x) ^2|x-\y_l| ^{-1}\,dx=\\
\Bigl(\int \rho ^\TF (x)\theta _m(x) ^2|x-\y_l| ^{-1}\,dx + O(Y)\Bigr)|\y_m-\y_l|^{-1},
\label{25-6-51}
\end{multline}
\begin{gather}
\int \rho _\Psi (x)\tilde{\theta}_m(x) ^2\,dx= N_m ^\TF + O(Y),
\label{25-6-52}\\
\shortintertext{with}
N_m ^\TF=\int \rho ^\TF (x) \theta _m(x) ^2\,dx,
\qquad Y\coloneqq   Q ^{\frac{1}{2}}\bar{r}^{\frac{1}{2}}
\label{25-6-53}\\
\intertext{(compare with (36)--(37) of Ruskai--Solovej~\cite{ruskai:solovej}) which yields}
\int \rho _\Psi (x)\bigl(1-\sum_{1\le m\le M} \tilde{\theta}_m(x)\bigr)\, dx
\le CY,
\label{25-6-54}
\end{gather}
and we prove that (\ref{25-6-52}) holds for $\theta _m$ as well (compare with (38) of Ruskai--Solovej~\cite{ruskai:solovej}).

The last term in (\ref{25-6-51}) is estimated by Proposition~\ref{prop-25-5-1} and estimate (\ref{25-4-55}) and the same replacement trick:
\begin{multline}
\int \rho _\Psi ^{(2)}(x,y)\theta _m(x) ^2\theta _l(y) ^2|x-y| ^{-1}\,dxdy\ge \\
\int \rho _\Psi ^{(2)}(x,y)\tilde{\theta}_m(x) ^2\theta _l(y) ^2
|x-y| ^{-1}\, dxdy \ge \\
\int \rho ^\TF (x)\rho _\Psi (y)\tilde{\theta}_m(x) ^2\theta _l(y) ^2
|x-y| ^{-1}\, dxdy-\\
C\Bigl( Q^{\frac{1}{2}}\int \rho _\Psi (x) \theta _l (x) ^2dx +
Y a ^{-1}\Bigr)|\y_m-\y_l| ^{-1}
\label{25-6-55}
\end{multline}
and repeating the same trick we get that it is larger than
\begin{gather}
\int \rho ^\TF(x)\rho ^\TF(y) |x-y| ^{-1}dxdy - C(Z-N)Ya ^{-1}-CY ^2a ^{-1}.
\label{25-6-56}\\
\intertext{Then we conclude that}
\blangle J_{ml}\Psi ,\Psi \brangle \ge J_{ml} ^\TF - CNYa ^{-1}
\label{25-6-57}
\end{gather}
with
\begin{multline}
J_{ml} ^\TF =
\int \rho ^\TF(x)\rho ^\TF(y)\theta _m(x)\theta _l(y)|x-y| ^{-1}\, dxdy \\
- Z_m\int \rho ^\TF(x)\theta _l(y)|x-\y_m| ^{-1}\,dx
- Z_l\int \rho ^\TF(x)\theta _m(y)|x-\y_l| ^{-1}\, dx \\
+ Z_mZ_l|\y_m-\y_l| ^{-1}
\label{25-6-58}
\end{multline}
and
\begin{equation}
|\blangle J_{0l}\Psi ,\Psi \brangle |\le C(Z-N) Y a ^{-1}+CY^2a ^{-1}
\label{25-6-59}
\end{equation}
(compare with (39)--(40) of Ruskai--Solovej~\cite{ruskai:solovej}) provided
\begin{multline}
\Bigl|\int \rho ^\TF (y) |x-y| ^{-1} \tilde{\theta}_m(x)\, dy-
N_m ^\TF|x-\y_m| ^{-1}\Bigr|\le\\
C(Z-N) |x-\y_m| ^{-1}
\label{25-6-60}
\end{multline}
for $|x-\y_m|\ge C\bar{r}$.

Let us note that the absolute value of the last term in the right-hand expression of (\ref{25-6-43}) does not exceed $Ca ^{-2}Y$ due to (\ref{25-6-52}). Now stability condition yields
\begin{equation}
J=\sum_{0\le m<l\le M} J_{ml} \le CYa ^{-2}+C(Z-N)Ya ^{-1}+C Y^2 a^{-1}.
\label{25-6-61}
\end{equation}
This inequality, (\ref{25-6-41}) and Proposition~\ref{prop-25-6-6} below yield that  $Z-N\le CY = C \bar{r}^{\frac{1}{2}} Q^{\frac{1}{2}}$.  Since $\bar{r}\asymp (Z-N)^{-\frac{1}{3}}$ we arrive to  $(Z-N)\le CQ^{\frac{3}{7}}$:

\begin{theorem}\label{thm-25-6-4}
 Let condition \textup{(\ref  {25-3-33})} be fulfilled. Then in the framework of free nuclei model with $M\ge 2$ the stable molecule does not exist unless
\begin{equation}
Z-N \le Z^{\frac{5}{7}-\delta}.
\label{25-6-62}
\end{equation}
\end{theorem}

\begin{remark}\label{rem-25-6-5}
Unfortunately, we do not prove that molecules exist. We are not aware of any
rigorous result of this type in the frameworks of our models.
\end{remark}

\begin{proposition}\label{prop-25-6-6}
Let \textup{(\ref{25-6-41})} be fulfilled. Then inequality \textup{(\ref{25-6-60})} holds and
\begin{equation}
J\ge \epsilon (Z-N) ^2a ^{-1}.
\label{25-6-63}
\end{equation}
\end{proposition}

\begin{proof} Note first that
\begin{multline}
\cE ^\TF \le \cE (\bar{\rho})=
\sum_j \cE (\rho _j ^\TF)+ J ^\TF (\bar{\rho})\le\\
\sum_j \cE ^\TF (\rho _j ^\TF)+C(Z-N) ^2a ^{-1}
\label{25-6-64}
\end{multline}
with $\bar{\rho}=\sum_j \rho _j ^\TF$ while
\begin{gather}
\sum_j \cE (\rho _j ^\TF)\le \sum_j \cE (\theta _j\rho ^\TF);
\label{25-6-65}\\
\shortintertext{then}
J ^\TF (\rho ^\TF)\le C(Z-N) ^2a ^{-1}
\label{25-6-66}\\
\intertext{and using (\ref{25-6-62}) we conclude that}
\D(\rho ^\TF -\bar{\rho},\rho ^\TF -\bar{\rho})\le C(Z-N) ^2a ^{-1}.
\label{25-6-67}
\end{gather}
Further,
\begin{multline}
\Lambda \coloneqq
\sum _j \D(\rho ^\TF \theta _j-\rho _j ^\TF, \rho ^\TF \theta _j-\rho _j ^\TF) \le \\
\D(\rho ^\TF -\bar{\rho},\rho ^\TF -\bar{\rho})+
C\bar{r} a ^{-1}\Lambda
\label{25-6-68}
\end{multline}
and combining with (\ref{25-6-66}) we conclude that
\begin{equation}
\Lambda \le C(Z-N) ^2a ^{-1}
\label{25-6-69}
\end{equation}
due to (\ref{25-6-41}). Combining with
$\rho _j ^\TF *|x| ^{-1}=N_j ^\TF |x-\y_j| ^{-1}$ for $|x-\y_j|\ge r_S$ we arrive to (\ref{25-6-60}). Further,
\begin{equation}
J ^\TF (\rho ^\TF)\ge J ^\TF (\bar{\rho}) -
C\Lambda ^{\frac{1}{2}}\bar{r}{\frac{1}{2}}(Z-N)a ^{-\frac{3}{2}} -
C\Lambda \bar{r} ^{-1}
\label{25-6-70}
\end{equation}
which together with (\ref{25-6-68}) and (\ref{25-6-69}) yields (\ref{25-6-63}).
\end{proof}

\begin{subappendices}
\chapter{Appendices}
\section{%
Electrostatic inequalities}%
\label{sect-25-A-1}

We know already that there are two sources of errors in the lower estimate: due to electrostatic inequality (\ref{25-2-1}) and semiclassical errors. For the first error in the case $\vec{B}=\const$ E.~Lieb, J.~P.~Solovej and J.~Yngvarsson~\cite{LSY2} provide the (almost) perfect estimate; the reader can find the proof based on the magnetic Lieb--Thirring inequality (and this inequality as well) in that paper (p. 122) which in the case of $\vec{B}=0$ becomes

\begin{theorem}\label{thm-25-A-1}
For the ground state $\Psi $ of \textup{(\ref{25-1-1})} with potential \textup{(\ref{25-1-4})}
\begin{equation}
\int \rho _\Psi ^{\frac{4}{3}}\, dx \le
CZ ^{\frac{5}{6}} N ^{\frac{1}{2}} \bigl(Z +N\bigr)^{\frac{1}{3}}
\label{25-A-1}
\end{equation}
otherwise.

In particular for $c ^{-1}N\le Z\le cN$ the right-hand expression does not exceed $CZ ^{\frac{5}{3}}$.
\end{theorem}

On the other hand, for $B=0$ there is a more precise inequality
due to V.~Bach~\cite{bach} and G.~Graf and J.~P.~Solovej~\cite{graf:solovej}:

\begin{theorem}\label{thm-25-A-2}
For the ground state $\Psi $ of \textup{(\ref{25-1-1})} with potential \textup{(\ref{25-1-4})}
\begin{multline}
\blangle \mathbf{H}\Psi ,\Psi \brangle \ge\\
\N_1(A-\nu )+\nu N -\frac{1}{2}\iint |x-y| ^{-1}|e(x,y,\nu) | ^2\, dxdy-
CN ^{\frac{5}{3} -\delta }
\label{25-A-2}
\end{multline}
with some exponent $\delta >0$.
\end{theorem}

We will discuss magnetic field case in more details in the Appendix to the next paper~\ref{book_new-sect-26}.

\section{Hamiltonian trajectories}%
\label{sect-25-A-2}

We are going to prove that for $W=W ^\TF$ in $M=1$ case the generic trajectory on the energy level $\nu $ is not periodic. We use some ideas from V.~Arnold~\cite{arnold:classical}, pages 37--38. Recall that in this case $W=W(r)$ ($r=|x-\y_1|$) and angular momentum $\vec{M}$ is a motion integral. Then any trajectory lies on some plane and if $M=|\vec{M}|>0$ it lies in $\{0<r<\bar{r}\}$ where $W$ is analytic and $W(\bar{r})=-\nu $.

\begin{claim}\label{25-A-3}
Let us assume that all the trajectories on the energy level $\nu $ are periodic.
\end{claim}
Then the rotation number
\begin{equation}
\Phi =
\int_{r_1} ^{r_2} \frac{M\,dr}{r ^2\sqrt{2\bigl(W(r)+\nu \bigr)-M ^2r ^{-2}}}
\label{25-A-4}
\end{equation}
showing the increment of the polar angle over a half-trajectory should be
$\pi k ^{-1}$ with $k\in \bN$ and should not depend on $M$ where
$r_1\le r_2$ are roots of $\bigl(W(r)+\nu \bigr)-M ^2r ^{-2}=\nu $. So,
$\Phi $ should be the same for all trajectories on the energy level $\nu$. One can see easily that for $M\to 0$ \ $\Phi $ tends to those of the Coulomb potential. So, $\Phi =2\pi $ for all trajectories on the energy level $\nu $.

Let $r_0$ be a root of
$F(r)=2\bigl(W(r)+\nu \bigr)+r ^4\bigl(W'(r)\bigr) ^2=0$.
One can see easily that $F({\bar r})>0$, $F(r)\to \infty $ as $r\to 0$ and
$F'(r)<0$ because
\begin{equation}
W'<0, \qquad W''+2rW'=r ^{-1}(r ^2W')')=r \Delta W>0.
\label{25-A-5}
\end{equation}
So, the unique root exists. Then $r=r_0$ is a circular motion with
$M=-r_0 ^3 W'(r_0)$.

Then (V.~Arnold~\cite{arnold:classical}, problem 2 at page 37)
\begin{equation*}
\Phi \to \pi \sqrt{W'/(3W'+rW'') ^{-1}}\bigr|_{r=r_0}=\Phi _0
\end{equation*}
for trajectories tending to circular. However, $3W'+rW''>W'$ due to (\ref{25-A-5}) and then $\Phi _0>\pi $. Contradiction to assumption (\ref{25-A-3}).

\section{Some spectral function estimates}
\label{sect-25-A-3}
\begin{proposition}\label{prop-25-A-3}
For Schr\"odinger operator with $W\in \sC ^\infty $ and for
$\phi \in \sC_0 ^\infty ([-1,1])$ the following estimate holds for any $s$:
\begin{gather}
|F(x,y)|\le C h ^{-3}\bigl(1+h ^{-1}|x-y|\bigr) ^{-s}, \label{25-A-6}\\
F(x,y)\coloneqq   \int \phi (\lambda )\,d_\lambda e(x,y,\lambda ).\label{25-A-7}
\end{gather}
\end{proposition}

\begin{proof} Let
$u(x,y,t)=\int e ^{-ih ^{-1}t\lambda }\,d_\lambda e(x,y,\lambda )$ be the
Schwartz's kernel of $e ^{-ih ^{-1}Ht}$.

Fix $y$. Note first that $\sL ^2$-norm\footnote{\label{foot-25-25} With respect to $x,t$ here and below.} of
$\phi (hD_t)\chi (t)\omega (x)u(x,y,t)$ is less than $Ch ^s$ as
$\chi \in \sC_0 ^\infty ([-\epsilon ,\epsilon ])$ and $\omega \in \sC ^\infty $
supported in $\{|x-y|\ge \epsilon _1\}$ ($\epsilon _1=C\epsilon $) due to the
finite speed of propagation of singularities.

We conclude then that $\sL ^2$-norm of $\phi (hD_t)\chi (t)\omega (x)u(x,y,t)$ is less than $C h ^s$ for $\omega \in \sC ^\infty $ supported in
$\{x\colon  |x-y|\ge C\}$.

Then $\sL ^2$-norm of $\partial _t^l \nabla ^\alpha \phi (hD_t)\chi (t)\omega (x)u$ does not exceed $C h ^s$. Then due to imbedding inequality
$\sL ^\infty $-norm of $\phi (hD_t)\chi (t)\omega (x)u$ does not exceed
$C  h ^s$. Setting $t=0$ and using this inequality and
$ |F(x,y)|\le Ch ^{-3}$ (due to paper~\ref{book_new-sect-4}) we get that $ |F(x,y)|\le Ch^s$ for $|x-y| \ge \epsilon _0$;

Now let us consider general $|x-y|=r\ge Ch$. Rescaling
$x-y \mapsto (x-y)r ^{-1}$ we need to rescale $h\mapsto hr ^{-1}$ and rescaling above inequality and keeping in mind that $F(x,y)$ is a density with respect to $x$ we get $|F(x,y)|\le C h ^sr ^{-3-s}$ which is equivalent to (\ref{25-A-6})--(\ref{25-A-7}).
\end{proof}
\end{subappendices}

\chapter*{Comments}

There are papers of physicists. L.H. Thomas and E. Fermi have suggested in 1927 that a large Coulomb system (atom or molecule) in the ground state looks like a classical gas but with the Pauli principle, so, leading to the first term $\cE^\TF$ in the asymptotics. The second term of asymptotics was conjectured by J.~M.C. Scott in 1952 as a contribution of those electrons which move very close to the nuclei. Next terms, Dirac and Schwinger corrections were conjectured in 1930 and 1980, respectively.

The mathematical rigorous papers one can separated into several  groups:

First, there are papers concerning only the Thomas--Fermi model (so, studying the Thomas-Fermi equation, may be, with some modifications, without any consideration of the quantum mechanical model, even if the latter was a source for the former). Most notably H. Brezis, H. and E. Lieb E. \cite{brezis:lieb},
R. Benguria~\cite{benguria}, R. Benguria, R. and Lieb E. H. \cite{benguria:lieb}.

The second group consists of the papers, justifying Thomas--Fermi model as an approximation to the quantum mechanical model:  E.~H.~Lieb and B. Simon, \cite{lieb:simon}, where the leading term was derived; also certain properties of the the Thomas--Fermi model were established.

Next, W.~Hughes \cite{hughes} and H.~Siedentop and R.~Weikart \cite{sied:wei:1,sied:wei:2,sied:wei:3}  justified the Scott correction term in the atomic case, while V.~Ivrii and M.~Sigal~\cite{ivrii:ground} justified it  in the molecular case.

Then, C.~Fefferman and L.~Seco \cite{feffer:seco} justified Dirac and Schwinger correction terms in the atomic case, while V.~Ivrii justified them in \cite{ivrii:MQT2} (even in the case of the relatively weak magnetic field).

The third group consists of the papers, related to the ground state energy problem: B.~Ruskai and J.~P.~Solovej~\cite{ruskai:solovej}, J.~P.~Solovej~\cite{solovej:neutrality} and L.~A.~Seco, I.~M.~Sigal,  and J. P. Solovej \cite{SSS}.

Finally, we already mentioned papers which provided the solid functional-analytical base for all this construction. Pretty complete survey could be found in C. L. Fefferman, V. Ivrii,  L. A. Seco,  and I. M. Sigal \cite{feffer:coulomb}.

\input Preprint-25.bbl
\end{document}